\newcommand{\ignore}[1]{}
\documentclass[12pt]{article}
\usepackage{float}
\usepackage[left]{lineno}
\usepackage{amsmath}
\usepackage{amssymb}
\usepackage{bm}
 \usepackage{graphicx}
\usepackage{braket}
\usepackage{authblk}
\usepackage{caption,subcaption}
\usepackage{enumitem}
\usepackage{wrapfig,lipsum,booktabs}
\usepackage[pdftex, pdfstartview={FitH}, pdfnewwindow=true,
colorlinks=false, pdfpagemode=UseNone]{hyperref}
\numberwithin{equation}{section}
\numberwithin{figure}{section}
\graphicspath{{fig/}}
\renewcommand{\baselinestretch}{1.1}
\usepackage{color}

\newcommand{\rcb}[1]{RICH: \textcolor{blue}{  #1 }}
\newcommand{\com}[1]{}

\newcommand{\esw}[1]{}

  \newcommand{\be}{\begin{linenomath*}\begin{equation}}
  \newcommand{\bea}{\begin{eqnarray}}
  \newcommand{\eea}{\end{eqnarray}}
  \newcommand{\ee}{\end{equation}\end{linenomath*}}
  \newcommand{\eeq}{\end{equation}}
  
  \newcommand{\<}{\langle\,}

  \renewcommand{\>}{\rangle}
\newcommand\dd{\partial}
\newcommand\nn{\nonumber \\}
\newcommand\mS{{\mathbb S}}
\newcommand\mR{{\mathbb R}}
\newcommand\mT{{\mathbb T}}

\renewcommand\ell{l}
\newcommand{\ndim}[1]{#1-d}

\newcommand{\figref}[1]{Fig.~\ref{#1}}
\newcommand{\secref}[1]{Sec.~\ref{#1}}
\newcommand{\tabref}[1]{Table~\ref{#1}}
\renewcommand{\eqref}[1]{Eq.~(\ref{#1})}
\newcommand{\eqrefrange}[2]{Eqns.~(\ref{#1}) to~(\ref{#2})}

\title{Lattice $\phi^4$ Field Theory on Riemann Manifolds: Numerical
  Tests for the 2-d Ising CFT on $\mS^2$}
\author[*]{Richard C. Brower}
\author[*]{Michael Cheng}
\author[\#]{George T. Fleming}
\author[\#]{Andrew D. Gasbarro}
\author[+@]{Timothy G. Raben}
\author[+]{Chung-I Tan}
\author[*]{Evan S. Weinberg}
\affil[*]{Boston University, Boston, MA 02215}
\affil[\#]{Yale University, Sloane Laboratory, New Haven, CT 06520} 
\affil[+]{Brown University, Providence, RI 02912}
\affil[@]{University of Kansas, KS 66045}

\begin{document}

\newpage

\ignore{
\rcb{TODO}
\begin{enumerate}
\item  Check all Figure and caption. They are often too verbose!
\item Should the Title be more general: Like 'Lattice $\phi^4$  Theory on Riemann
Manifold: Numerical Tests on 2-d  Riemann Sphere''
\item Please everyone look for missing and relevant references!
\item Change $s$ to $L$?
\item We use $i,j$ for sites on general manifold and  and $x,y$ the sphere.
\item All $U_{2n}$  normalized be 0 for the Gaussian limit and 1 for
  the ordered limit.
\item  Multiple uses of  $V_{ij}$,$V^D_{ij}$ $V_{nn^*}$ is confusing!
\item Change $l$ or l to $\ell$.
\item Check dimensionless variable.. Check on indices $i,j$
\item  Bare paramater $\mu_0$?
\end{enumerate}
}

\date{}

\maketitle

\begin{abstract}
  We present a method for defining a lattice realization of the
  $\phi^4$ quantum field theory on a simplicial complex in order to
  enable numerical computation on a general Riemann manifold. The
  procedure begins with adopting methods from traditional Regge
  Calculus (RC) and finite element methods (FEM) plus the addition of
  ultraviolet counter terms required to  reach the  renormalized
  field theory in the continuum limit.  The construction is tested numerically for
  the two-dimensional $\phi^4$ scalar field theory on the Riemann two-sphere,
  $\mS^2$, in comparison with the exact solutions to the two-dimensional Ising
  conformal field theory (CFT).  Numerical results for the Binder
  cumulants (up to 12th order) and the two- and four-point correlation
  functions are in  agreement with the exact $c = 1/2$ CFT
  solutions.
\end{abstract}

\setlength{\parskip}{0in}
\thispagestyle{empty}
\setcounter{page}{-1}
\pagebreak
\tableofcontents
\thispagestyle{empty}
\setcounter{page}{0}
\pagebreak
\setcounter{page}{1}
\setlength{\parskip}{.2in}
\newpage


\section{\label{sec:intro}Introduction}

Lattice Field Theory (LFT) has proven to be a powerful
non-perturbative approach to quantum field theory~\cite{Appelquist:2013sia}.
However the lattice regulator has generally been restricted to flat Euclidean
space, $\mathbb R^d$, discretized on hypercubic lattices with a
uniform ultraviolet (UV) cut-off $\Lambda_{UV} = \pi/a$ in terms of
the lattice spacing $a$.  Here we propose a new approach to enable
non-perturbative studies  for a  range of problems on curved Riemann
manifolds. There are many applications that benefit from this.
 In the study of  Conformal Field Theory (CFT), it is useful to make a Weyl transform from flat Euclidean space $\mR^d$ to a compact spherical manifold $\mathbb S^{d}$, or in {\em Radial Quantization }\cite{Brower:2012vg}, a Weyl transformation to the
cylindrical boundary, $\mathbb R \times \mathbb S^{d-1}$, of
$AdS^{d+1}$. Other applications that could benefit from extending LFT
to curved manifolds include two-dimensional condensed matter systems such
as graphene sheets~\cite{Brower:2012zd}, 
four-dimensional gauge theories for beyond the standard model (BSM) strong
dynamics~\cite{Appelquist:2016viq,Kuti:2015awa}, and perhaps even quantum
effects in a  space-time near massive systems such as black holes.

Before attempting a non-perturbative lattice construction, one should
ask if a particular renormalizable field theory in flat space is even
perturbatively renormalizable on a general smooth Riemann manifold.
Fortunately, this question was addressed with an avalanche of
important research~\cite{Luscher:1982wf,Jack:1983sk,Jack:1984vj,Buchbinder:1989zz}
in the 70s and 80s.  A rough summary is that any UV
complete field theory in flat space is also perturbatively
renormalizable on any smooth Riemann manifold with diffeomorphism
invariant counter terms corresponding to those in flat
space~\cite{Buchbinder:1989zz}. Taking this as given, in spite of our
limited focus on $\phi^4$ theory, we hope this paper is the beginning
of a more general non-perturbation lattice formulation for these UV
complete theories on any smooth Euclidean Riemann manifolds.

The basic challenge in constructing a lattice on a sphere---or indeed
on any non-trivial Riemann manifold---is the lack of an infinite
sequence of finer lattices with uniformly decreasing lattice spacing~\cite{Brower:2012vg,Brower:2012mn,Brower:2014daa}.  For
example, unlike the hypercubic lattice with toroidal boundary
condition, the largest discrete subgroup of the isometries of a sphere
is the icosahedron in 2d and the hexacosichoron, or 600 cell, in
3d.  This greatly complicates constructing a suitable bare
lattice Lagrangian that smoothly approaches the continuum limit of the
renormalizable quantum field theory when the UV cut-off is
removed. Here we propose a new formulation of LFT
on a sequence of simplicial lattices converging to a general smooth
Riemann manifold. Our strategy is to represent the geometry of the
discrete simplicial manifold using Regge Calculus~\cite{Regge:1961px}
(RC) and the matter fields using the Finite Element Method
(FEM)~\cite{StrangFix200805} and Discrete Exterior Calculus
(DEC)~\cite{2005math8341D,Arnold2006,Miller:2013gy,Crane:2018}.  Together these
methods define a lattice Lagrangian which we conjecture
is  convergent in the classical (or tree) approximation. However, the convergence
fails at the quantum level due to ultraviolet divergences in
the continuum limit.  To remove this quantum obstruction, we compute
counter terms that cancel the ultraviolet defect order by order
in  perturbation theory. We will refer to the resultant
lattice construction as the 
 {\bf Quantum Finite Element (QFE)}   method and give a first numerical test
in 2d on $\mS^2$ at the Wilson-Fisher CFT fixed point.

While our current development of a QFE Lagrangian and numerical 
tests are  carried  out for the simple case of a \ndim{2} scalar $\phi^4$
theory projected on the Riemann sphere we attempt a more general
framework.    Since the map to the
Riemann sphere, $\mR^2 \rightarrow
\mS^2$,  is a  Weyl transform, the CFT is  guaranteed~\cite{Rychkov:2016iqz} to be exactly
equivalent to the $c$ =1/2 Ising CFT in flat space and therefore presents a convenient
and rigorous test of convergence to the continuum theory.  More
general examples can and will be pursued mapping
conformal field theories in
flat Euclidean space $\mathbb{R}^{d+1}$ either to
$\mathbb{R}\times \mathbb{S}^{d}$,  appropriate for radial quantization,
\be 
ds^2_{flat} =\sum^{d+1}_{\mu =
  1} dx^{\mu}dx^{\mu} = e^{2t}(dt^2 + d\Omega^2_{d})
\xrightarrow{Weyl} (dt^2 + d\Omega^2_{d}) \; ,
\label{eq:Radial}
\ee
with $t = \log(r)$ or to the  sphere, $\mS^{d}$, 
\be
ds^2_{flat} =\sum^{d}_{\mu = 1} dx^{\mu}dx^{\mu} = e^{\sigma(x)} d\Omega^2_{d}
\xrightarrow{Weyl} d\Omega^2_{d} \; .
\label{eq:Riemann}
\ee
Current tests of the QFE method for the \ndim{3} Ising model in
radial quantization, $\mR^3 \rightarrow \mR \times \mS^2$, are underway, so we take
the opportunity here to give a brief introduction to both geometries.
By considering the sphere $\mS^d$ as a dimensional reduction of the
cylinder $\mR \times \mS^d$ by taking the length of the cylinder to
zero, both cases are conveniently presented together in
Appendix~\ref{app:FreeTheory}.

 The organization of the article is as
 follows. In~\secref{sec:scalar_fields} we review the basic Regge Calculus/Finite
 Element Method framework as a discrete form of the exterior
 calculus and show its failure for a quantum field theory beyond the
 classical limit due to UV divergences.   The reader is referred to the literature~\cite{StrangFix200805,2005math8341D,Arnold2006,Miller:2013gy} for more details
 and to  Ref.~\cite{Brower:2016vsl} for the extension to Dirac
 fermions.
 In~\secref{sec:quantum_corrections} we address
 the central issue of counter terms in the interacting $\phi^4$ theory 
 required  to restore the isometries on  $\mS^2$ in the continuum
 limit.  \secref{sec:numerical_CT} compares our Monte Carlo
 simulation for fourth and sixth order Binder cumulants with the
 exactly solvable $c =1/2$ CFT. In \secref{sec:numerical_correlators} we
extend this analysis to the two-point and four-point correlation
 functions. We fit the operator product expansion (OPE) as a test case
 of how to extract the central charge, OPE couplings, and operator
 dimensions.


\section{\label{sec:scalar_fields}Classical Limit for Simplicial Lattice Field Theory}

The scalar $\phi^4$ theory provides the simplest example.  On a smooth Riemann  manifold,  $(\mathcal{M},g)$, the action,
\be
S= \int_{\mathcal{M}} d^d x \sqrt g \left[ 
 \tfrac{1}{2} g^{\mu \nu} \partial_\mu \phi(x) \partial_\nu \phi(x)
 +\tfrac{1}{2} (m^2 + \widetilde \xi_0 Ric) \phi^2(x)
  + \lambda \phi^4(x)  + h \phi(x) \right] \; ,
\label{eq:contaction}
\ee
is manifestly invariant under diffeomorphisms:
$x' = f(x) \equiv x'(x)$. We include the coupling to the scalar Ricci
curvature ($ \widetilde \xi_0 Ric$), where
$\widetilde \xi_0 = (d-2)/( 4 (d-1))$ and an external constant
(scalar) field $h$.  $Ric = (d - 1)(d - 2)/R^2$ on the sphere $\mS^{d}$
of radius $R$. The field, $\phi(x)$, is an {\bf absolute} scalar or in the
language of differential calculus a  {\bf 0-form}
with a fixed value at each point $P$ in the
manifold, independent of co-ordinate system: $\phi'(x') = \phi(x(x'))$
 for $x' = f(x)$ .  For future reference to the discussion
in~\secref{sec:Hibert} on the discrete exterior calculus, we identify $\mbox{vol}_d =\sqrt{g}\; 
 d^dx  = \sqrt{g} dx^1 \wedge \cdots \wedge dx^d$,    as the volume {\bf d-form}. 
%

\begin{figure}[ht]
\centering
\includegraphics[width=0.7\textwidth]{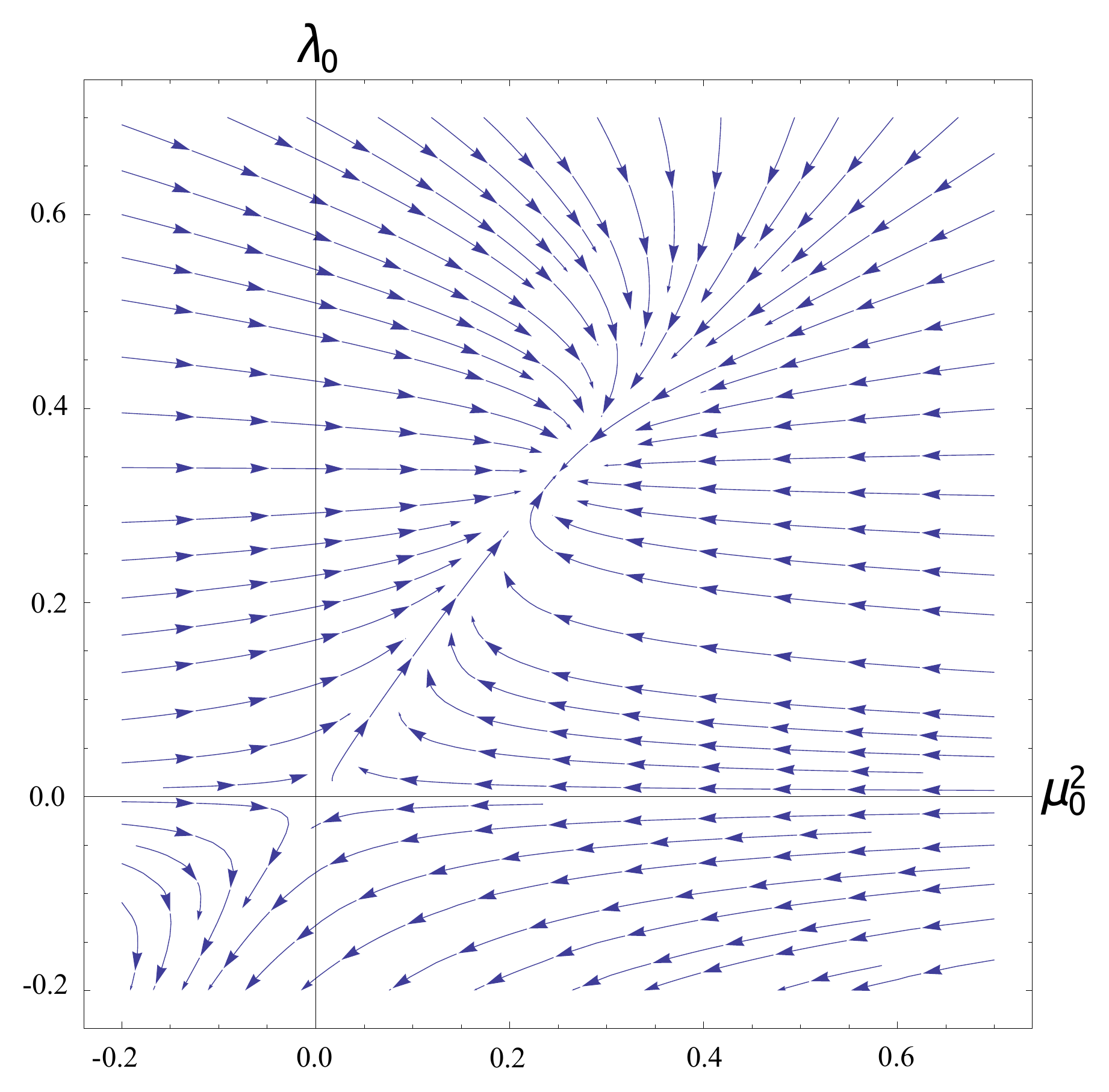}
\caption{\label{fig:WF} The  phase plane of
  $\phi^4$ for $d < 4$,  depicting the renormalization flow
  from the repulsive 
weak coupling ultraviolet (UV)  fixed point  at $(\mu^2_0,\lambda_0) = (0,0)$ to
the infrared (IR) Wilson Fisher  fixed point at $(\mu^{2}_*
,\lambda_*)$,   parameterized in terms of the bare parameters, $\lambda_0 \sim
  \lambda$ and $\mu^2_0 \sim - m^2$.  }
\end{figure}

The  scalar $\phi^4$ theory in $d = 2, 3$ has a (super-)renormalizable UV weak coupling fixed point
and a strong coupling Wilson-Fisher conformal fixed point in the infrared (IR), as illustrated in the
phase diagram in~\figref{fig:WF}.  In passing, we also note 
its similarity to \ndim{4} Yang Mills theory with a sufficient number of massless
fermions to be in the  conformal window~\cite{Kuti:2015awa}, which is a
central motivation for this research.  Both theories are UV
complete, perturbatively renormalizable at weak coupling, and have a 
strong coupling conformal fixed point in the IR. The  mass terms ($m^2_0 \phi^2 $  or $m_0 \bar \psi \psi$, respectively)  must be tuned  either to or near the
critical surface to reach the  conformal  or mass deformed theory.

\begin{figure}[ht]
\centering
 \includegraphics[width=.7\textwidth]{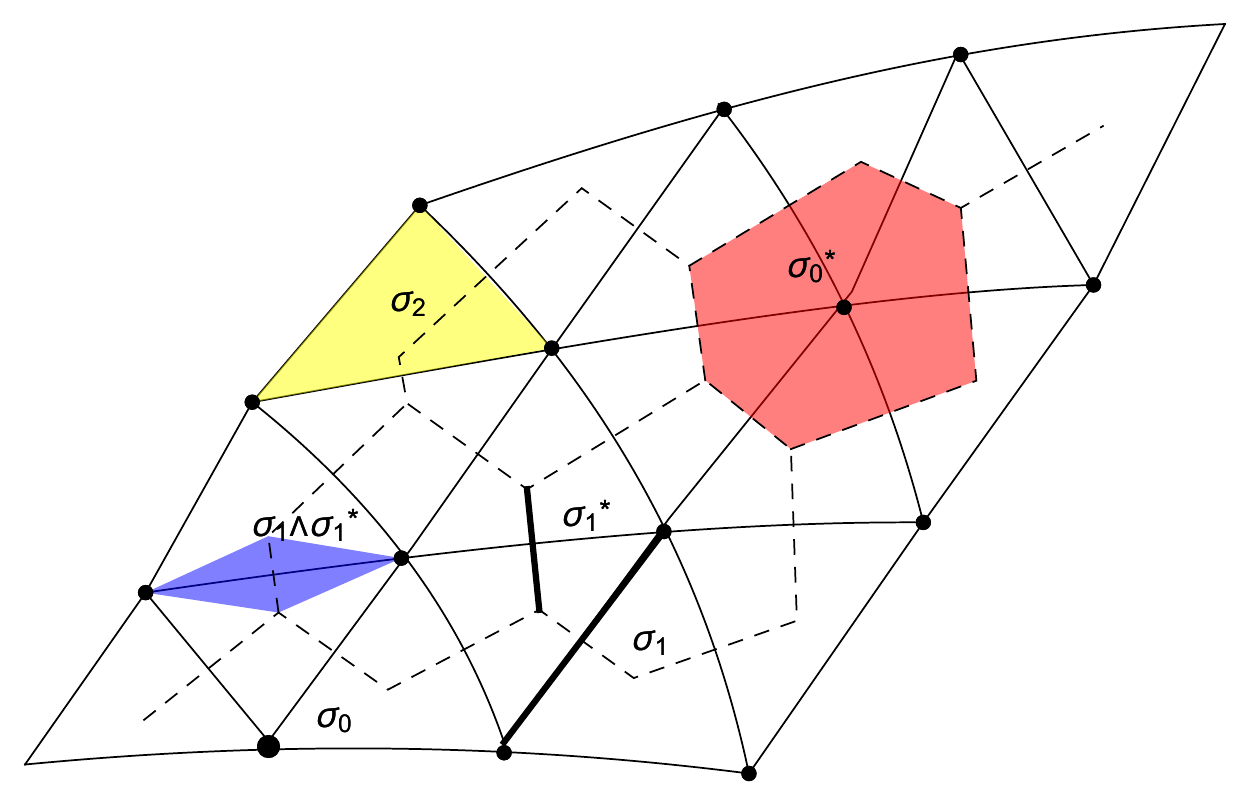}
  \caption{\label{fig:lattice} A \ndim{2} simplicial complex with points
    $(\sigma_0)$, edges ($\sigma_1$) and triangles $(\sigma_2)$ (illustrated in yellow). At
    each vertex $\sigma_0$ there is a dual polytope, $\sigma^*_0$
    (illustrated in red), and at each link, $\sigma_1$, there is a
    dual link $\sigma^*_1$ and its associated hybrid cell
    $\sigma_1 \wedge \sigma^*_1$ (illustrated in blue).  }
\end{figure}

To formulate a lattice action, we introduce a simplicial complex, or
triangulation in 2d, as illustrated in~\figref{fig:lattice}, and
a discrete action,
\be
S = \frac{1}{2} \sum_{\langle i j \rangle}K_{ij}  \frac{
(\phi_i - \phi_j)^2}{l^2_{ij}} +  \sum_i \sqrt{g_i} ( \tfrac{1}{2}m^2_i \phi_i^2 +
\lambda_i \phi^4_i ) + h \sum_i \sqrt{g_i} \phi_i
\ee
where $i$ labels all  vertices and the sum $\langle i j \rangle$ runs over all links
with proper length $l_{ij}$.  The technical requirement  is
to fix the weights, ($K_{ij}, \sqrt{g_i}, m^2_i ,
\lambda_i$) as functions of bare couplings and the
target manifold,   on a sequence of increasingly fine tessellations so
that in the  limit of vanishing  lattice spacing, $a = O(l_{ij})$,  the
 quantum path integral  converges to the continuum
renormalized quantum theory  with a minimal set of  fine tuning
parameters. For the $\phi^4$ theory,
 there is one parameter to tune in the approach to the critical surface: the relevant mass parameter $\mu_0^2
\rightarrow \mu_*^2(\lambda_0)$, illustrated in~\figref{fig:WF}.  

The construction of our  {\bf QFE}  simplicial lattice action can be broken  into three steps: 
\begin{itemize}[noitemsep, topsep=0pt]
\itemsep0em
\item [i.] {\bf Simplicial Geometry:} The smooth Riemann manifold,
  $(\mathcal{M},g)$, is replaced  by
   simplex complex, $(\mathcal{M}_\sigma,g_\sigma)$ with piecewise
   flat cells  (see \secref{sec:Geometry}).
\item[ii.] {\bf Classical Lattice Action:} The continuum field
  $\phi(x)$ is
  replaced by a discrete sum over  FEM basis elements,
$\phi(x) \rightarrow W^i(x) \phi_i$ (see~\secref{sec:Hibert}).
\item[iii.] {\bf QFE: Quantum Corrections:}  Quantum counter terms  are
added to the discrete lattice  action to  cancel  defects at the UV
cut-off (see~\secref{sec:OneLoopCT}).
\end{itemize}


\subsection{\label{sec:Geometry}Geometry and Regge Calculus}
The Regge Calculus approach to constructing  a discrete
approximation to a Riemann manifold, $(\mathcal{M},g)$,
proceeds as follows.  First  the  manifold, $\mathcal{M}$, is replaced  by a
simplicial complex, $\mathcal{M}_\sigma$, composed of
elementary simplicies:  triangles in 2d as illustrated in~\figref{fig:lattice},  tetrahedrons in 3d,
etc. This graph defines the topology of the manifold in the language
of {\em Category Theory}~\cite{spanier2012algebraic}. Next a
discrete  metric is introduced as a set of edge lengths on the graph: $g_{\mu\nu}(x) \rightarrow   g_\sigma =
\{l_{ij}\}$.  Assuming a piecewise  flat interpolation  into the
interior of each simplex, we now have the 
Regge representation of a Riemann manifold  $(\mathcal{M}_\sigma,g_\sigma)$
that is continuous but not differentiable.  The
curvature is given by a singular  distribution at the boundary of the each
simplex;  in 2d, concentrated  at the vertices,
and in higher dimensions at the $d\!-\!2$-dimensional ``hinges.'' The integrated
curvature over the defect is  easily computed by parallel transport
of the tangent vector around each defect. 
 
Each simplex is parameterized by using $d+1$ local barycentric coordinates, $0 \leq \xi^i
\leq 1$. Using the constraint, $\xi^0 +\xi^1  + \cdots + \xi^d =
1$ to eliminate $\xi^0$, our   $\phi^4$  action
in~\eqref{eq:contaction} on this piecewise flat Regge
manifold is   given as a  sum over each simplex,
\be
S \rightarrow S_\sigma =  
\sum_{\sigma \in \mathcal{M_\sigma}} \int_\sigma d^d \xi \sqrt{G_\sigma}
\left[ \tfrac{1}{2}  G^{ij}_\sigma \partial_i \phi(\xi) \partial_j \phi(\xi) +   \tfrac{1}{2}  (m^2
  + \widetilde \xi_0 Ric)  
  \phi^2(\xi) + \lambda \phi^4(\xi) \right] \; .
\label{eq:scalaraction3}
\ee
 We may choose an isometric embedding into a sufficiently high
dimensional flat Euclidean space 
with vertices at $\vec y =  \vec{r}_n$ so a point in the
interior of each simplex and its induced
metric are given by
\be
\vec{y} = \sum_{n=0}^d \xi^n \vec{r}_n = \sum_{i=1}^d \xi^i
\vec{l}_{i0} + \vec{r}_0 \quad , \quad G_{ij} = \frac{\dd {\vec y}}{\dd {\xi^i}}  \cdot \frac{\dd {\vec y}}{\dd {\xi^j}} \equiv  \vec{l}_{i0} \cdot
\vec{l}_{j0} \; .
\ee 
This defines  the volume element, $\sqrt{G_\sigma} =
\sqrt{\det[G_{ij}]}$ and inverse metric, $G^{ij}_\sigma$, as well.

Since we are not considering dynamical gravity, this piecewise flat
metric is frozen (i.e. quenched) and chosen to conform as closely as
possible to our target manifold for the quantum field theory.  For
example this can be achieved by computing the edge length $l_{ij}$,
from the proper distance between points on the original manifold
$x_i,x_j$  or from the Euclidean distance
  $|\vec r_i - \vec r_j| $ in the isometric embedding space,
  $\vec r_i= \vec y(x_i)$ to first order relative to the local
  curvature of the target manifold.  At this stage, the dynamical
quantum field $\phi(x)$ is still a continuum function on the
piecewise flat Regge manifold.

\subsection{\label{sec:Hibert}Hilbert Space and Discrete Exterior Calculus}
The second step is the approximation of the matter field $\phi(x)$ as 
an expansion,
\be
\phi(x) \rightarrow \phi_\sigma(\xi) = E^0_\sigma(\xi)\phi_0 + E^1_\sigma(\xi) \phi_1 +
\cdots + E^d_\sigma(\xi) \phi_d \;, 
\label{eq:FEbasis}
\ee
into a finite element basis on each simplex $\sigma_d$. The simplest
form  is a piecewise linear function, $E^i(\xi) = \xi^i$, on each
simplex. In this case we are using essentially the
same linear approximation for both the metric, $g_{\mu\nu}(x)$, and
matter, $\phi(x)$, fields. To evaluate the FEM action, we  simply plug the
expansion in~\eqref{eq:FEbasis} into~\eqref{eq:scalaraction3} and perform the integration.
For the kinetic term, this is particularly simple because the gradients of the barycentric coordinates are
constant. For 2d, this  gives the well known
form on each triangle,
\bea
I_{\sigma} &=& \frac{l_{31}^2 + l_{23}^2 - l_{12}^2}{8 A_{123}} (\phi_1
- \phi_2)^2 + (23) + (31)  \nn
&=& \frac{1}{2} A^{(3)}_{12}(\phi_1 - \phi_2)^2 + (23) + (31)  
\eea
 where $ A^{(3)}_{12}$ is the 
area of the triangle formed by the sites 1, 2, and the circumcenter, $\sigma_2^*(123)$.
The free scalar action on the entire simplicial complex is now found 
by summing over triangles,
\be
\sum_{\sigma} I_\sigma[\phi] =  \frac{1}{2}\sum_{\langle i j \rangle}
A_{ij} \left( \frac{ \phi_i - \phi_j} {l_{ij}} \right)^2  \label{eq:scalaraction4} \; .
\ee
Each link $\langle ij \rangle$ receives two contributions---one from
each triangle that it borders---resulting in the total area for the
hybrid cell, $A_{ij} =l_{ij}|\sigma^*_1(ij)|/2 = \left|
  \sigma_1(ij)\wedge \sigma_1^*(ij) \right|$, as illustrated
in~\figref{fig:lattice}.

For higher  dimensions, a natural generalization of the kinetic term is
 \be
S_{\sigma}[\phi] =\frac{1}{2} \sum_{\<i,j\>} V_{ij} \frac{(\phi_i
  -\phi_j)^2}{l^2_{ij}} + \frac{1}{2} \sum_i m \sqrt{g_i} \phi^2_i \; ,
\label{eq:DECscalar}
\ee
where
$ V_{ij} = |\sigma_1(ij) \wedge \sigma_1^*(ij)| =l_{ij} S_{ij}/d $ is
the product of the length of the link ($l_{ij}$) times the volume of
the $d\!-\!1$-dimensional ``surface'', $S_{ij} =|\sigma_1^*(ij)|$, of the dual polytope normal
to the link $\<i,j\>$.  A mass term has also been included weighted by
the dual lattice volume $\sqrt{g_i} = |\sigma^*_0(i)|$.   This elegant form was  recommended in the seminal
papers on random lattices by Christ, Friedberg and Lee~\cite{Christ:1982ci,Christ:1982ck,Christ:1982zq} and subsequently
in the FEM literature by the application of the simplicial Stokes'
theorem for Discrete Exterior Calculus (DEC). {\bf  However, we
note this  generalization is not equivalent to linear FEM for
  $d > 2$} (See Ref.~\cite{Brower:2016vsl}).

To appreciate the DEC approach~\cite{2005math8341D,Arnold2006,Miller:2013gy,1802.04506},  it is useful to expand a little on the
geometry of a simplicial  complex, $\cal S$, and its Voronor{\"i} dual,
$\cal S^*$. 
 A pure simplicial complex $\cal S$ consists of a set of
$d$-dimensional simplices (designated by $\sigma_d$) ``glued''
together at shared faces (boundaries) consisting of $d-$1-dimensional
simplices ($\sigma_{d-1})$, iteratively giving a sequence:
$\sigma_d \rightarrow \sigma_{d-1}\rightarrow \cdots \sigma_1
\rightarrow \sigma_0$. 
 This hierarchy is specified by the boundary operator,
\be
\partial \sigma_n(i_0 i_1 \cdots  i_n) = \sum^n_{k = 0} (-1)^k
\sigma_{n-1}(i_0 i_1\cdots \widehat i_k\cdots  i_n)\; ,
\label{eq:boundary}
\ee
where $\widehat i_k$ means to exclude this site.  Each simplex
$\sigma_n(i_0 i_1\cdots i_n) $ is an anti-symmetric function of its
arguments. The signs in~\eqref{eq:boundary} keep track of the
orientation of each simplex.  It is trivial to check that the boundary operator
is closed: $\partial^2\sigma_n = 0$. On a finite simplicial
lattice $\partial$ is  a matrix and its transpose, $\partial^T$, is
the co-boundary operator.  
The circumcenter Voronor{\"i} dual  lattice, $\mathcal S^*$, is  composed of polytopes,
$\sigma^*_0 \leftarrow \sigma^*_1 \leftarrow \cdots \leftarrow
\sigma^*_d$,
where $\sigma^*_n$ has dimension $d-n$ as illustrated in
Fig.~\ref{fig:lattice}. 
 A crucial property of this circumcenter
duality is {\bf orthogonality}.   Each simplicial element
$\sigma_n \in S$ is orthogonal to its dual polytope
$\sigma^*_n \in S^*$. As a  consequence,  the volume,
\be
|\sigma_n \wedge
\sigma^*_n|   = \frac{n! (d - n)!}{d!} |\sigma_n| |\sigma^*_n|  \; ,
\label{eq:SimplicialVolume}
\ee
 of the hybrid cell,  $\sigma_n \wedge \sigma^*_n$, is  a simple product.
 Hybrid cells, constructed from
simplices $\sigma_n$ in $\cal S$ and their orthogonal dual $\sigma^*_n$ in
$\cal S^*$, give a proper tiling of the discrete d-dimensional
manifold with the special case $|\sigma_0(i)| = 1$ and $|\sigma^*_0(i)| = \sqrt{g_i}$.  This is a first
 modest step into discrete homology and De Rham cohomology on a simplicial complex.

The discrete analogue of differential forms,
 $\omega_k$, is a pairing or map,
\be
\< \omega_k, \sigma_k\> \equiv \omega_k(\sigma_k) = \omega_k(i_0 i_1
\cdots i_k) \; ,
\ee
to the numerical value of a field from sites in the case of 0-forms, from links in the case of
1-forms, and from k-simplices in the case of k-forms. Of course this is
familiar to lattice field theory, associating scalars ($ \omega_0 \sim
\phi_x$), gauge fields ($\omega_1 \sim A_\mu dx^\mu$) and field strengths
($\omega_2\sim F_{\mu\nu} dx^\mu \wedge dx^\nu$) with sites, links
and plaquettes respectively.   This enables us to define the discrete
analogue of the exterior derivative  ${\bf d} \omega_k$ of a k-form 
by   replacing the  continuum Stokes' theorem by the discrete map
or DEC Stokes' theorem:
\be 
\int_{\sigma_{k+1}}{\bf d}\omega(x) =\int_{\partial\sigma_{k+1}}w(x) \quad 
\rightarrow \quad 
 \< {\bf d} \omega_k, \sigma_{k+1}  \> =  \<  \omega_k, \partial \sigma_{k+1}  \>
\label{eq:discretederivative}
\ee
The discrete exterior derivative  is automatically closed (${\bf d d} = 0$)
because the boundary operator  is closed ($\partial \partial = 0$).
Applying~\eqref{eq:discretederivative} to the discrete exterior derivative 
of a scalar (0-form) field trivially gives, 
\be
 \< {\bf d} \phi, \sigma_1(ij) \>  =   \<\phi, \partial \sigma_1(ij)
\> =  (\phi_i - \phi_j )\; ,
\ee
the standard finite  difference  approximation on each link.

Next we need to define the discrete analogue of the  Hodge star ($*$).
This is the first time an explicit dependence on the
metric is introduced.  In the continuum, a  k-form is  
an anti-symmetric tensor, $\omega_k(x) = (k!)^{-1} \omega_ {\mu_1,
  \mu_2 \cdots \mu_{k}} dx^{\mu_1} \wedge  dx^{\mu_2} \wedge 
\cdots \wedge dx^{\mu_k} $,   in a orthogonal basis of 1-forms dual to 
tangent vectors: $dx^\mu (\partial_\nu) = \delta^\mu_\nu$.
The Hodge star takes a k-dimensional  basis  into its orthogonal complement, 
\be
*\!(dx^{\mu_1} \wedge dx^{\mu_2} \wedge 
\cdots \wedge dx^{\mu_k} ) =  dx^{\mu_{k+1}} \wedge dx^{\mu_{k+2}} \wedge 
\cdots \wedge  dx^{\mu_n}
\label{eq:HodgeStartOp}
\ee
where $  \mu_1, \mu_2, \cdots, \mu_n $ is an even
permutation of $(1,2,\cdots, n) $.
\ignore{~\footnote{REMOVE LATER: See
  \href{https://en.wikipedia.org/wiki/Hodge\_star\_operator}{https://en.wikipedia.org/wiki/Hodge\_star\_operator}
  or see Carroll's Bosk or the remark at
\href{https://physics.stackexchange.com/questions/284271/tensor-vs-tensor-densities}{https://physics.stackexchange.com/questions/284271/tensor-vs-tensor-densities}}. 
}
However the wedge product  (or equivalently the Levi-Civita symbol) is
not a tensor but a  weight 1 tensor  density~\cite{Carroll:1997ar}.  The true volume k-form (or tensor),
$\mbox{vol}_k  = \sqrt{g_k} dx^{\mu_1} \wedge  \cdots \wedge
dx^{\mu_k} $, requires a factor of $\sqrt{g_k}$ which under the Hodge
star operation in~\eqref{eq:HodgeStartOp} gives the identity, $\sqrt{g_{n\!-\!k}}\;*\!(\mbox{vol}_k)
=  \sqrt{g_k} \;\mbox{vol}_{n\!-\!k}  $, where we have
used orthogonality to factor $\sqrt{g} = \sqrt{g_k} \sqrt{g_{d-k}}$, that is, between the
  plane and its dual.  Consequently on the simplicial 
complex, it is reasonable that the proper definition of the discrete  Hodge star,
\be
\< \omega^*_k, \sigma^*_k \> \; |\sigma_k|   = \< \omega_k, \sigma_k
\> \; |\sigma^*_k|   \; ,
\label{eq:HodgeStar} 
\ee
replaces these factors $\sqrt{g_k} $,
$\sqrt{g_{n\!-\!k}}$  by finite volumes $|\sigma_k|$,
$|\sigma^*_k|$ respectively. The Hodge star identity in~\eqref{eq:HodgeStar} uniquely fixes the dual field
values, $\< \omega^*_k, \sigma^*_k \>$ and the action of a  discrete
co-differential, $\bm{\delta}  = * {\bf d} *$  through Stokes theorem in~\eqref{eq:discretederivative} on $\cal S^*$.

\begin{figure}
  \centering 
\includegraphics[width=.5\textwidth]{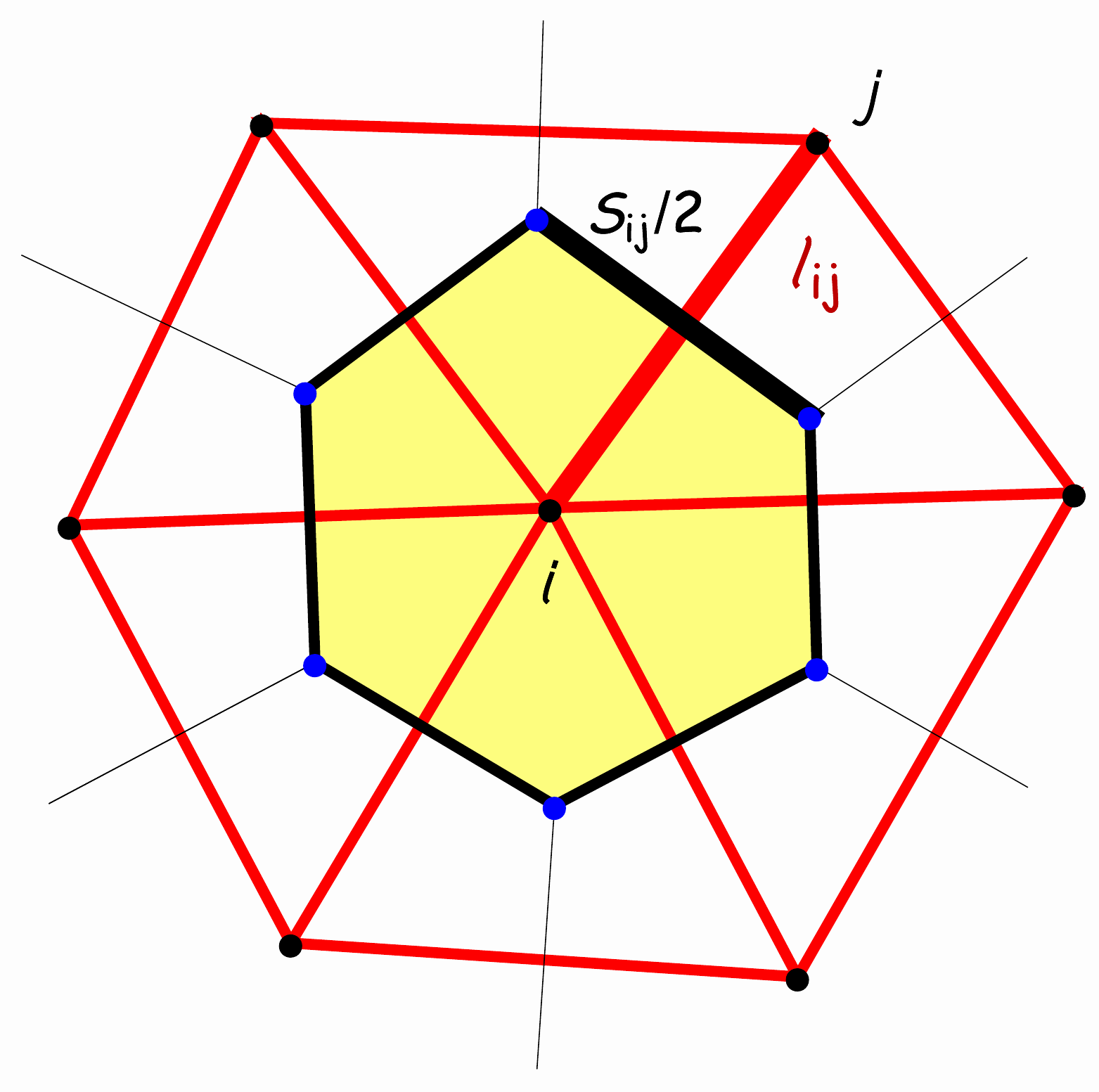}
\caption{\label{fig:LapBelop} The discrete Laplacian at a site $i$ is given by the sum on 
 all links  $\<i,j\>$ (in red)   weighed by gradients $(\phi_i -
  \phi_j)/l_{ij}$ multiplied by the surface  $S_{ij} = 2
  V_{ij}/l_{ij}$ (in black) and normalized by the dual volume
  $|\sigma^*_o(i)| = V_i$ (in yellow). }
\end{figure}

Putting this all
together, we consider the DEC Laplace-Beltrami operator on scalar fields, 
$(\bm{\delta}  +   {\bf d}  )^2\phi = ({\bf d} \bm{\delta}  +   \bm{\delta}{\bf d}  )\phi = *  {\bf d} * {\bf d} \phi$, is  
\be
* {\bf d} * {\bf d} \phi(i) = \frac{|\sigma_0(i)|}{|\sigma^*_0(i)|} \sum_{j \in\<i,j\>} 
                      \frac{|\sigma^*_1(ij)|}{|\sigma_1(ij)|} (\phi_i	-\phi_j)
= \frac{1}{\sqrt{g_i}} \sum_{j \in \<i,j\>} \frac{ V_{ij}}{l_{ij}} \frac{\phi_i
	-\phi_j}{l_{ij}} \; ,
\label{eq:BL}
\ee
which corresponds to the action in~\eqref{eq:DECscalar}.  For
2d this is illustrated in~\figref{fig:LapBelop}, as the sum of
fluxes through the boundaries $\partial \sigma^*_0(i)$ with surface
area, $S_{ij} /(d-1)!= V_{ij}/l_{ij} = |\sigma_1(ij)
\wedge\sigma^*_1(ij)|/l_{ij}$.  This is identical to linear finite
elements in 2d.  Besides providing an alternate approach to
linear finite elements for constructing the discrete Hilbert space for
scalar fields in $d>2$ dimensions, it provides a useful geometric
framework for fields with spin.  We note however that additional
considerations are still needed for non-Abelian gauge
fields~\cite{Christ:1982ck}, Dirac Fermions~\cite{Brower:2016vsl} and
Chern-Simons terms~\cite{Sen:2000ez}. The best geometrization of
simplicial field theories and the error estimates thereof are an active
research topic.

To complete the simplicial action, we  need to add the potential term.  This  may be constructed in a variety of
ways.  If one follows strictly the linear finite element prescription, the expression
for the local polynomial potential (e.g. mass and quartic terms) will
not be local but rather point split on  each simplex $\sigma_d$. 
For example, in 2d, after expanding 
$\phi(\xi) = \sum^d_{i =0}\xi^i \phi_i$ and evaluating the integral over the linear elements, the contribution for the quadratic term 
on a single triangle, $\sigma_2(123)$, is 
\be
\int_\sigma d^2\xi \sqrt{G_\sigma}   \phi^2_\sigma(\xi) = \frac{A_{123}}{6} (  \phi^2_1  +\phi^2_2  + \phi^2_3  +
\phi_1 \phi_2 + \phi_2 \phi_3 + \phi_3 \phi_1 ) \; .\esw{This is very close to text in the fermion paper, do we want to re-write it a bit?} 
\ee
The general expression for a  homogeneous polynomial over a d-simplex
with volume $V_d = |\sigma_d|$ is given as as sum over distinct partitions of $n$:
\be
\int_{\sigma_d} dV_d \; (\phi(\xi))^n = \frac{V_d \; d! \; n!}{(n+d)!} \sum_{(\sum_i k_i=n)} \phi_0^{k_0}
\phi_1^{k_1}\cdots\phi_d^{k_d} \;  .
\ee
Nonetheless in the spirit of
dropping higher dimensional operators in a derivative expansion, we choose local 
terms approximating the potential at each vertex weighted by the volume ,  $\sqrt{g_i}  = |\sigma^*_0(i)| $, of the dual simplex $\sigma^*_0(i)$.

This approximation leads to our complete simplicial action, combining the DEC Laplace-Beltrami operator for the kinetic term and the local approximation for the boundary,
\be
S = \frac{1}{2} \sum_{\langle i j \rangle} \frac{V_{ij}}{l_{ij}^2}
(\phi_i - \phi_j)^2 + \lambda_0 \sum_i \sqrt{g_j}  \left( \phi_i^2 -
  \frac{\mu^2_0}{2\lambda_0} \right)^2 + h \sum_i \sqrt{g_i} \phi_i
\label{eq:simplicial_action} \; ,
\ee
where $\mu^2_0  = - m^2_0/2$. We will use this action for our discussion of $\mS^2$.



\begin{figure}[t]
\centering
\includegraphics[width=0.32\textwidth]{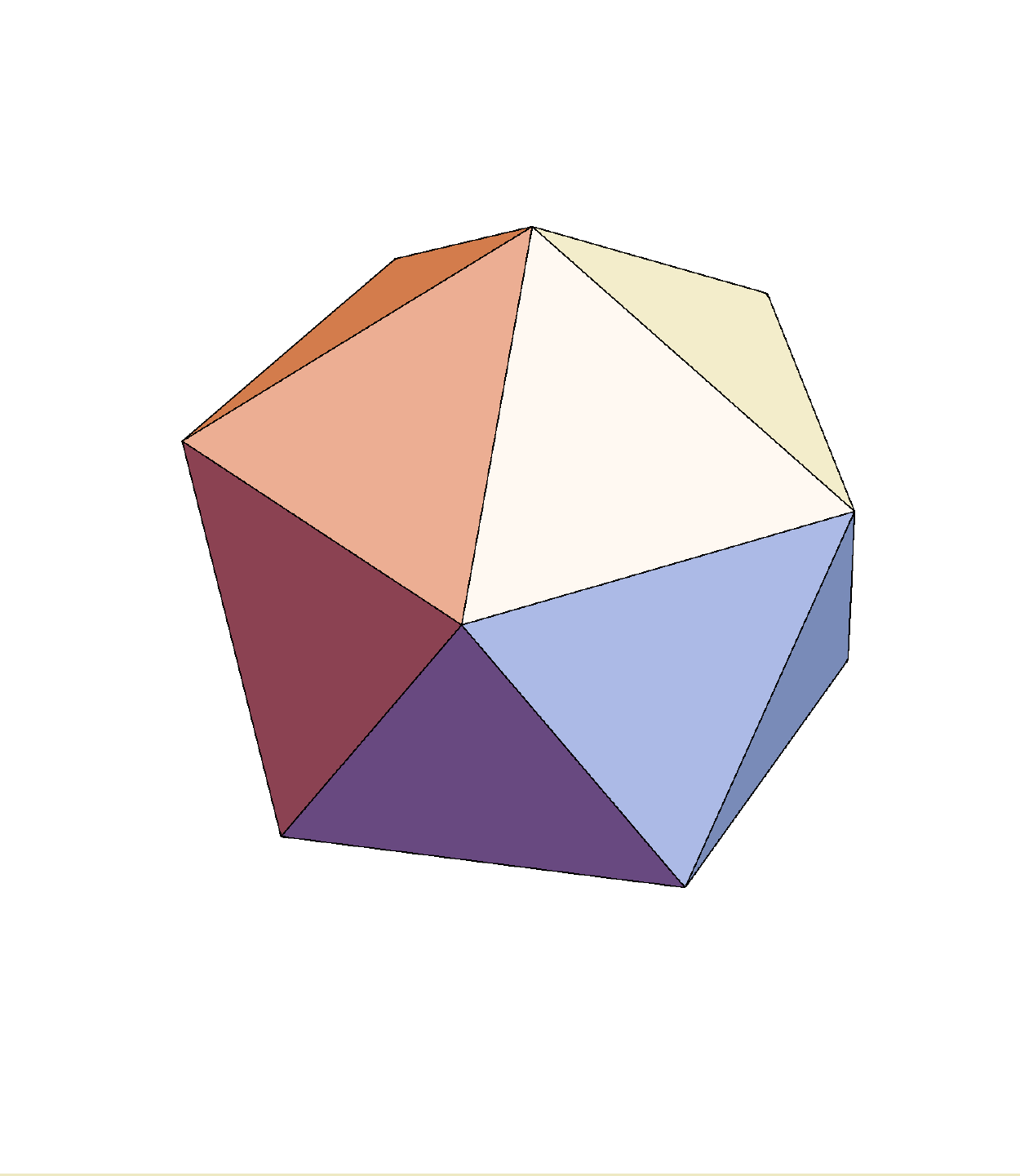}
\includegraphics[width=0.32\textwidth]{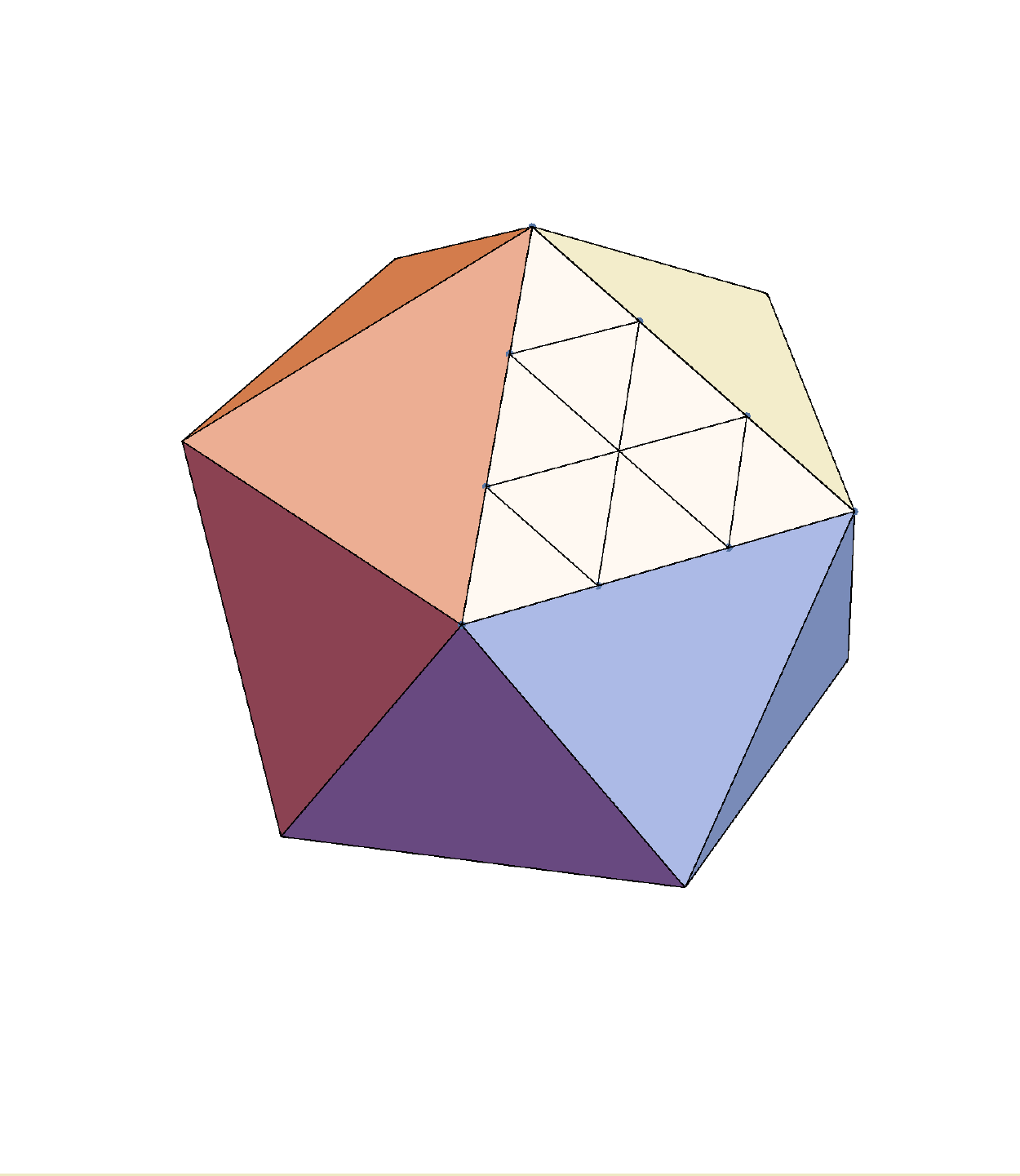}
\includegraphics[width=0.32\textwidth]{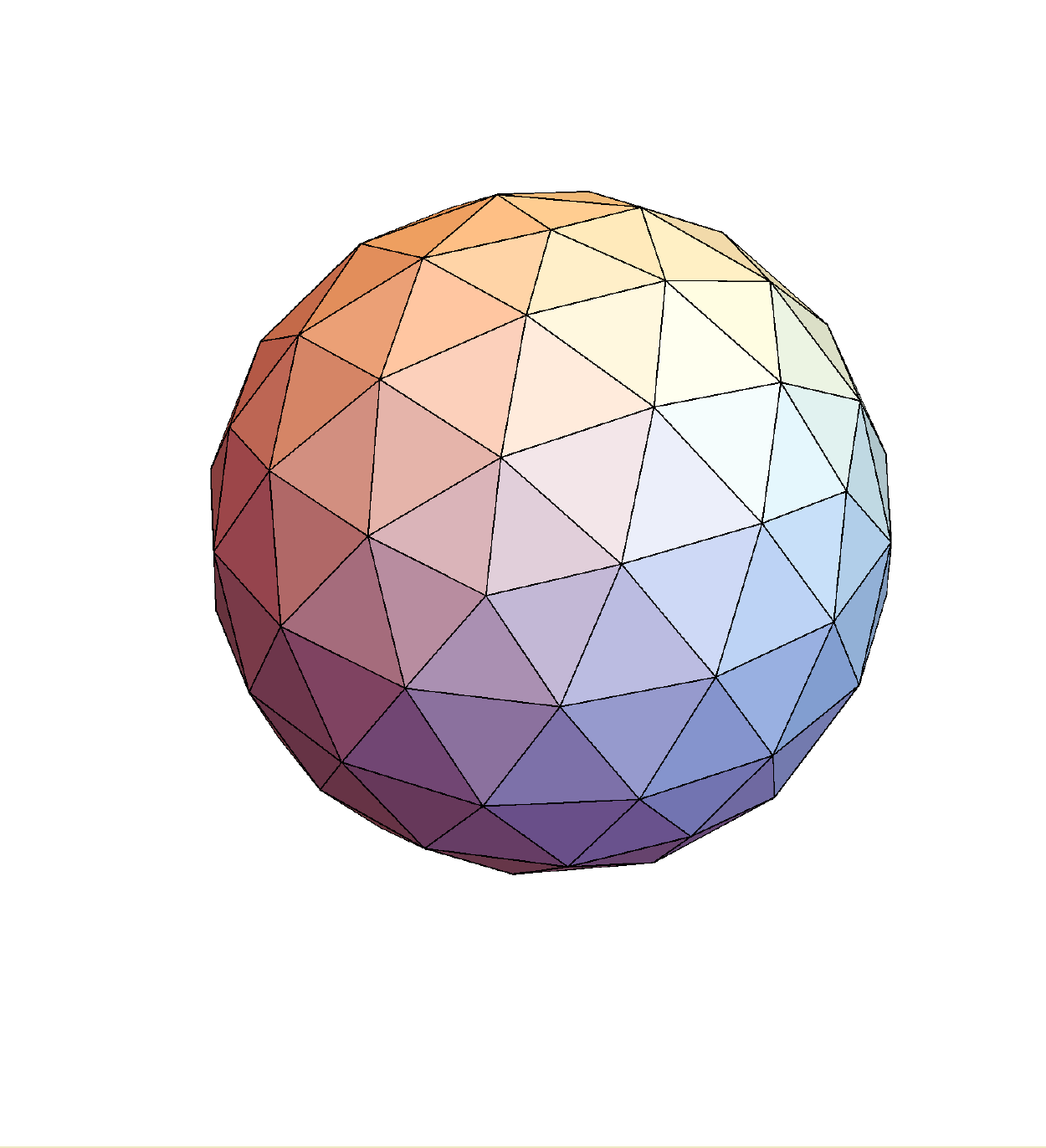}
\vskip -1.0 cm
\caption{\label{fig:icos} The  $L=3$ refinement of
the icosahedron with $V = 2 + 10L^2 = 92$ vertices or sites. The icosahedron on
the left is refined in the middle with $L^2 = 9$ equilateral triangles on 
each face, and  then on the right the new vertices are projected  onto the unit
sphere. The resulting simplicial  complex preserves the icosahedral symmetries. }
\end{figure}

Finally we should acknowledge that there  are many other
alternatives in the FEM literature worthy of consideration which may offer improved
convergence and faster restoration of the continuum
symmetries.
Our goal here is to find  the simplest discrete action on simplicial
lattice capable of reaching the correct continuum theory with no more fine tuning
than is required  on the hypercubic lattice.

\subsection{Spectral Fidelity on \texorpdfstring{$\mS^2$}{S2}}
We represent $\mathbb S^2$ embedded in $\mathbb R^3$
parameterized by \ndim{3} unit vectors:
\be
\hat r = (r_x,r_y,r_z) \in \mathbb  R^3 \quad, \quad  r^2_x +r^2_y +
r^2_z = 1 \; .
\label{eq:SphereInR3}
\ee
 In order to preserve the largest available discrete
subgroup of $O(3)$, we start with the regular icosahedron illustrated  on the left in~\figref{fig:icos} and subsequently divide
each of the 20 equilateral triangles into $L^2$ smaller equilateral
triangles. Then we project the vertices radially outwards onto the
surface of the circumscribing sphere, dilating and distorting each
triangle from its equilateral form, but preserving exactly
the icosahedral symmetries.  The total numbers of faces, edges,
and vertices are $F = 20 L^2$, $E = 30 L^2$ and $N = 2 + 10 L^2$,
respectively, satisfying the Euler identity $F -E + N = 2$.

The images of the vertices on the sphere are then connected by new
links consistent with the unique Delaunay triangulation~\cite{Delaunay:1934}. In the
triangulation, each vertex is connected to five or six neighboring vertices by
edges $\<x,y\>$. The  lengths  are set to the secant lengths $l_{xy}
= |\hat r_x - \hat r_y|$ in the embedding space between the vertices
on the sphere, which approximates the geodesic length to $O(l^2_{xy})$. 
The extension to a lattice for the $\mR \times \mS^2$ cylinder 
introduces a uniform (periodic) lattice perpendicular to the spheres
at $t = 0,1,\cdots, L_t -1$. The sequence of refinements as
$L \rightarrow \infty$ divides the total curvature into vanishingly
small defects at each vertex. As an alternative formulation, we could
replace the flat triangles with spherical
triangles~\cite{Brower:2016moq}, introducing spherical areas, geodesic
lengths, moving the curvature uniformly into the interior of each
triangle. In this formulation, each triangle would have a vanishing
deficit angle in the continuum limit.  Both discretizations are
equivalent at $O(a^2)$, and as such we prefer flat triangles due to
their relative simplicity.

The first test of our construction is to  look at the spectrum of the
free theory,
\be
S_0 = \frac{1}{2} \phi_{x} M_{x, y} \phi_{y}  =
\frac{A_{xy}}{2 l^2_{xy} }(\phi_{x} - 
\phi_{y})^2 
+ \frac{m^2_0}{2} \sqrt{g_x}\phi^2_{x},
\ee
where $x,y = 1,..., N$ enumerates the lattice sites on a finite simplicial
lattice.  
The spectrum is computed from the generalized eigenvalue condition
\be
M_{xy} \phi_n(x) = E_n  \sqrt{g_x}\phi_n(x) = (E^{(0)}_n + m^2_0) \sqrt{g_x}\phi_n(x),
\ee
where the eigenvalues at zero mass are  $E^{(0)}_n$.  Each distinct eigenvalue $E_n$ has right/left generalized eigenvectors, $\phi_n(x)/\phi^*_n(x)$, with the orthogonality and completeness relations,
\be
\sum_x  \sqrt{g_x}\phi^*_n(x) \phi_m(x) = \delta_{nm} \quad, \quad
\sum_n \phi^*_n(y) \phi_n(x) = \delta_{xy}/ \sqrt{g_x} \; .
\ee
 The spectral decomposition for the free propagator  is
\be
G_{xy}(m^2_0) \equiv \left[\frac{1}{M}\right]_{xy} = \sum_n
\frac{\phi_n(x) \phi^*_n(y)}{E^{(0)}_n + m^2_0} 
\label{eq:Greens}
\ee
 Alternatively we may define   a Hermitian form, $\widetilde M =  g^{-1/4}
Mg^{-1/4}$, with complex conjugate  right and left
eigenvectors,  $g_x^{1/4}\phi_n(x) = \<x|n\> $ and $\phi^*_n(x)  = \<n|x\>$, 
respectively. In Dirac notation the completeness and orthogonality are
given by  ${\bf 1} = \sum_x |x\>\<x|$ ,
${\bf 1} = \sum_n |n\>\<n|$ and $ \delta_{x,y} =\<x | y\>$,  $
\delta_{n,m} = \<n | m\> $ respectively.   
\begin{figure}[ht]
\centering
\includegraphics[width=0.45\textwidth]{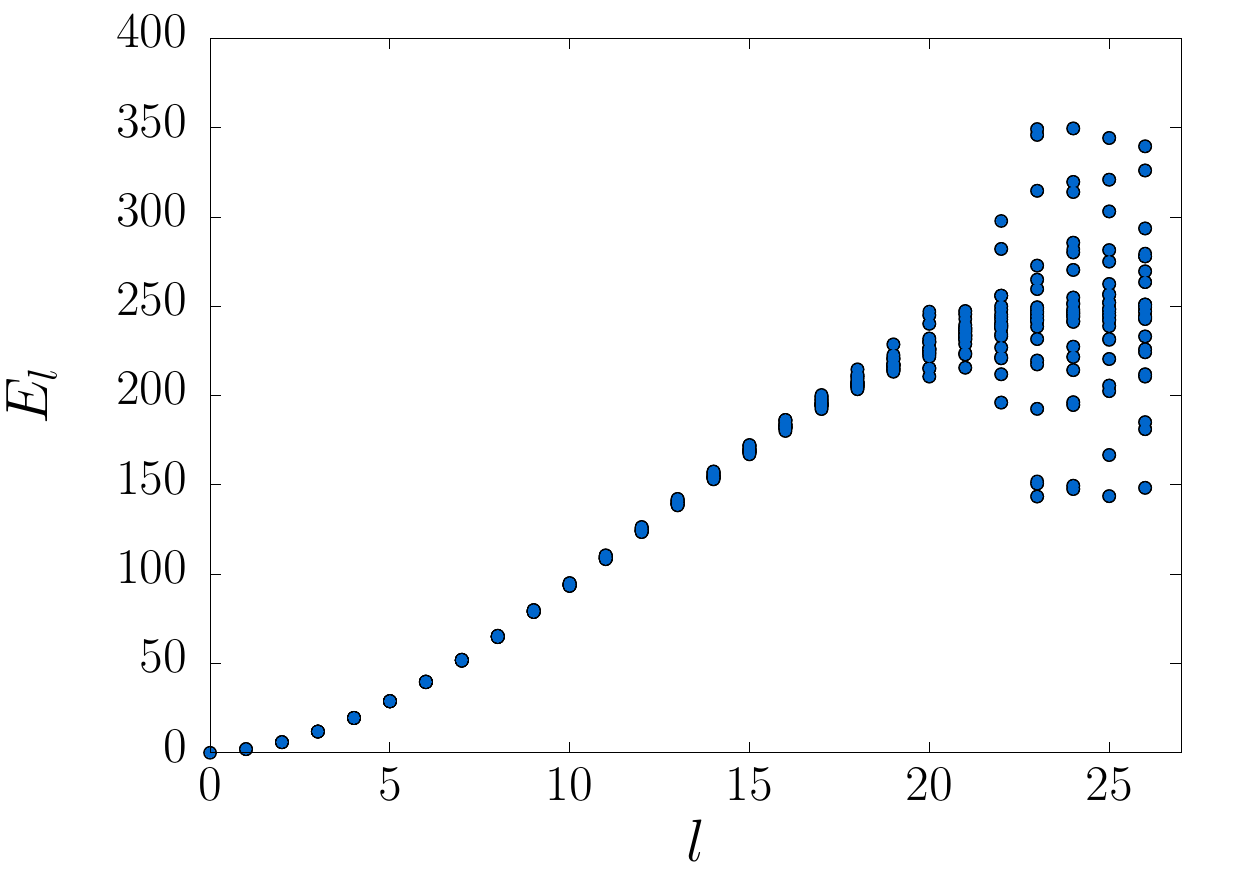}
\includegraphics[width=0.45\textwidth]{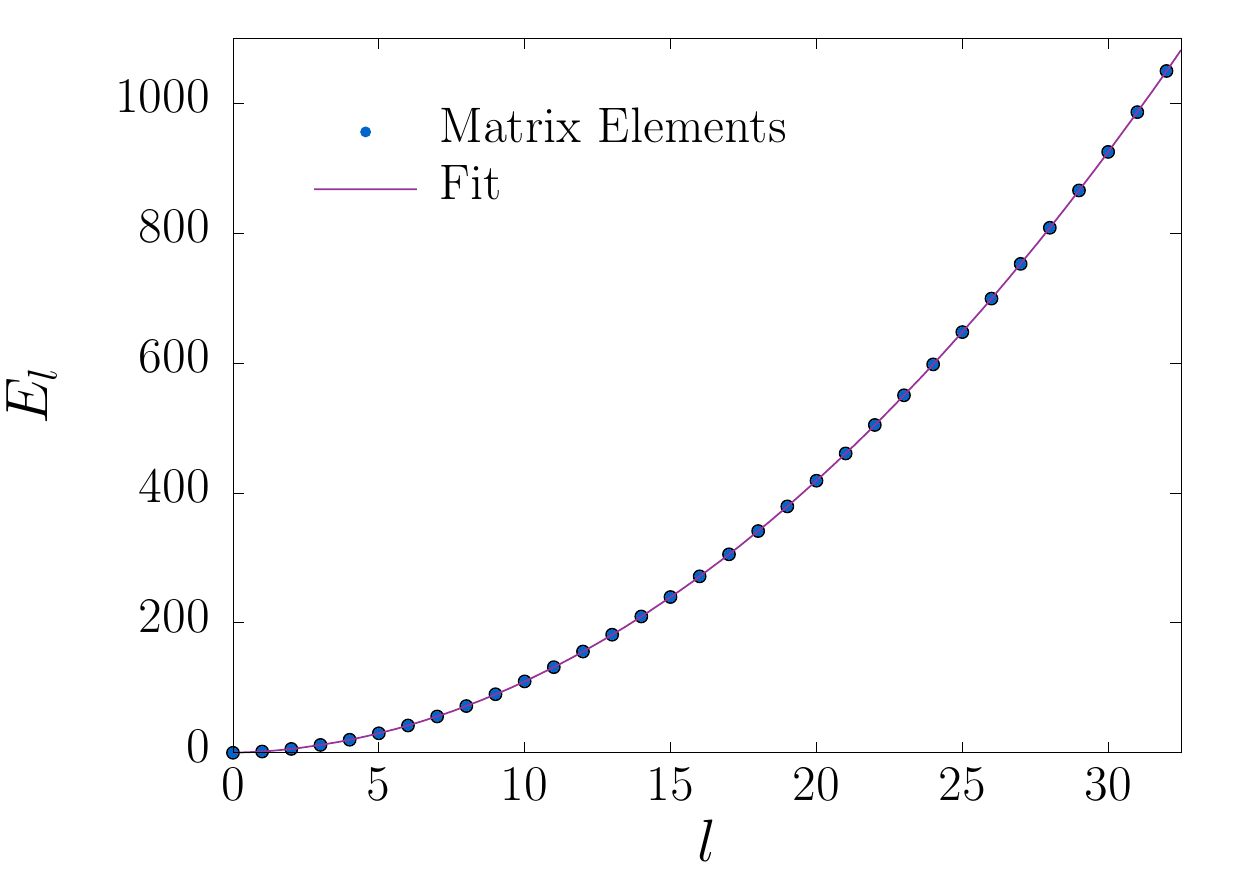}
\caption{\label{fig:FEMdiag} Left: The  $2l+1$  spectral  
 values for  $m \in [-l,l]$ are plotted against $l$ for $L
  = 8$.   Right: The spectral values averaged over $m$  fitted to $l + 1.00002 \; l^2
  - 1.27455\times 10^{-5} \; l^3 - 5.58246 \times 10^{-6} \; l^4$
  for $L =  128$   and $l \le 32$. }
\end{figure}

Returning to the example of the FEM sphere, $\mS^2$, we have verified
that the generalized eigenvalues  which lie well below the cut-off are
well  fitted  by the continuum spectrum, $E_{lm} = l (l +
1)$, with the  $2l+1$ degeneracy $m = -l, \cdots l$. Indeed, 
any finite eigenvalue approaches its continuum
value as $1/L^2$ in the  limit of infinite refinement $L \rightarrow
\infty$.  
In  addition, we note that  the right eigenvectors are well
approximated by the continuum spherical harmonics, $ Y_{lm}(\hat
r_x)$, evaluated at the lattice sites $\hat r_x$.
This is illustrated in~\figref{fig:FEMdiag},
where the eigenvalues are  estimated by  
computing diagonal matrix elements, 
\be
E_{l,m}  = \frac{Y^*_{lm}(\hat r_x) M_{xy} Y_{lm}(\hat r_y)}{ \sum_x  \sqrt{g_x} Y^*_{lm}(\hat
r_x) Y_{lm}(\hat r_x)} \; ,
\ee
against $l$. On the left of~\figref{fig:FEMdiag},  the lack of $2l +1$
degenerate multiplets  on a coarse lattice with $L = 8$,  as we approach the cut-off at large $l \sim
O(L)$, shows the breakdown of
rotational symmetry. On the right of Fig.\ref{fig:FEMdiag}, the degeneracy of the spherical representation   holds to high accuracy
and the average of the $2 l + 1$ eigenvalues at each $l$ level gives the
correct continuum dispersion relation, $l(l+1)$, to $O(10^{-5})$ for $l \ll L = 128$.

We will refer to the exact convergence of fixed eigenvalues and their
associated eigenvectors to the continuum as the
cut-off is removed as {\bf spectral
  fidelity}. This is a theoretical consequence of FEM convergence
theorems for {\bf shape regular} linear elements as the diameter goes
uniformly to zero~\cite{StrangFix200805}.  We do not provide a proof
but will assume that this property holds for our
implementation if we apply the DEC~\cite{Arnold2006} to the
Laplace-Beltrami operator.

\begin{figure}[ht]
\centering
\includegraphics[width=0.6\textwidth]{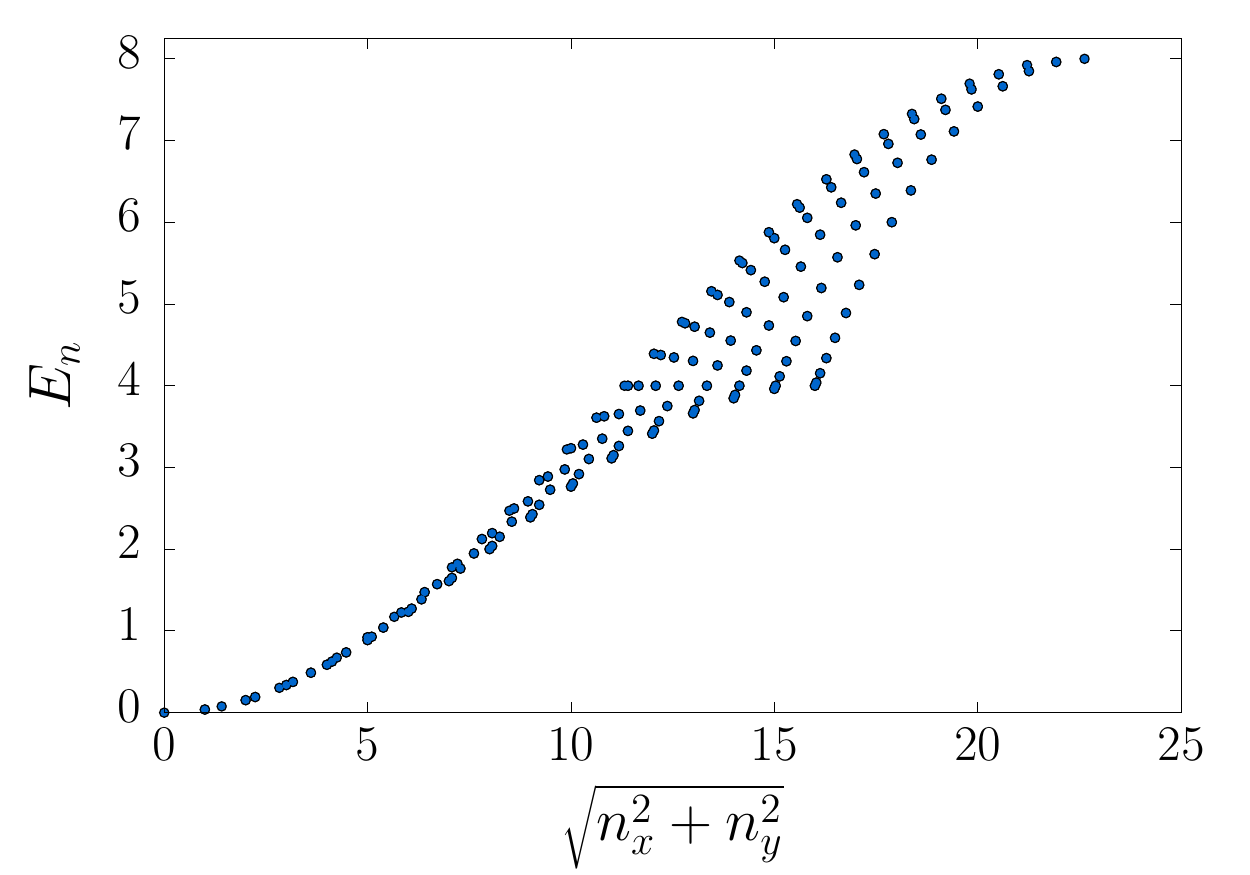}
 \caption{\label{fig:2DtorusEV} The zero mass spectrum $E_n = 4  (\sin^2(
   k_x/2)+\sin^2(k_y/2))$, $k_\mu = \frac{n_\mu \pi}{L}$, $n_\mu \in
   \left[-L/2,L/2-1\right]$ for the \ndim{2} lattice Laplacian  on a
   regular   $32 \times 32$ square lattice plotted against $\sqrt{n^2_x + n^2_y}$.}
 \end{figure}

It is useful to compare the low spectrum of our FEM
operator on $\mS^2$, shown in~\figref{fig:FEMdiag}, to the
corresponding spectrum on the hypercubic refinement of the torus
$\mT^2$, shown in~\figref{fig:2DtorusEV}. For a finite $L^d$ toroidal
lattice, the exact hypercubic spectra is given by
\be
E_n = \sum_\mu 4 \sin^2( k_\mu/2) + m^2_0 \simeq m^2_0+  \sum_\mu [ k^2_\mu - 
\frac{1}{12} k^4_\mu + \cdots],
\label{eq:HypSpectrum}
\ee 
where the discrete eigenvalues are enumerated by $ k_\mu = 2\pi n_\mu/L$ for integer $n_\mu
\in [- L/2, L/2-1]$.  The eigenvectors can be found by a Fourier analysis, and are given by
\be
\phi_n(x) = \frac{1}{\sqrt{N}} e^{-i k \cdot x} \quad, \quad
\phi^*_n(x) =\frac{1}{\sqrt{N}} e^{i k \cdot x} \; .
\label{eq:HyperEV}
\ee
Converting to dimensionful variables ($m, p$) and holding a physical 
mass fixed   ($m \equiv  m_0/a$),  the dispersion relation is
$a^2 E_n = m^2 + p^2 -  \frac{1}{12}\sum_\mu a^2p^4_\mu + \cdots$.
The Lorentz breaking term vanishes as $a^2 = O(1/L^2)$. 
Again {\bf spectral fidelity} holds for  any fixed spectral value $E_n$ as the lattice converges to the continuum. However, as we approach
the cut-off, it is increasingly distorted.

Clearly the spectral fidelity on the hypercubic torus is comparable to
the FEM spectra on $\mS^2$. This is not surprising; indeed the square lattice
can also be viewed as a FEM realization. One simply divides each
square into two right angle triangles and notices that the formula in~\eqref{eq:DECscalar} implies a zero contribution
on the diagonal links.  This generalizes to higher dimensional
hypercubic lattices when using the DEC form. 
 Of course, the major difference between 
 the square  lattice on $\mR^2$ and the simplicial sphere $\mS^2$ 
is  the former  breaks one  rotational
isometry of the continuum but conserves two discrete subgroups of
translations, whereas the later breaks all three isometries of
$O(3)$ down to a fixed finite subgroup independent of
the refinement.

\subsection{Obstruction to Non-Linear Quantum Path Integral } 

Having demonstrated the spectral fidelity of our construct{ion on $\mS^2$
at the Gaussian level, we turn next to the more difficult problem  of an
interacting quantum theory, starting with the FEM action given
in~\eqref{eq:simplicial_action}.  We have
performed extensive Monte Carlo simulations for the path integral
given the  FEM action in~\eqref{eq:simplicial_action} on $\mS^2$ with a conclusive result:
the FEM action does not converge to a spherically symmetric theory as
you approach  the continuum and fails to  have a well defined critical surface. 
Attempting to locate the critical surface, we monitor the Binder
cumulant. The results are given on the left-hand side
of~\figref{fig:NoCT}. As we increase the cut-off (or
$L \rightarrow \infty$) while tuning the only relevant coupling, $\mu^2_0$,
the fourth Binder cumulant should stabilize to the known, exact value
$U^*_{4 } = 0.8510207(63)$ for the Ising CFT (see~\secref{sec:numerical_CT} for details).  Instead we see that there
is an alarming instability at large $L = O(100)$, which we believe is due to the
lack of well-defined critical surface with a second order phase transition
required to define a continuum limit.  A more graphic indication of
this failure is present when we consider the average value of
$\< \phi^2_i\>$ as a function of position on  the sphere as shown on the
right-hand side of~\figref{fig:NoCT}.  Due to a spatial variation in
the UV cut-off, this is not uniform, with variation that can be seen
by comparing the regions close to and far form the 12 exceptional
five-fold vertices of the original icosahedron.  As we will show, this
failure is also evident in a lattice perturbation expansion.  
At small $\lambda_0$ a spherically asymmetric contribution to $\< \phi^2_i\>$ is
given by a UV divergent one-loop diagram.

\begin{figure}[t]
\centering
\includegraphics[width=0.58\textwidth]{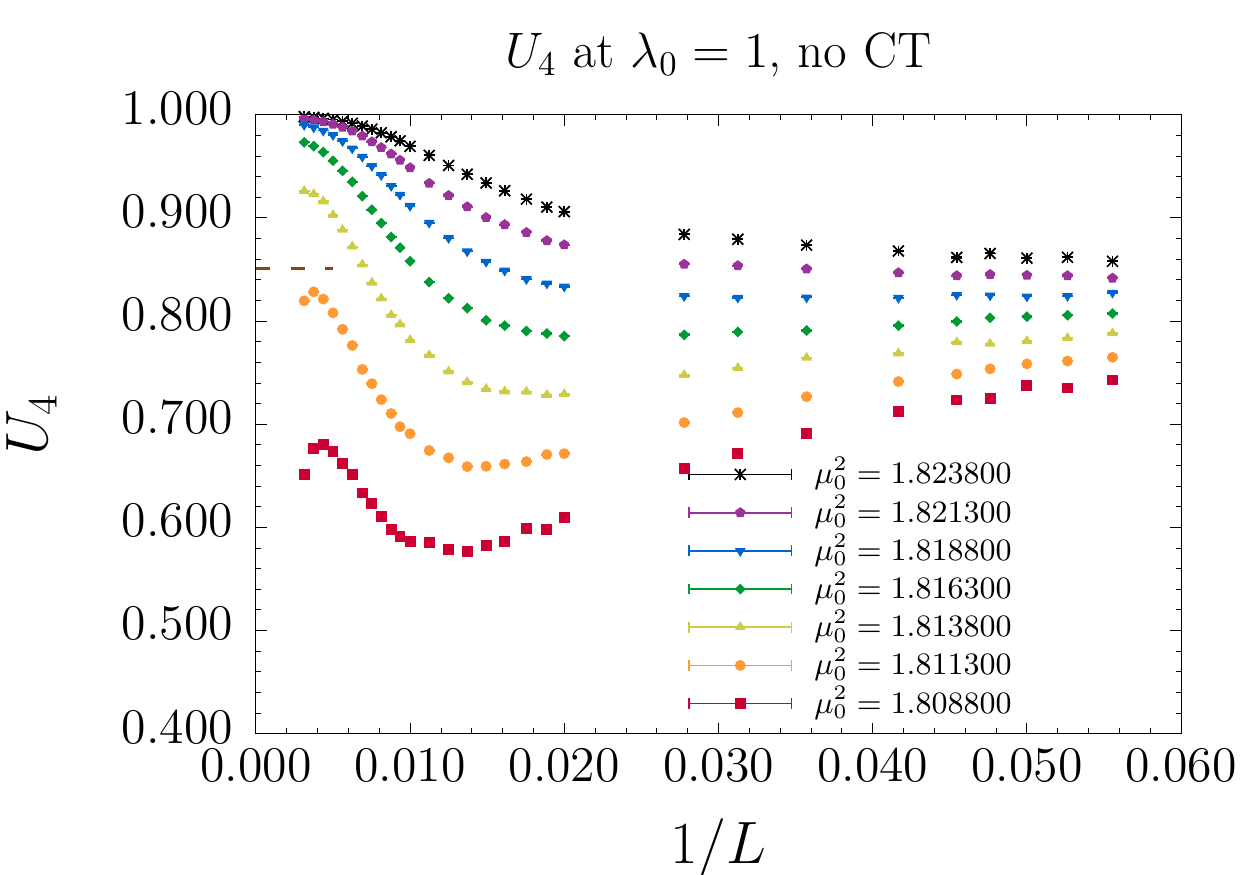}
\includegraphics[ width=0.380\textwidth] {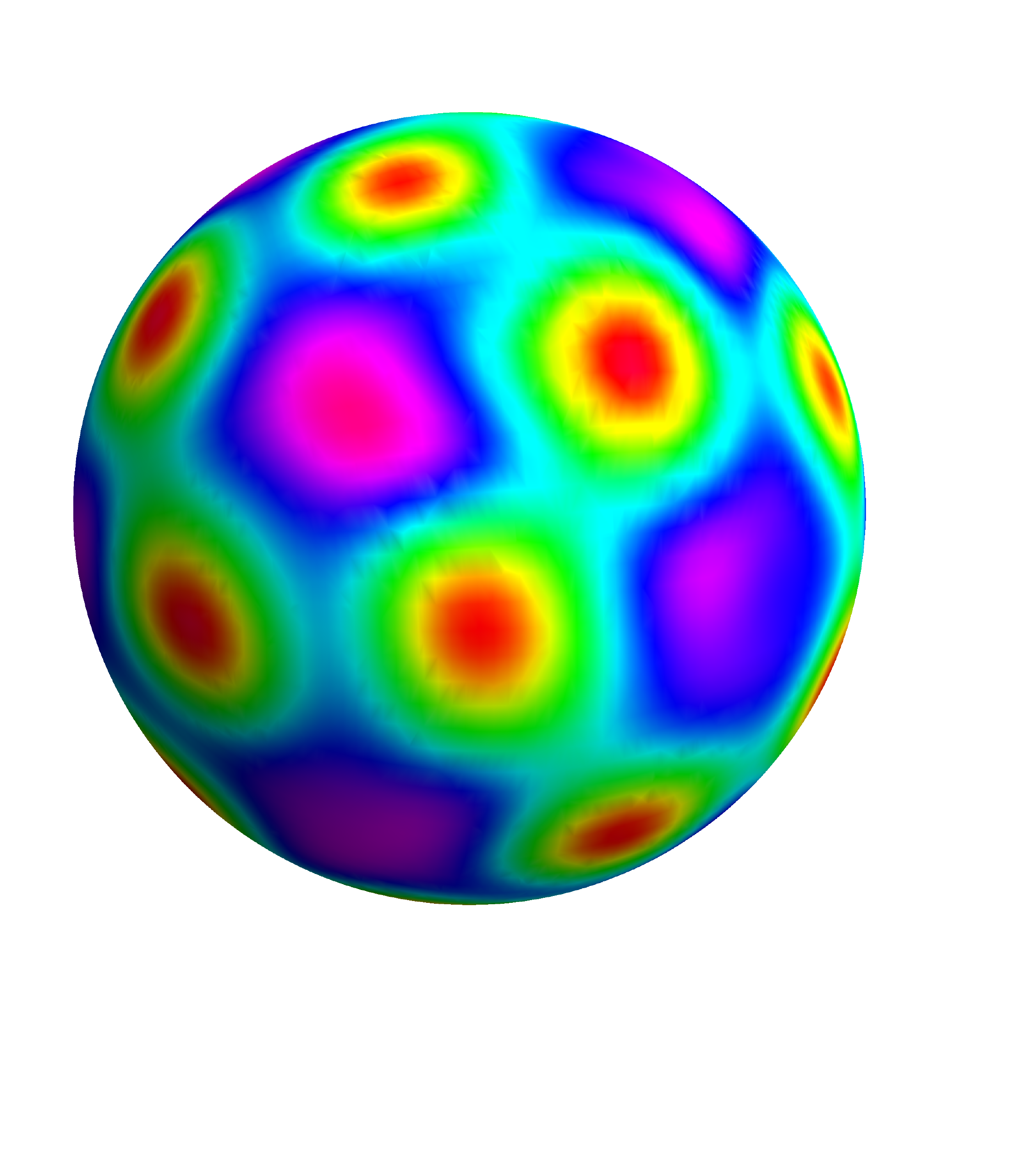}
 \caption{\label{fig:NoCT} On the left the Binder cumulants for the FEM
  Lagrangian with no QFE counter term. On the right the  amplitude of
  $\< \phi^2_i\>$  in simulations with the  unrenormalized FEM
  Lagrangian.  }
 \end{figure}

 In conclusion, in the application of FEM to quantum field theory, we
 have encountered a fundamentally new problem. The FEM methodology has
been developed to  give a discretization for non-linear PDEs that converges to the correct
 continuum solution as the simplicial complex is refined.  In the
 context of a Lagrangian system for a quantum field, this implies a
 properly implemented FEM should therefore guarantee convergence to
 all smooth classical solutions of the EOM as the cut-off is removed. However, quantum field
 theory is more demanding. The path integral for an interacting
 quantum field theory is sensitive to arbitrarily large fluctuations---even in perturbation theory---on all distance scales, down to the
 lattice spacing, due to ultraviolet divergences.  This amplifies
 local UV cut-off effects on our FEM simplicial action.  

 Our solution is to introduce a new Quantum Finite Element (QFE)
 lattice action that includes explicit counter terms to regain the
 correct renormalized perturbation theory. We conjecture that, for any
 ultraviolet complete theory, if the QFE lattice Lagrangian is proven
 to converge to the continuum UV theory to all orders in perturbation
 theory, this is sufficient to define its non-perturbative extension
 to the IR.  We believe this is a plausible conjecture consistent with
 our experience on hypercubic lattices for field theories in $\mR^d$,
 but it is far from obvious. It needs careful theoretical and
 numerical support to determine its validity.  In~\secref{sec:numerical_CT} and~\secref{sec:numerical_correlators}
 we give extensive numerical test for $\phi^4$ on $\mS^2$ on the
 critical surface. In particular the new Monte Carlo simulation of the
 Binder cumulant $U_4$ on the right hand side of~\figref{fig:Q_vs_musq} including the quantum counter term gives a
 critical value of the Binder cumulant,
 $U_{4 ,\mathrm{cr}} = 0.85020(58)(90)$, in agreement with the
 continuum value, $U^*_{4} = 0.8510$, to about one part in $10^3$.
 While this is promising, the limitations due to statistics and the restriction of our studies to the simplest scalar \ndim{2} CFT is duly acknowledged.


\section{\label{sec:quantum_corrections}Ultraviolet Counter Terms on the Simplicial Lattice}

To remove the quantum obstruction to criticality for the FEM simplicial action, we begin
by asking if we can add counter terms to the
action~\figref{eq:simplicial_action} to reproduce the renormalized
perturbation expansion order by order in the continuum limit at the UV
weak coupling fixed point.  On a hypercubic lattice, it has been
proven for $\phi^4$ theory~\cite{Rothe:1992nt} that this can be
achieved by taking the lattice UV cut-off to infinity, holding the
renormalized mass and coupling fixed.  Here we will suggest how this
can be achieved on a simplicial lattice for 2d and 3d.
While we do not attempt a  proof, the similarity with the
hypercubic example strongly suggests that a proof could be found at
the expense of increased technical difficulty.

The $\phi^4$ theory is super-renormalizable in 2d and 3d,
with one-loop and two-loop divergent diagrams, respectively, only
contributing to the two-point function. The one-loop diagram is
logarithmically divergent in 2d and linearly divergent in
3d. The two-loop diagram is  UV finite in 2d, but
logarithmically divergent in 3d. The divergent contributions
renormalize the mass via the one-particle irreducible (1PI)
contribution to the self-energy,
\be
\Sigma(x,y) = - 12\lambda_0 G_0(x,y) \delta(x-y) + 96\lambda^2_0 G^3_0(x,y) \;,
\ee
as illustrated in the first two panels in~\figref{fig:OneTwoLoop}. 
In both 2d and 3d the three-loop diagram, and all higher-order diagrams, are UV finite.
In 4d there is a logarithmically divergent contribution to the four-point function
which contributes to the quartic coupling $\lambda_0$; however the
complete non-perturbative theory is the trivial free theory with no
interesting IR physics.


\subsection{Lattice Perturbation Expansion}

The perturbation expansion for the $\phi^4$ theory on our FEM lattice (or indeed any lattice) starts
with the partition function,
\be
Z(m_0,\lambda_0) = \int {\cal  D}\phi_i  e^{- \textstyle
  \frac{1}{2} \phi_i M_{i,j} \phi_j    -
  \lambda_i \phi^4_i } \; ,
\ee
by expanding in the quartic term.  In our FEM  representation,  the
action is 
\be
 S[\phi_i] =\frac{1}{2} \phi_i M_{i,j} \phi_j    +
  \lambda_i \phi^4_i = \frac{1}{2} [\phi_i K_{i,j} \phi_j +  m_i^2  \phi^2_i]  +
  \lambda_i \phi^4_i \; .
\ee
 For convenience, both bare parameters $m_i^2 \equiv \sqrt{g_i} m_0^2$ and $\lambda_i \equiv \sqrt{g_i} \lambda_0 $ include the factor of the local dual volume $\sqrt{g_i}$. The  quadratic form ($M_{i,j}$)
includes both the DEC Laplace-Beltrami operator ($K_{i,j}$) and the
bare mass ($m_0$).

Following the standard Feynman rules for a perturbative expansion in $\lambda_0$, we compute the
propagator, $\< \phi_{i} \phi_{j}\> = Z^{-1} \int {\cal D}\phi \; \phi_{i}
\phi_{j} e^{ - S}$, to second order:
\begin{flalign}
\< \phi_{i} \phi_{j}\> =  \left[M^{-1}\right]_{i,j}
&+ \left[M^{-1}\right]_{i, i_1} (- 12 \lambda_{i_1} \left[M^{-1}\right]_{i_1, i_1} )\left[M^{-1}\right]_{i_1, j}  \nonumber \\
&+ \left[M^{-1}\right]_{i,i_1} (96 \lambda_{i_1} \lambda_{i_2} \left[M^{-1}\right]^3_{i_1,i_2} )
\left[M^{-1}\right]_{i_2,j} \; .
\end{flalign}
After amputating the external lines, we find the inverse propagator
$\widetilde M_{ij}( m_0,\lambda_0) =  M_{i,j}  +  \Sigma_{ij}
(m_0,\lambda_0) $, where 
\be
\Sigma_{ij} (m_0,\lambda_0) =  - 12 \lambda_i
\left[M^{-1}\right]_{ii} \delta_{ij} +96 \lambda_i \lambda_j \left[M^{-1}\right]^3_{i,j}  +
O(\lambda^3_0) \; .
\label{eq:1PI}
\ee
is the 1PI simplicial self-energy.

\begin{figure}[ht]
\centering
\includegraphics[width=0.45\textwidth]{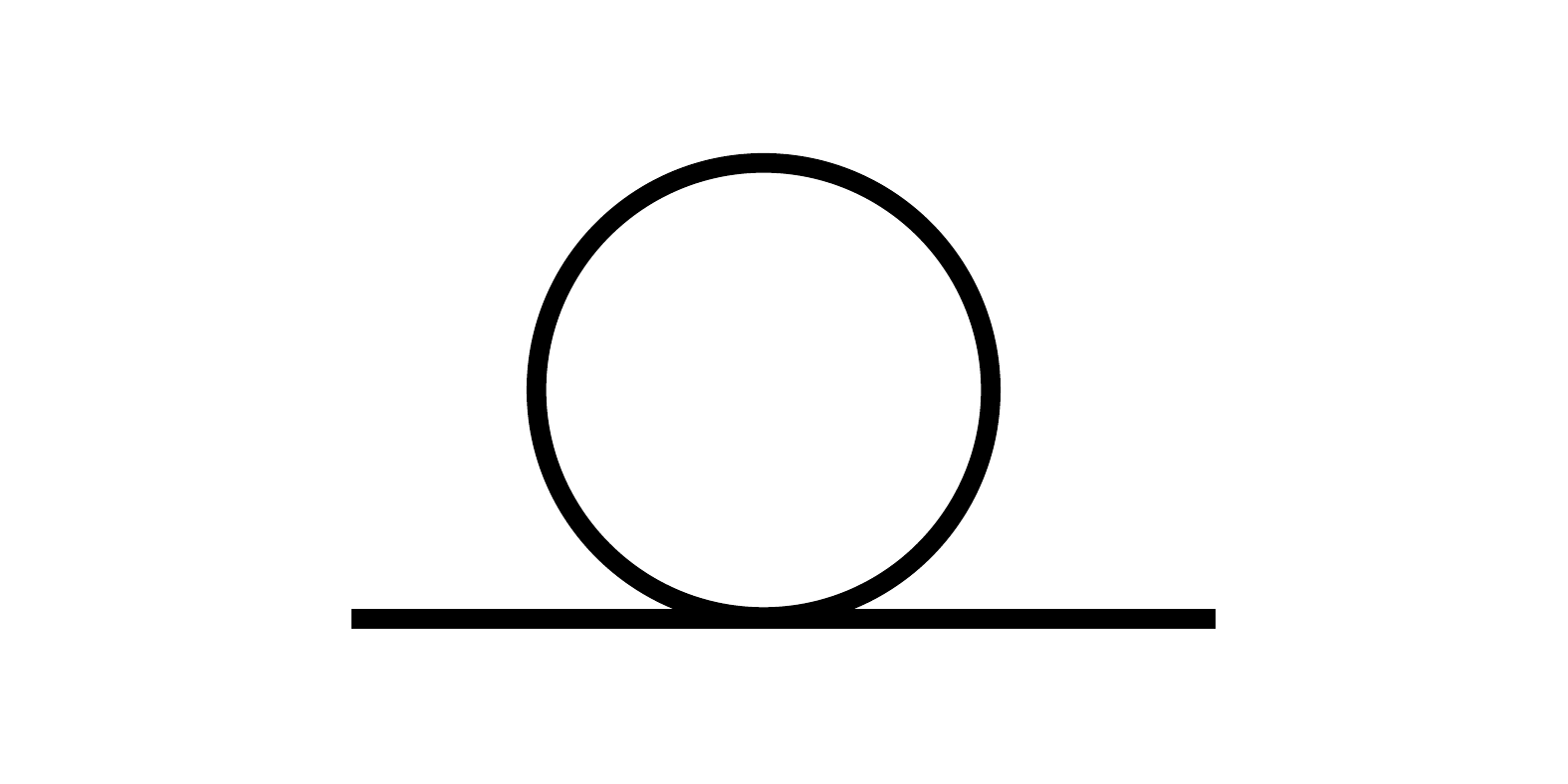}
\includegraphics[width=0.45\textwidth] {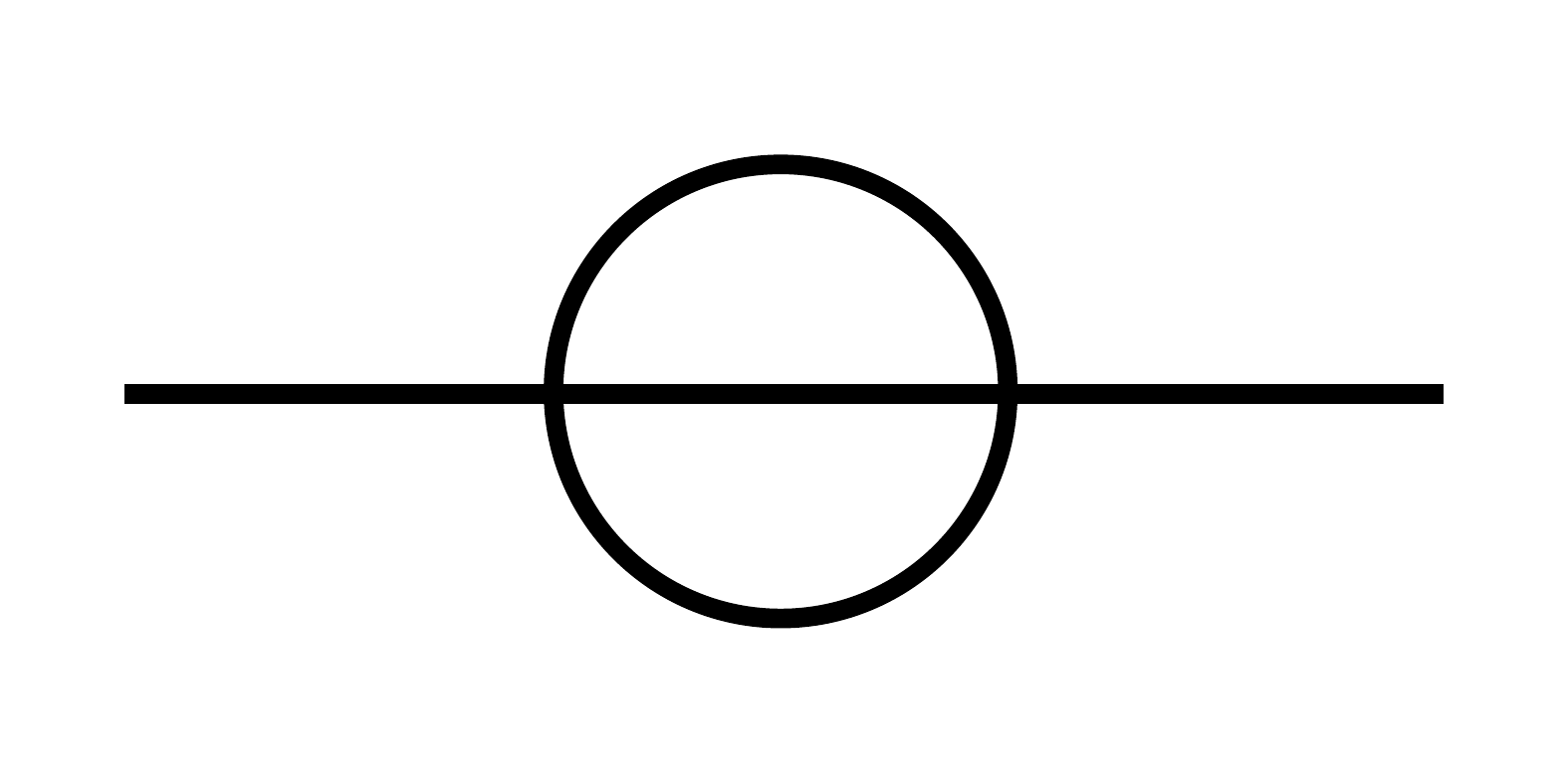}
\includegraphics[width=0.45\textwidth] {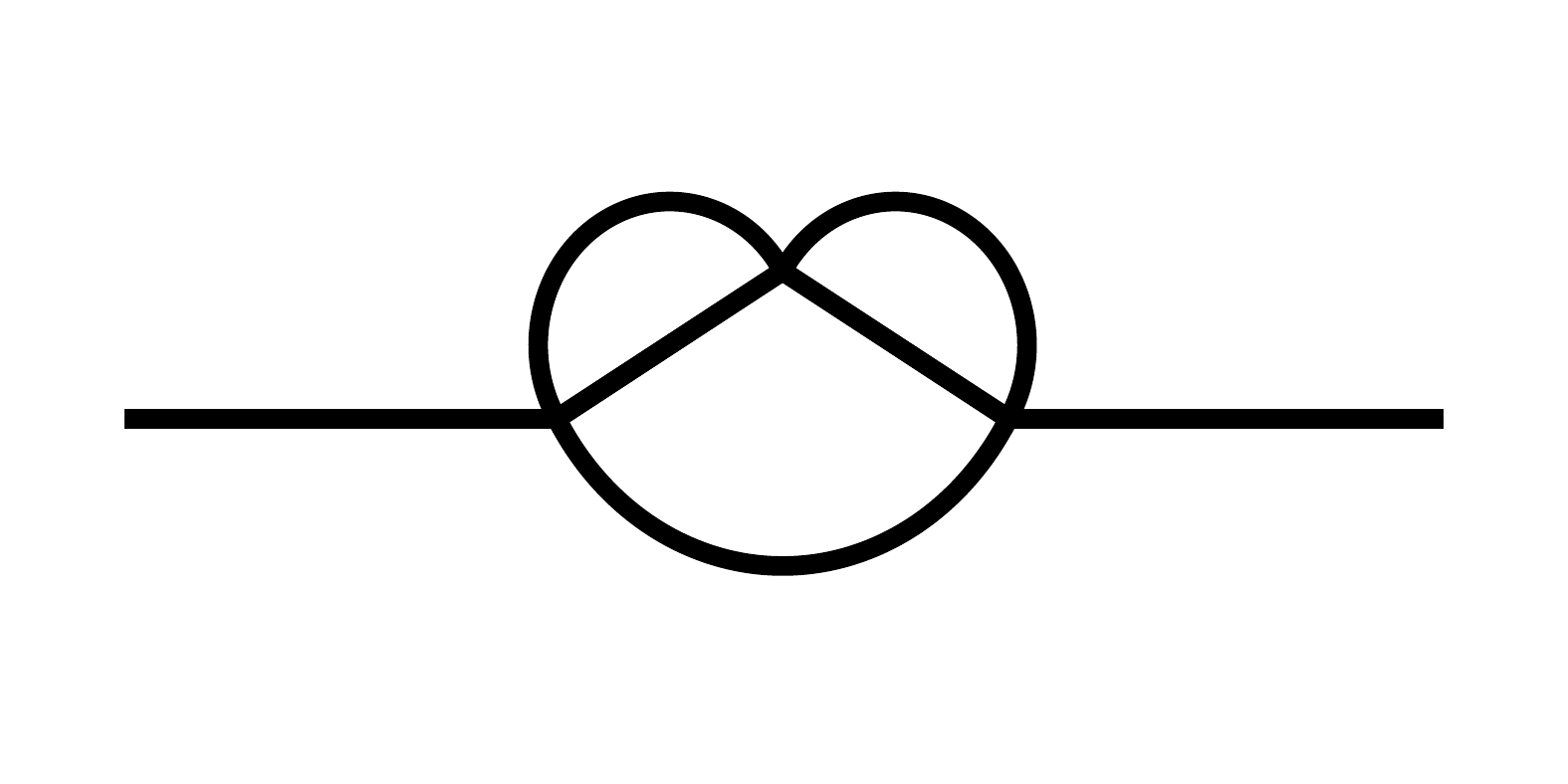}
 \caption{\label{fig:OneTwoLoop} The one-loop diagram
is logarithmically divergent in 2d and linearly divergent in
3d, whereas the two-loop is finite and logarithmically divergent
for 2d and 3d, respectively. The three-loop diagram is
finite in both 2d and 3d.}
 \end{figure}
%


\subsection{\label{sec:OneLoopCT}One Loop Counter Term}

Since there is no analytic
spectral representation of the free FEM Green's function, we compute
the one loop diagram  in co-ordinate space by numerical evaluation of
the propagator.  To be concrete,
for 2d on $\mS^2$ and for 3d on $\mR \times \mS^2$,
the Gaussian  term is 
\be
\phi_{x,t_1} M_{x,t_1; y,t_2} \phi_{y,t_2}  =  \phi_{x,t_1}
K_{x,y} \phi_{y,t_1}+\sqrt{g_x} ( \phi_{x,t}  -
\phi_{x,t \pm 1} )^2
+ m^2_0 \sqrt{g_x} \phi^2_{x,t} \; ,
\ee
where the sites are now labeled by $i = (x,t)$. The \ndim{2} geometry, $\mS^2$, can be viewed as a special
case with a single sphere at $t = 0$.  The integer $t$
labels each sphere along the cylinder, and $x$ indexes the sites on
each sphere. $K_{x,y}$ is non-zero on only nearest neighbor links
$\<x,y\>$.  To take the continuum limit we
need to define the normalization convention of our lattice constants:
\be
\sum^{N-1}_{x=0} \sqrt{g_x} = N  \quad \mbox{and} \quad \sum_{\<x,y\>}
 K_{x,y} = (2/3) E \;,\label{eq:normalization}
\ee
where in this specific context $E$ refers to the number of edges on the simplicial graph. The extra factor of $2/3$ is introduced to compensate, on average, for
the six nearest neighbors per site on the sphere  relative to a
conventional square lattice with four nearest neighbors. In flat
\ndim{2} space  this factor of 2/3 for the triangular lattice gives
the same dispersion relation, $E = m^2_0 + k^2 + O(k^4)$, as the square lattice. 

In lattice perturbation theory on a \ndim{2} FEM simplicial lattice, we expect
the logarithmically divergent  one-loop term to give a site-dependent mass shift,
\be
m^2_0 \rightarrow m^2_0 + \Delta m_x^2 \;.
\ee 
The numerical computation of the one-loop diagram on $\mS^2$  does
indeed have the appropriate  form,
\be
\left [ M^{-1} \right]_{xx} \approx
\frac{\sqrt{3}}{8 \pi} \log \left(\frac{1}{m^2 a_x^2}
\right) = \frac{\sqrt{3}}{8 \pi} \log (N) + \frac{\sqrt{3}}{8 \pi}
\log \left(\frac{a^2}{a_x^2}\right) + O(1/N) \; . 
\label{eq:counterterm}
\ee
\begin{figure}[t]
\centering
\includegraphics[width=0.7\textwidth]{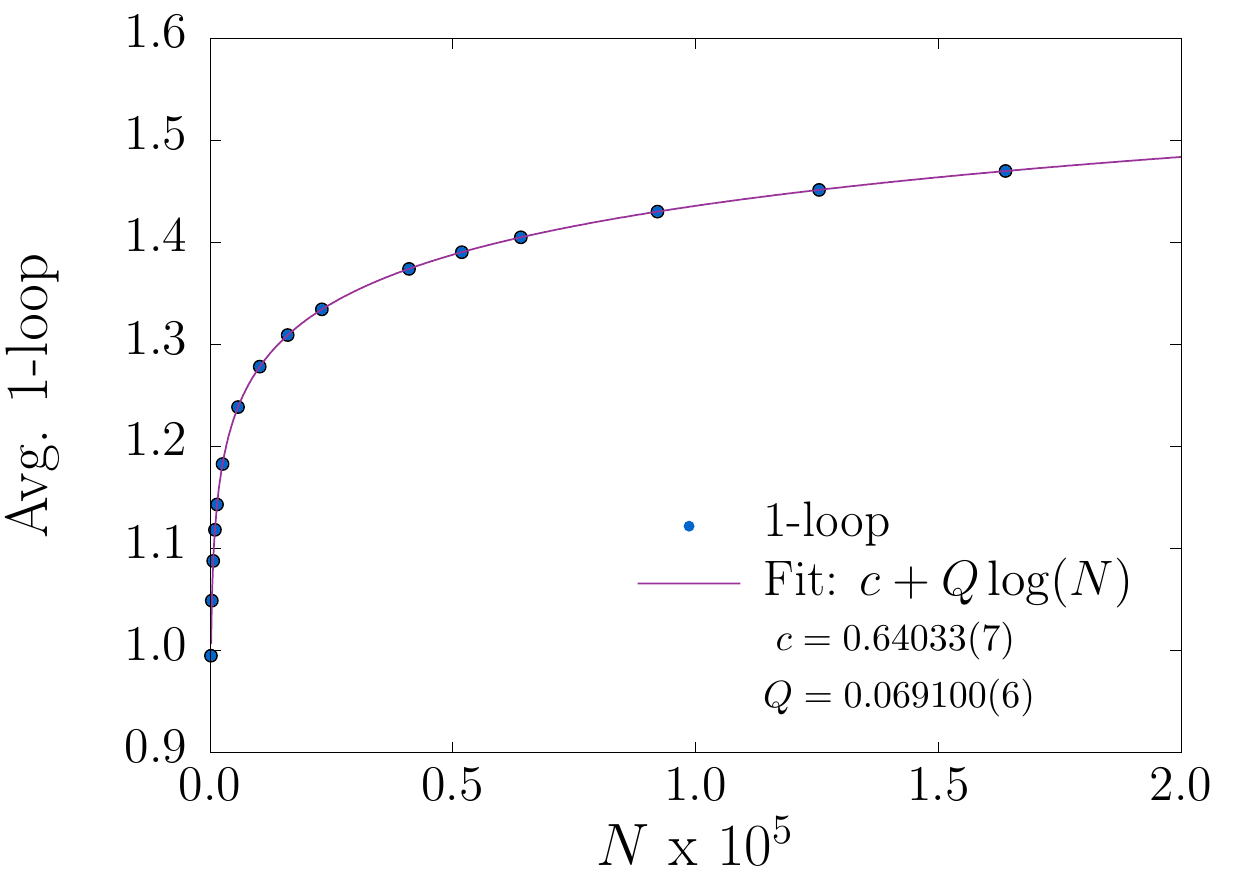}
\caption{\label{fig:CTexact} The numerical fit of
the lattice one loop  log divergent determining coefficient 
$Q$  compared to the  continuum value: $Q = \sqrt{3}/( 8
\pi) = 0.0689161\cdots$.}
\end{figure}
The logarithmic divergence is regulated by the IR mass, $m^2$. Of course on the lattice,  at fixed dimensionless
bare mass, $m^2_0 =a^2 m^2$ and bare coupling, $\lambda_0$, there is
no actual UV divergence. The renormalized perturbation expansion is defined  by fixing the physical 
mass ($m$), which in the
continuum limit ($N\rightarrow \infty$) corresponds to  a vanishing
 effective lattice spacing, $a^2m^2= O(1/N)$. On a hypercubic lattice, 
there is a universal cut-off ($\pi/a_x = \pi/a$), whereas on a simplicial
lattice  the one-loop term  separates into two terms on the 
right side of~\eqref{fig:CTnumerical}: (1)  a position independent
divergence
and (2) a finite (scheme dependent)  constant for each site $x$.

We have checked numerically, to high
accuracy, that the first divergent term is independent of position. This is a crucial
observation which we believe is a general consequence of the FEM
prescription. Spatial dependence for a divergent term would imply the
need for a divergent counter term to  restore spherical symmetry.
This departs from the standard cut-off procedure on
regular lattices and would be very difficult if not impossible to
implement. Moreover, we show in \figref{fig:CTexact}  that the fit coefficient is
accurately determined to be $\sqrt{3}/(8 \pi)$, which
is the exact continuum value. 

 A heuristic argument for the value of this  constant 
follows from  our $\mS^2$ tessellation as a nearly regular triangular lattice.
In flat space, relative to a square lattice, the density of  states 
on a  triangular  lattice is $N^{-1} \sum_n \rightarrow \int d^2k~\rho_T(k)$, where
$\rho_T = \sqrt{3}/(8 \pi^2)$. The extra factor of
$\sqrt{3}/2$ is due to  the area $|\sigma^*_0| $
of the dual lattice hexagon relative to the dual lattice square on a
hypercubic lattice.  However, since this constant defines the local
charge  in the \ndim{2} Coulomb law and our linear elements do not
  implement  local flux conservation, this is surprising result. Nonetheless
in~\secref{sec:univlogdiv} we will prove this
based on the FEM principle of {\bf spectral fidelity} and
the renormalization group for  the mass anomalous dimension.
To the best of our knowledge this is a novel extension of FEM
convergence theorems.


\begin{figure}[t]
\centering
\includegraphics[width=0.7\textwidth]{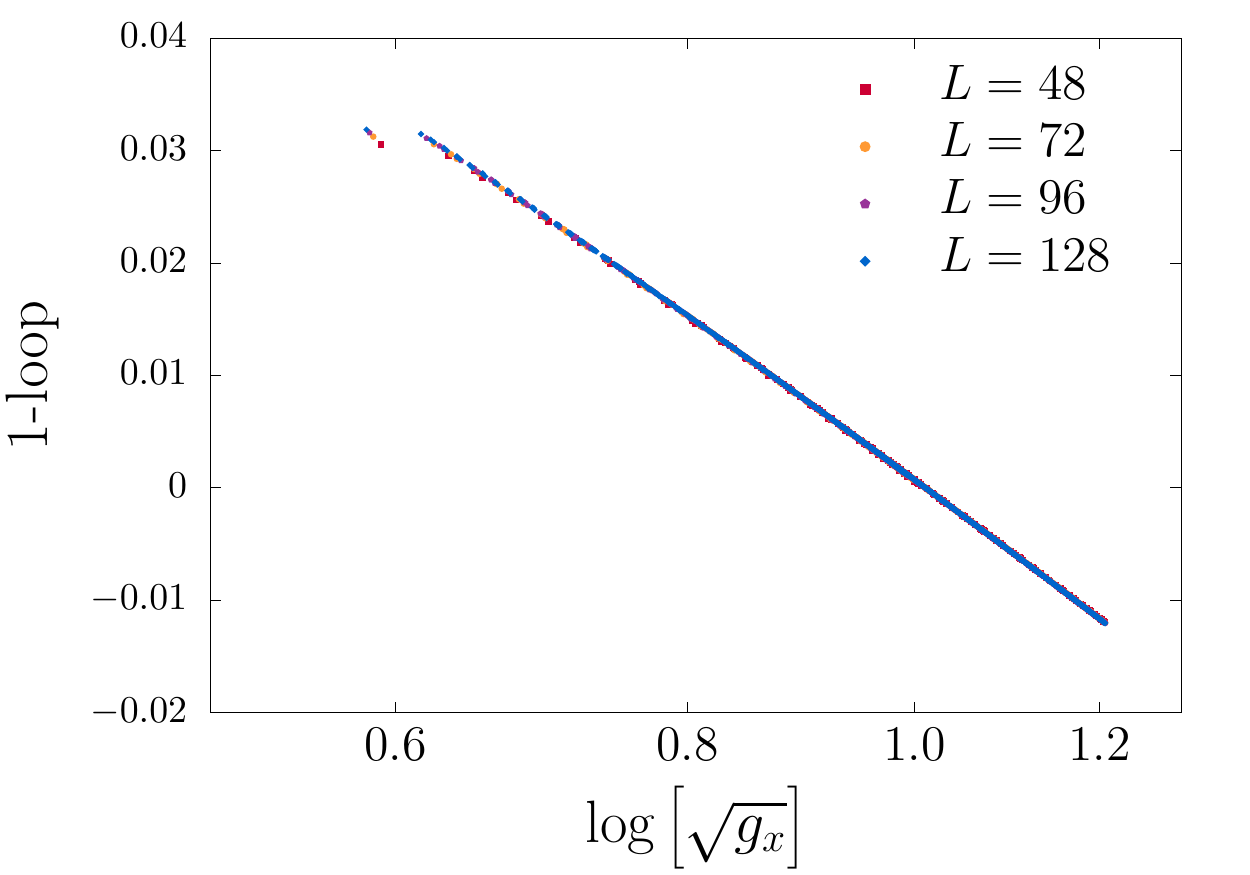}
\caption{\label{fig:CTnumerical} One Loop counter term
plotted against $\log(\sqrt{g_x})$.}
\end{figure}
Next consider the finite spatial dependence term in~\eqref{eq:counterterm}.  As
illustrated  in~\figref{fig:CTnumerical}, it is almost
exactly a linear function of $\log[\sqrt{g_x}]$.  Again there is
a heuristic explanation for this.  An almost
perfect analytic expression can be found by
considering the dilatation incurred by the projection of  our triangular tessellation of the
icosahedron   onto the sphere.  The icosahedron is a
manifold with a flat metric except for 12 conical singularities at the
vertices. Our choice of the simplicial complex began with  flat equilateral triangles on
each of the 20 faces of the original icosahedron. The radial projection onto
the sphere is a Weyl transformation to constant curvature with 
conformal factor (or Jacobian of the map),
\be 
\sqrt{g(x)} = e^{\sigma(x)} =\frac{(x^2_1 + x^2_2 + 1 -
  R_c^2)^{3/2}}{\sqrt{1 - R_c^2}} \; ,
\ee
where $R_c$ is the circumradius for one of the 20 icosahedral faces and $x_1,x_2$ the
flat coordinates on that face. This formula gives a line coinciding
 with the linear data in~\figref{fig:CTnumerical}.
 Although this is an extremely   good approximation to the counter term, it is
not perfect since it neglects the conical singularities in the map
near the 12   exceptional vertices of the original
icosahedron. These exceptional points are  visible  in the upper left
corner  of~\figref{fig:CTnumerical}. 
 We find that the true numerical computation does give better
 results and, moreover, it  is a general method not requiring our careful triangulation procedure.   
 A more irregular procedure should also be allowed. The lesson 
is that the simplicial metric, $g_{ij} = \{ l_{ij} \}$, on a given Regge manifold not only defines the intrinsic geometry 
but also a co-ordinate system breaking diffeomorphism explicitly.  The
set of lengths includes both a definition of curvature and the discrete
co-ordinate system. The counter term must compensate for this
arbitrary choice. 

The QFE one loop counter term is introduced to cancel the  finite position
dependence  in~\eqref{eq:counterterm} that violates rotational
invariance  to order $\lambda_0$. To project out the spherical component of a local 
scalar density, $\rho_x$, we average over the rotation group $R \in SO(3)$,
\be
\frac{1}{\mbox{vol}O(3)} \int dR \; \rho(R \hat r) = \frac{1}{4 \pi} \int
  d\Omega \; \rho(\theta, \phi) \simeq N^{-1}\sum_x \sqrt{g_x} \rho_x \; .
\label{eq:project}
\ee
Applied to $\rho_x = \left[ M^{-1} \right]_{xx}$, the subtracted
lattice Green's function is 
\be
\delta G_{xx}   = \left[M^{-1}\right]_{xx} - \frac{1}{N} \sum_{x=1}^N \sqrt{g_x} \left[ M^{-1} \right]_{xx} .
\ee
This removes completely the logarithmic divergence, leaving the position dependent finite counter term which adds a contribution to the FEM action, 
\be
S_{FEM} \rightarrow S_{QFE} = S_{FEM} +  6 \sum_x\sqrt{g_x} \lambda_0 \; \delta G_{xx} \; \phi^2_x
\;  .
\label{eq:ct1}\
\ee

In 2d $\phi^4$ theory this is the only UV divergent term. 
We also computed the two-loop contribution to the 1PI simplicial self
energy~\eqref{eq:1PI} which in co-ordinate space is given
by  the third power of the Green's function,
$\left[M^{-1}\right]^3_{xy}$, between the vertex at $x$ and $y$. The
position dependence,

\be
\delta \left[ G_x \right]^3=  \sum_y ( \left[M^{-1}\right]^3_{xy} - \frac{1}{N}
\sum_{x=1}^N \sqrt{g_x} \left[ M^{-1} \right]^3_{xy} )\; ,
\ee 
is
found again  by subtracting the average contribution on the sphere.
The result is plotted in~\figref{fig:TwoLoop2d}, against
$\log[\sqrt{g_x}]$ which demonstrates that it vanishes rapidly in the 
continuum limit as  $1/L^2 \sim 1/N$ consistent with
na\"ive {\em power counting} in continuum perturbation theory.
\begin{figure}[ht]
\centering
\includegraphics[width=0.7\textwidth]{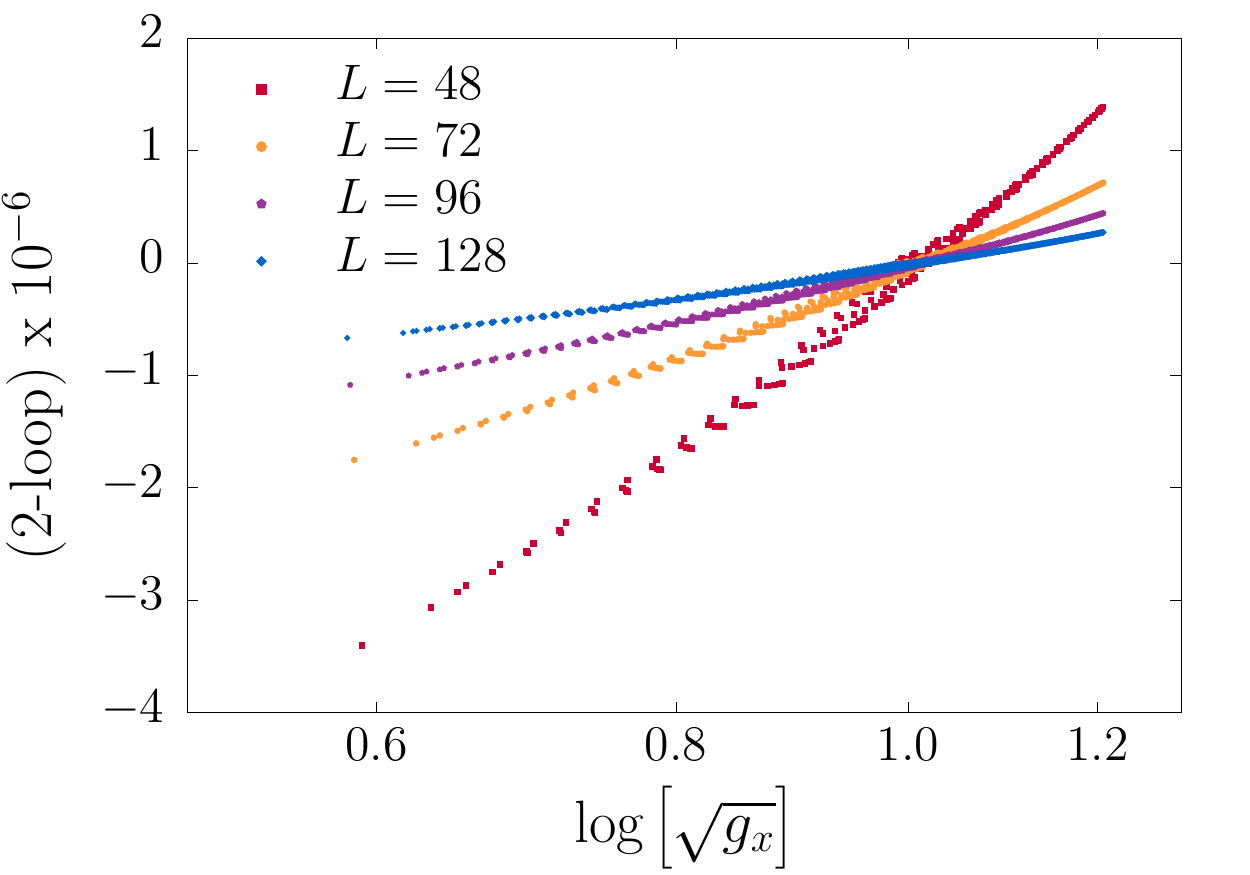}
\caption{\label{fig:TwoLoop2d} The contribution of
the two loop term in 2d.  }
\end{figure}
While we do not attempt a general analysis of the simplicial lattice
perturbation expansion, it is plausible based on analogous studies on
the hypercubic lattice~\cite{Reisz:1987da,Reisz:1987px} that after
canceling cut-off dependence in the one loop UV divergent term the
entire expansion restores the renormalized perturbation series in the
continuum limit.  However our test on the \ndim{2} Riemann  sphere
simplicial lattice goes further. Namely we
give numerical evidence that this \ndim{2} QFE lattice action, when tuned to
the the critical surface, approaches the non-perturbative continuum
theory in the IR for the Ising CFT field theory.


\subsection{\label{sec:univlogdiv}Universal Logarithmic Divergence}

The success of our QFE action prescription
depends on the coefficient of the  logarithmic
divergence being exact in the continuum limit.  On $\mS^2$ or
any {\bf maximally symmetric} space, this  implies the coefficient is 
also independent of position. At first, this
observation  appears surprising. After all, the UV
divergence is sensitive to the local, short distance cut-off which is
not uniform.  We claim this is a consequence of the observed spectral fidelity  of
the DEC Laplace-Beltrami operator and renormalization group (RG) for
a logarithmic divergence  relating UV divergences to the IR regulator.
Adapting the argument to a general \ndim{2} Riemann manifold is
in principle straight forward, matching the local divergence to the one
loop renormalization of the continuum theory~\cite{Jack:1983sk}  at each point $x$.

Let us first   apply this renormalization group (RG) argument to the hypercubic lattice,
where we already know and understand the answer. Suppose that the
one-loop diagram has a logarithmic term,
\be
 G_{xx} (m) \simeq  c_x log (1/m^2 a^2_x) + O(a^2 m^2) \; .
\label{eq:LogCoef}
\ee
with an unknown position dependent coefficient $c_x$.
To isolate this coefficient, we take the logarithmic derivative, finding
\be
\gamma_1(m^2) = -  m \frac{\partial }{\partial m} G_{xx} (m^2)=  2  \iint^{\pi/a}_{-\pi/a}
\frac{d^2p} {(2 \pi)^2}\frac{ m^2}{(4 a^{-2} \sum_\mu
  \sin^2(a p_\mu/2)  +  m^2  )^2}  \; .
\ee
so that $\gamma_1(m^2) = 2 c_x + O(a^2 m^2)$.  This is the one-loop
contribution to the anomalous dimension of the $\phi^2$ operator. (To
clarify scaling, we have re-introduced the lattice spacing $a$ so that
$p = k/a, m = m_0/a$ have mass dimensions.)  By  power counting, 
the integral resulting from the logarithmic
derivative is UV finite in the continuum limit. We may now introduce a
new cut-off, $\Lambda_0$,  separating the IR from the UV: $m \ll \Lambda_0\ll
\pi/a$. To estimate the integral, we can take the continuum  limit
$\pi/a \rightarrow \infty$,
\be
\gamma_1(m^2) \simeq  2 \iint^{\Lambda^2_0}_0
\frac{dp^2} {4 \pi }\frac{m^2}{(p^2  +  m^2  )^2}  +
O(m^2/\Lambda^2_0) =\frac{1}{2 \pi}
\frac{1}{1 + m^2/\Lambda^2_0} + O(m^2/\Lambda^2_0) \; ,
\ee
followed by taking the IR regulator to zero: $m^2/ \Lambda^2_0 \rightarrow 0$.
This proves that the coefficient $c_x = 1/(4\pi)$  is independent of
position and identical to the continuum
value on $\mR^2$. Of course, this is not surprising for flat space. 

Now we can apply the same reasoning to the simplicial sphere $\mS^2$, 
beginning with the spectral representation of the
Green's function, 
\be
 G_{xx} (m^2) = \sum_{n}\frac{ \phi^*_n(x) \phi_n(x) } {E^{(0)}_{n} +
   m^2_0} \; .
\ee
On the basis of FEM {\bf spectral fidelity}, assume the low
eigenspectrum for $l \le L_0$ is well approximated to $O(1/N)$  by  spherical harmonics,
\be
E_n \simeq a^2 R^{-2} l(l+1) +m^2_0  \quad ,\quad \phi_n(x) \simeq 
\frac{\sqrt{4\pi}}{\sqrt{N}} Y_{lm}(\hat r_x)\; ,
\ee
where $R/a$ is the radius of the sphere in units of the lattice
spacing. Matching the area of the sphere with $N$ triangles we have
$4 \pi R^2 = a^2 N \sqrt{3}/2 $. For $R = 1$ the spectrum is $l(l+1)$,
but with our convention of setting $a = 1$, we find
$R = \sqrt{3}N/(8 \pi)$.  Next, we fix the eigenvector normalization
by requiring $\sum_x \sqrt{g_x} |Y_{00}|^2 = N/4 \pi$.

 By splitting the spectral sum at the cut-off $L_0$,  and using the
 addition formula, $4 \pi \sum_m Y^*_{lm}(\hat r_x) Y_{lm}(\vec
 r_y) = (2 l +1) P_l(r_x \cdot r_y) $  for $l <
 L_0$, we have
\be
 G_{xy} (m^2_0) \simeq \frac{\sqrt{3}}{8 \pi}\sum^{L_0}_{l = 0}\frac{ (2
   l + 1) P_l(r_x \cdot r_y) } {l(l + 1) +
  \mu^2}
+ \sum^N _{n =(L_0 + 1)^2} \frac{ \phi^*_n(x) \phi_n(y) } {E^{(0)}_{n} +
   m^2} \; ,
 \label{eq:splitting}
\ee
where we introduced $\mu^2 =R^2(m^2_0/a^2)$. 
The first term is indeed rotationally invariant.  At $x = y$, the asymptotic 
limit of the first term is $\sqrt{3}\log(L_0)/( 8 \pi)$, which is the desired behavior. As with the infinite, flat lattice, we wish to convert the sum of the first term to an integral in the limit $N \rightarrow \infty$: 
\be
\sum_l \rightarrow \int dE \rho(E)  \quad, \quad \rho(E) =
\frac{dl}{dE_n}  = \frac{R^2}{2 l +1}
\ee
so that 
\be
G_{xx} (m^2_0) \simeq \frac{\sqrt{3}}{8 \pi}\sum^{L_0}_{l = 0}\frac{ (2
   l + 1)  } {l(l + 1) +
  \mu^2}  \rightarrow\frac{\sqrt{3}}{8 \pi}\int^{\Lambda^2_0}_0 dE^{(0)} \frac{1}{E^{(0)}  +m^2_0}  
\ee
with $\Lambda^2_0 = a^2L_0(L_0 +1)$. Again, if we take
a logarithmic derivative, we have a UV convergent integral
that is saturated by  rotationally invariant contributions up to power corrections at $O(m^2_0/\Lambda^2_0)$,
\be
\gamma(m_0^2) =  - m_0 \frac{\partial }{\partial m_0} G_{xx} (m^2_0)\simeq 
\frac{\sqrt{3}}{ 4\pi}  \int^{\Lambda^2_0}_0  dE^{(0)}   \frac{m^2_0}{(E^{(0)}  +m^2_0)^2}
=  \frac{\sqrt{3}}{4 \pi}  \frac{1}{1 + m^2_0/\Lambda^2_0}  \, .
\label{eq:S2log}
\ee
In the limit $O(m^2_0/\Lambda^2_0) \rightarrow 0$, we isolate the
coefficient in~\eqref{eq:LogCoef} and prove that the divergence  is both independent of
position on the sphere with its coefficient identical
to the continuum one loop diagram on the
sphere. The 
essential assumption we have made is the {\bf spectral fidelity} property of the discrete DEC Laplace-Beltrami  operator. In
Appendix~\ref{app:FreeTheory} we provide the 
exact continuum propagator on the sphere,
\bea
G(\theta,\mu) &=&  \frac{1}{4 \pi}\sum^\infty_{l =0} \frac{ (2 l + 1) P_l(\cos(\theta))}{(l+1/2)^2 +
  \mu^2} =   \frac{2}{4 \pi}\int^\infty_{0}\frac{\cos(\mu  t)  dt }{\sqrt{2
    \cosh(t)- 2 \cos(\theta) }} \nn
 & \simeq&  \frac{1}{4 \pi} [ \log[\frac{32}{1 -  \cos(\theta)}]  - 14\zeta(3) \mu^2]  + O(\theta^2, \mu^4)\, ,\label{eq:contprop2d}
\eea
and identify the coefficient of the logarithmic divergence by 
relating the angle to the lattice cut-off by  $\theta^2  \sim 1/L^2_0
= O(1/N)$. 


\subsection{Comments on Structure  of the  Counter Terms}

In spite of our very limited example of the {\bf QFE counter terms} on
$\mS^2$, we hazard a general interpretation which we hope will guide
extensions to  any smooth Riemann manifold.  This is based 
in part on  current study of counter terms in \ndim{3} $\phi^4$ theory on $\mR \times\mS^2$ . Nonetheless,
we acknowledge the fact that our examples so far
may have very special features not shared more generally.

The first special feature is that the $\mS^2$ manifold is   an example of a
maximally symmetric manifold. This property is} shared by flat space, spheres, and
anti-de Sitter space with constant curvature.  On a d-dimensional
maximally symmetric manifold, there is a full set of $d (d+1)/2$
isometries implying that all points have identical metric properties.
A consequence is in these cases, the counter terms can be easily
defined by projecting out the average over the full set of isometries
as in~\eqref{eq:project} above.  While these maximally symmetric
manifolds are actually of paramount interest, we are also confident that this
restriction is not necessary.  On a general manifold, the procedure
would be to compute the continuum UV divergence at each site and
subtract it  before defining the counter term.

The second special feature of the one-loop diagram in 2d in $\phi^4$
theory is that it is  both local on the lattice and logarithmically
divergent  in the continuum. In 3d the situation
changes. First the  one-loop term in the continuum is
  linearly divergent.  However on the lattice such a divergence is
 not manifest because it enters as a  dimensional factor of $1/a$,
 which in the bare lattice action is chosen implicitly to be  $1/a =
O(1)$. Also  all UV power divergent   diagrams are  finite in the IR so that
there is no need for a dimensionless  mass regulator $m_0 = a m$. 
We have computed the one-loop diagram on the simplicial cylinder $\mR \times \mS^2$, and found
numerically that the counter term is almost identical to the \ndim{2}
case. This is easy to understand. The
spectral decomposition of the \ndim{3} Green's function, 
\begin{align}
 G_{xt,yt'} (m^2_0) \simeq& \frac{\sqrt{3}}{8 \pi}\sum^{L_0}_{l = 0} \sum_k
 \frac{ (2 l + 1) P_l(r_x \cdot r_y) e^{i \omega_k (t -t')}  } {l (l+1) + 
   \omega^2_k+ m^2_0 } \nonumber\\
&\quad\quad\quad\quad+ \sum^N _{n >(L_0 + 1)^2} \sum_k \frac{ \phi^*_n(x) \phi_n(y)e^{i
    \omega_k (t -t') }} {E^{(0)}_{n} +  \omega^2_k+
   m^2_0 } \; ,\label{eq:splitting3D}
\end{align}
is a natural generalization to the case on $\mS^2$ given
in~\eqref{eq:splitting}. We see that the breaking of rotational invariance on
$\mR \times \mS^2$ is very similar to the case of $\mS^2$ for each
mode $\omega_n$ and as such the overall breaking term will
also be very similar to the \ndim{2} case.

Next, in 3d, there is a new, logarithmically divergent two-loop term.
Again, a comparison between 2d and 3d is
interesting.  The one-loop term in Hamiltonian form can be viewed
simply as a problem of normal ordering:
$\phi^4(x) = :\phi^4(x): + \; \phi^2(x) \< \phi^2(x) \>_0$. The diagram
has no unitarity cuts and can be simply dropped when defining a normal
ordered perturbation expansion. In 3d, the two-loop term
introduces a non-local contribution with essential unitarity cuts that
must be preserved in a renormalized perturbation theory. 
On the lattice, the continuum position space two-loop diagram is
replaced by the element-wise third power of the Green's function,
\be
\lambda^2_0 G(t_1 - t_2, \cos(\theta_{xy})) \rightarrow \lambda^2_0 \big[M^{-1}\big]^3_{x,t_1; y,t_2 } \; .
\ee
Again, we must subtract the rotationally invariant contribution at
fixed $t_1 -t_2$ and
$|\hat r_x - \hat r_y|^2 = 2 - 2 \cos(\theta_{xy})$. We have found
numerically that although the rotationally invariant part has a power
fall-off dictated by na\"ive dimensionality, the residual breaking term,
$\delta \mu^2_{xy,t_2-t_1} \phi_{x,t_1} \phi_{y,t_2}$, falls off
exponentially in {\emph{lattice}} units.  Consequently, in
{\emph{physical}} units, it  is exponential in $ (t_2 - t_1)/a$ and
$ |\hat r_x - \hat r_y|/a$ and therefore to leading order may be  replaced
by a local counter term in the QFE action.  We postpone further
analysis of counter terms to  future publications.

We are also considering alternative methods that avoid  the difficulty
of computing individual UV divergent perturbative diagram.  For example, 
as a proof of concept, we have implemented the Pauli-Villars (or
Feynman-Stuekelberg) approach.
 It introduces an  intermediate scale of Pauli-Villars
mass, $M_{PV} \ll \pi/a$ separating the IR from the UV  and protects the UV
diagram from the dependence of the variable cut-off of the simplicial
lattice. This separation of scales is another way to exploit the {\bf
  spectral fidelity} of the low spectrum, but this time to all orders in
perturbation theory. The implementation amounts to the
addition of a ghost field giving a simplicial equivalent of the
continuum propagator,
\be
\frac{1}{p^2} \rightarrow \frac{1}{p^2} - \frac{1}{p^2+ M^2_{PV}} =
\frac{1}{p^2 + p^4/M^2_{PV}} \; .
\ee
Now, reaching the  continuum
requires a double limit: $a \rightarrow 0$ at fixed $a M_{PV}$,
followed by $a M_{PV} \rightarrow \infty$.  We have implemented this
on $\mS^2$ by adding a quadratic PV term,  $ - M^{-2}_{PV} \phi_x K_{xz}
K_{z y} \phi_y/\sqrt{g_z} $, to the FEM action in~\eqref{eq:simplicial_action},
seeing qualitatively that this
does work. However, in addition to the cost of the 
double limit, in the context of
our current $\phi^4$ simulations, the PV term has the technical
disadvantage that it prevents the use of the very efficient cluster
algorithm of Ref.~\cite{Brower:1989mt}. 

We are also exploring extensions of the renormalization group approach in our demonstration
of the one-loop logarithmic divergences in 2d in~\secref{sec:univlogdiv}.
For asymptotically free theories in
4d,  such as the  non-Abelian gauge theory, the Lagrangian, $F^2/g^2$, has a
dimensionless coupling with a logarithmic divergence. We anticipate
the use of the RG approach and perhaps the scale setting properties of
Wilson flow~\cite{Luscher:2010iy} to correct the scheme dependence of
a simplicial lattice. As demonstrated in the classic paper by
A. Hasenfratz and P. Hasenfratz~\cite{Hasenfratz:1980kn} for pure
non-Abelian gauge theory, the lattice scheme
dependence is a one-loop effect which on a simplicial lattice should  be replaced by 
a local site dependent multiplicative  counter term for  $F^2$.


\section{\label{sec:numerical_CT}Numerical Tests of UV Competition on \texorpdfstring{$\mS^2$}{S2}}

Here we present our first test of our QFE simplicial lattice
construction for the non-perturbative study of quantum field theories
on curved manifolds. The Monte Carlo simulation of the   \ndim{2} scalar $\phi^4$ theory on the Riemann sphere,
$\mathbb{S}^2$, must agree within statistical and systematic
uncertainties with the exact solution of
the Ising or c = 1/2  minimal CFT in the continuum limit. The first test
is to fine-tune the mass parameter to the critical surface illustrated
in~\figref{fig:WF}  and to compare the Monte Carlo computation of bulk
Binder cumulants with analytic values.
In~\secref{sec:numerical_correlators}, we extend our tests to the detailed form
 of the conformal two-point and four-point correlators.


\subsection{\label{sub:moments}Ising CFT  on the Riemann \texorpdfstring{$\mS^2$}{S2}  }

Let us begin by defining the Euclidean correlations functions
\be
G_n(x_1,\cdots, x_n) = \< \phi(x_1) \cdots  \phi(x_n) \> = \frac{1}{Z[0]}
\frac{\delta}{\delta J(x_1)} \cdots \frac{\delta }{\delta J(x_n)}  Z[J] \big|_{J = 0} \; ,
\ee
 for our scalar field theory as a derivative expansion of the partition
 function,  $Z[J] = \int [D\phi]\exp[ - S[\phi]  + \int d^dx 
 J(x) \phi(x)]$, in the current $J(x)$.  Likewise replacing the
current by a constant {\em magnetic field}, $ J(x) = h \sqrt{g(x)}$,  the derivatives  of the
partition function give the magnetic moments,
\be
m_n = \< M^n \> =  \int d^2x_1  \cdots d^2 x_n
\langle \sqrt{g(x_1)} \phi(x_1) \cdots   \sqrt{g(x_n)} \phi(x_n)
\rangle
\label{eq:moments}
\ee
where $M = \int d^2x \sqrt{g}\phi(x)$.  
From these moments, defining
homogeneous quotients, $Q_{2n} =  \< M^{2n }\> /  \< M^2 \>^n$,
the first three  Binder cumulants are~\cite{Mon:1997}
\bea
U_4 & = & \frac{3}{2} \left( 1 - \frac{1}{3}Q_4 \right) \nn
U_6  &=&  \frac{15}{8} \left( 1  -
          \frac{1}{2}Q_4 + \frac{1}{30}Q_6\right) \nn
U_8 & = & \frac{315}{272} \left( 1 - \frac{2}{3} Q_4  + \frac{1}{18}
          Q_4^2 + \frac{2}{136} Q_6 - \frac{1}{630}Q_8 \right) \; .
\eea
These are just the  connected moment expansion of the
free energy, 
\be
F[h] = \log[\int [D\phi]e^{\textstyle - S[\phi]  + h \int d^dx \sqrt{g}
 \phi(x)} ]
\label{eq:FreeEnergy} \; ,
\ee
divided by appropriate factor of $\< M^{2n}\>$. Therefore they
vanish in the Gaussian limit. The  overall normalization has been chosen
so that  the Binder cumulants are  unity  in the ordered
phase.  We designate the exact value of Binder cumulants
for the critical Ising CFT by $U^*_{2n}$. 

On the sphere $\mS^2$ the exact value of the  Binder
cumulants $U^*_{2n}$ can be computed from the solution for the conformal n-point
functions  on $\mR^2$.   This a consequence of the 
special property for conformal correlators under a  Weyl transformation  of the flat metric: $g_{\mu\nu}(x) = \Omega^2(x)
\delta_{\mu\nu}$ on  $\mS^2$. For example the CFT correlators of a primary operator $\phi$ with dimension
$\Delta$ obey the identity \cite{Rychkov:2016iqz}
\be
\<\phi(x_1) \cdots \phi(x_n)\>_{g_{\mu\nu}} 
= \frac{1}{\Omega(x_1)^{\Delta}}  \cdots\frac{1}{\Omega(x_n) ^{\Delta}}\< \phi(x_1)  \cdots \phi(x_n)  \>_{flat}  \; .
\label{eq:ConformalMap}
\ee
 As also pointed out in
 Ref~\cite{Rychkov:2016iqz}, even the Weyl anomalies will cancel in
 homogeneous ratios of CFT correlators. Consequently as noted in 
Ref.~\cite{Deng:2003kiw} homogeneous moments in the Binder cumulants
are computed by integration over the n-point function on the
compact manifold.  

In our current application,  the Weyl transformation is a
stereographic projection  from $\mR^2$ to the Riemann sphere, $\mS^2$, 
which can be explicitly parameterized by
\be
w = (r_x +i r_y)/(1 + r_z) = \tan(\theta/2) e^{i\phi} \quad ,\quad \hat r = (\sin\theta, \cos\phi, \sin\theta \sin
\phi, \cos \theta) \; ,
\label{eq:StereoProj}
\ee
where $w$ is the complex co-ordinate $w = x + iy$ in the $\mR^2$ plane and
$\hat r$ is unit vector in $\mR^3$  in~\eqref{eq:SphereInR3}.  The
resulting metric on $\mS^2$ is 
\be
ds^2_{\mS^2} = \frac{2}{(1+w\bar w)^2} dw d\bar w =\cos^2(\theta/2)
ds^2_{\mR^2} \; . 
\label{eq:ProjectiveMap}
\ee
with $\Omega^2(\theta) =   \cos^2(\theta/2)$.   After the Weyl rescaling the two-point
function on $\mS^2$ is 
\be
\label{eq:2pt_function_sphere}
\left\langle \phi(\hat r_1) \phi(\hat r_2) \right\rangle = \frac{1}{(2 - 2 \cos\theta_{12})^{\Delta}} ,
\ee
where $\theta_{12}$ is the angle between the two radial vectors, $\hat
r_1, \hat r_2$, on the surface of the sphere. Incidentally a pedestrian proof
 uses standard trigonometric identities to 
show that $\Omega(\theta_1)|w_1 - w_2|^2 \Omega(\theta_2)  = |\hat r_1 - \hat
r_2|^2 = 2(1 - \cos\theta_{12})$. We have chosen
the normalization of $\phi$ so that the numerator is unity. 
Just as Poincare invariance on the plane implies that  correlators are
a function of the length (or Euclidean distance on the plane, $|w_1
-w_2|$), rotational invariance on the sphere fixes the correlator  to
be  a function of the geodesic distance, $\theta_{12}$.  

On $\mR^2$  the conformal n-point correlation
functions of the \ndim{2} Ising model can be constructed, in
principle, to any order \cite{Kadanoff:1969, Luther:1975wr,
  Polyakov:1984yq, Belavin:1984vu, Dotsenko:1984nm} and in practice
have been computed up to sixth order \cite{Kadanoff:1969,
  Luther:1975wr, Burkhardt:1987}.
To normalize the ratios $Q_{2n}$ we use the analytic result,
\be
m_2
= \int \frac{d\Omega_1}{4 \pi} \int \frac {d\Omega_2}{4 \pi} \<\phi(\Omega_1) \phi(\Omega_1) \>  
= \int_{-1}^1 \frac{1}{2(2-2\cos\theta_{12} )^{1/8}} d\cos \theta_{12}
= \frac{2^{11/4}}{7} \; ,
\label{eq:second_moment}
\ee
with $\Delta = 1/8$ for the Ising model. This allows  us to find the relationship between the normalization
of the continuum  field  $\phi(x)$ and the normalization of our discretized 
field $\phi_x$ in a QFE calculation.  Similarly,  higher moments can be computed numerically as integrals over
the n-point correlators on $\mS^2$. 

The integral of the four-point function on the sphere was performed  by  Deng and Bl{\"o}te in Ref.~\cite{Deng:2003kiw}.  
The computation of $m_4$ is na\"ively an eight-dimensional integral, but after utilizing
rotational invariance the integration is reduced to a five-dimensional
integral. The numerical four-point integral evaluation in
Ref.~\cite{Deng:2003kiw} used  1000
independent Monte Carlo estimates, yielding an estimate of $m_4 = 1.19878(2)$,
leading to a prediction of the critical value for the fourth Binder cumulant
$U^*_{4} = 0.8510061(108)$ after error propagation.

We computed both the four- and six-point integrals numerically using the
\texttt{MonteCarlo} method of Mathematica's \texttt{Integrate[]} function.
For the four-point integral, we set \texttt{AccuracyGoal$\to$4}, which yields a
Monte Carlo distribution with standard deviation approximately $10^{-4}$.  We
repeated the Monte Carlo estimation of the integral 100 times.  The quoted
error is the standard error on the mean of this sample distribution.  We find
$m_4 = 1.1987531(116)$ and the corresponding Binder cumulant is $U^*_{4} = 0.8510207(63)$. 
We again used the Mathematica \texttt{MonteCarlo} integrator to compute $m_6$.
We set \texttt{AccuracyGoal$\to$3}, which yields a Monte Carlo distribution
with standard deviation approximately $10^{-3}$.  We repeated the Monte Carlo
estimation 50 times. We find $m_6 = 1.632851(253)$, and a corresponding
critical Binder cumulant value of $U^*_{6} = 0.7731441(213)$.


\subsection{\label{sub:FSS}Finite Scaling Fitting Methods}
To compute these cumulants, we must obtain
 estimates for the  even moments of the average
magnetization.  On the lattice we define the moments of the field on the simplicial complex by
\be
\label{eq:simplicial_moments}
m_n = \<( N^{-1} \sum_x \sqrt{g_x} \phi_x )^n  \>
= \frac{1}{N^n} \sum_{x_1,\dots,x_n}
\left \langle \sqrt{g_{x_1}} \phi_{x_1} \cdots\sqrt{g_{x_n}} \phi_{x_n} \right \rangle
\ee
where $\langle\dots\rangle$ denotes the ensemble average and the spatial
average is shown explicitly by a summation.  Numerically, these moments are
easy to compute by a weighted sum of the field over the complex  on each field
configuration, with the weights $\sqrt{g_x}$ defined to be the areas of the
Vorono{\"i} cells \cite{Voronoi:1908} as computed on the flat triangles and
normalized to one per site on average, i.e. our convention $\sum_{x=1}^{N}
\sqrt{g_x} = N$ in~\eqref{eq:normalization}.

In our Monte Carlo calculations, the finite size of our simplicial complex
will break conformal invariance.  If the bare lattice parameters are tuned
sufficiently close to the critical point, finite-size scaling (FSS) relations
can be used to extract CFT data by fitting the volume dependence of moments or
cumulants of average magnetization \cite{Blote:1995zik}.
We give the key details below, and in Appendix \ref{app:cumulants} we give further details of the FSS
analysis used to  parameterize our numerical data as we take the
infinite volume limit. 

 It begins
by expanding the free energy  \textit{in a finite
  lattice  volume} $N$ around the critical point $(g_\sigma^*, g_\epsilon^*,
g_\omega^*)$.  By $Z_2$ symmetry the critical point for the
$Z_2$ odd operator is $g^*_\sigma = 0$. 
Of course, we cannot directly vary the renormalized couplings $(g_\sigma,
g_\epsilon, g_\omega, \cdots)$ but instead can vary our bare couplings $(h,
\mu^2_0, \lambda_0)$ defined in our cut-off theory.  The expansion then
for the  $Z_2$ odd operator takes the form
\be
g_\sigma  =  h \alpha_1 + h^3
\alpha_3/3! +  h^5 \alpha_3/5!\cdots \; ,
\ee
 and without loss of generality, we can rescale $h$ so that $\alpha_1 = 1$.
There is mixing between the two even operators so that 
\bea
\label{eq:coord_change}
\left( \begin{array}{c}
g_\epsilon - g_\epsilon^* \\
g_\omega - g_\omega^*
\end{array} \right) & = & \left( \begin{array}{cc}
R_{\epsilon\mu} & R_{\epsilon\lambda} \\
R_{\omega\mu} & R_{\omega\lambda}
\end{array} \right) \left( \begin{array}{c}
\mu^2_0 - \mu_*^2 \\
\lambda_0 - \lambda_*
\end{array} \right) + \cdots \; .
\eea
Since our current simulations were performed  at a single fixed $\lambda_0$,
we cannot directly determine the coordinate transformation $R$.  Instead, we define 
the derived quantities.
\begin{eqnarray}
\overline{a}_{k1} & = & a_{k1} R_{\epsilon\mu} , \quad
\overline{b}_{k1}  =  b_{k1} R_{\omega\mu} , \quad
\overline{c}_{k1}  =  c_{k1} R_{\epsilon\mu}, \\
\mu_a^2 & = & \mu_*^2 - \frac{R_{\epsilon\lambda}}{R_{\epsilon\mu}} \left( \lambda_0 - \lambda_* \right) , \quad
\mu_b^2 = \mu_*^2 - \frac{R_{\omega\lambda}}{R_{\omega\mu}} \left( \lambda_0 - \lambda_* \right) \; ,
\end{eqnarray}
where $a_{k1}$, $b_{k1}$, and $c_{k1}$ are defined in~\eqref{eq:RG_moment_F} and~\eqref{eq:RG_moment_G} in Appendix~\ref{app:cumulants}. Future simulations varying the bare coupling near to the Wilson Fisher
fixed points will improve the finite volume parameterization.
The result of the finite volume and scaling expansion  is
to parameterize our data by
\bea
m_2 &=& \mathbb{L}^{-2\Delta_\sigma} \left[
  a_{20} + \overline{a}_{21} ( \mu^2_0 - \mu_a^2 ) \mathbb{L}^{d-\Delta_\epsilon}
  + \overline{b}_{21} ( \mu^2_0 - \mu_b^2 ) \mathbb{L}^{d-\Delta_\omega} 
\right] + c_{20} \mathbb{L}^{-d} + \cdots \; , \nn
m_4 &=& \mathbb{L}^{-4\Delta_\sigma} \left[
  a_{40} + \overline{a}_{41} ( \mu^2_0 - \mu_a^2 ) \mathbb{L}^{d-\Delta_\epsilon}
  + \overline{b}_{41} ( \mu^2_0 - \mu_b^2 ) \mathbb{L}^{d-\Delta_\omega} 
\right] + c_{40} \mathbb{L}^{-3d} \nonumber \\*
&& + \alpha_3 \mathbb{L}^{-2d} m_2 + 3 m_2^2 + \cdots \; , \nn
m_6 &=& \mathbb{L}^{-6\Delta_\sigma} \left[
  a_{60} + \overline{a}_{61} ( \mu^2_0 - \mu_a^2 ) \mathbb{L}^{d-\Delta_\epsilon}
  + \overline{b}_{61} ( \mu^2_0 - \mu_b^2 ) \mathbb{L}^{d-\Delta_\omega} 
\right] + c_{60} \mathbb{L}^{-5d} \nn
&& + 15 m_4 m_2 - 30 m_2^3 + \alpha_3 \mathbb{L}^{-2d} \left( m_4 - 3 m_2^2\right)
+ 3 \alpha_5 \mathbb{L}^{-4d} m_2 + \cdots \; .
\label{eq:m_n_parameters}
\eea
where we introduce the length parameter ${\mathbb  L} = \sqrt{N}$ in
accord with the FSS analysis summarized in Appendix~\ref{app:cumulants}. The ellipses are a reminder that our expansion drops higher order terms
in the fitting expressions, leading to a systematic error in the included parameters.
For the $d=2$ Ising model, $\Delta_\sigma = 1/8$ and $\Delta_\omega = 4$, so the
$\overline{b}_{21}$ term scales like $\mathbb{L}^{-9/4}$ whereas the $c_{20}$ term
scales like $\mathbb{L}^{-2}$.  Interestingly, the \textbf{leading} irrelevant
correction to scaling is \textbf{not} due to a conformal quasi-primary
operator $\omega$, but rather a breaking of conformal symmetry due to a finite
volume.  We know of no other example of a CFT where this occurs.

Finally, to compare our simplicial calculations to the precise analytic
results described in~\secref{sub:moments}, we can fit our estimates of the
magnetization moments and use the fitted parameters to estimate the critical Binder
cumulants. Note that even though we chose to approach the critical surface along a line of constant
$\lambda_0$ and were therefore unable to determine the linear coordinate transformation
$R$ of~\eqref{eq:coord_change} leading to some of that dependence being absorbed
into redefinitions of fit parameters, \textit{e.g.} $\overline{a}_{21}$
\textit{vs.}\ $a_{21}$, our estimates of the critical Binder cumulants $U_{2n}^*$ are free
of this ambiguity.


\subsection{\label{sub:MonteCarlo}Monte Carlo  Results Near the Critical Surface}

For the Monte Carlo simulation, we use the embedded dynamics algorithm of
Brower and Tamayo~\cite{Brower:1989mt}, mixing a number of embedded
Wolff cluster updates~\cite{Wolff:1988uh}
\begin{wraptable}{r}{8.0cm}
\begin{tabular}{c|c}
Lattice size & Wolff/local update ratio \\
\hline
$L < 15$ & 4 \\
$15 \le L \le 42$ & 5 \\
$43 \le L \le 99$ & 6 \\
$100 \le L \le 200$  & 7 \\
$201 \le L \le 364$ & 8 \\
$L > 364$ & 9
\end{tabular}
\caption{\label{tab:wolff_updates_per_sweep}The number of  Wolff
cluster updates totaling $\mathcal{O}(N)$ spins  near criticality.}
\end{wraptable}
 with local updates consisting of one Rosenbluth-Teller sweep \cite{Metropolis:1953am} and one over relaxation sweep
\cite{Adler:1981sn, Whitmer:1984he, Brown:1987rra, Creutz:1987xi}.  
\tabref{tab:wolff_updates_per_sweep} shows our empirically determined number of
Wolff cluster updates per local update such that on average $\mathcal{O}(N)$
spins are flipped between local updates.
We can approximately locate the critical surface of the Wilson-Fischer fixed
point by computing the Binder cumulants at fixed $\lambda_0$, 
varying the volume $N = \mathbb{L}^2 = 2 + 10 L^2$ and the bare mass
$\mu^2_0$, searching for the
region of parameter space where $U_4 \approx 0.851$.  We can understand the
approximate behavior of the Binder cumulants by substituting the FSS
expressions in~\secref{sub:FSS} and re-expanding \cite{Binder:2001ha},
giving
\be
\label{eq:Binder_critical}
U_{2n}(\mu_0^2) =  U_{2n}^*
+ \overline{A}_{2n} \left( \mu_0^2 - \mu_a^2 \right) \mathbb{L}^{d-\Delta_\epsilon} 
 + \overline{B}_{2n} \left( \mu_0^2 - \mu_b^2 \right) \mathbb{L}^{d-\Delta_\omega} 
+ C_{2n} \mathbb{L}^{2 \Delta_\sigma - d} + \cdots \; .
\ee
Since $\Delta_\epsilon = 1$, the $\overline{A}_{2n}$ term will cause the
cumulant to diverge from its critical value if $\mu_0^2$ is not tuned to
$\mu_a^2$ as $\mathbb{L} \to \infty$.  Importantly, the direction of the divergence
will depend on the sign of $\left( \mu_0^2 - \mu_a^2 \right)$.  At smaller $\mathbb{L}$,
the $C_{2n}$ term will tend to dominate since $\Delta_\sigma = 1/8$ and
$\Delta_\omega = 4$ as previously noted.  It is our experience that~\eqref{eq:Binder_critical} should not be used to actually fit the Binder
cumulant data to accurately determine $U_{2n}^*$ since there tend to
be delicate cancellations that occur in the expansion of the ratio.  Instead,
the moments should be analyzed separately using the parameterization 
in~\eqref{eq:m_n_parameters}.

\begin{figure}[ht]
\includegraphics[width=0.49\textwidth]{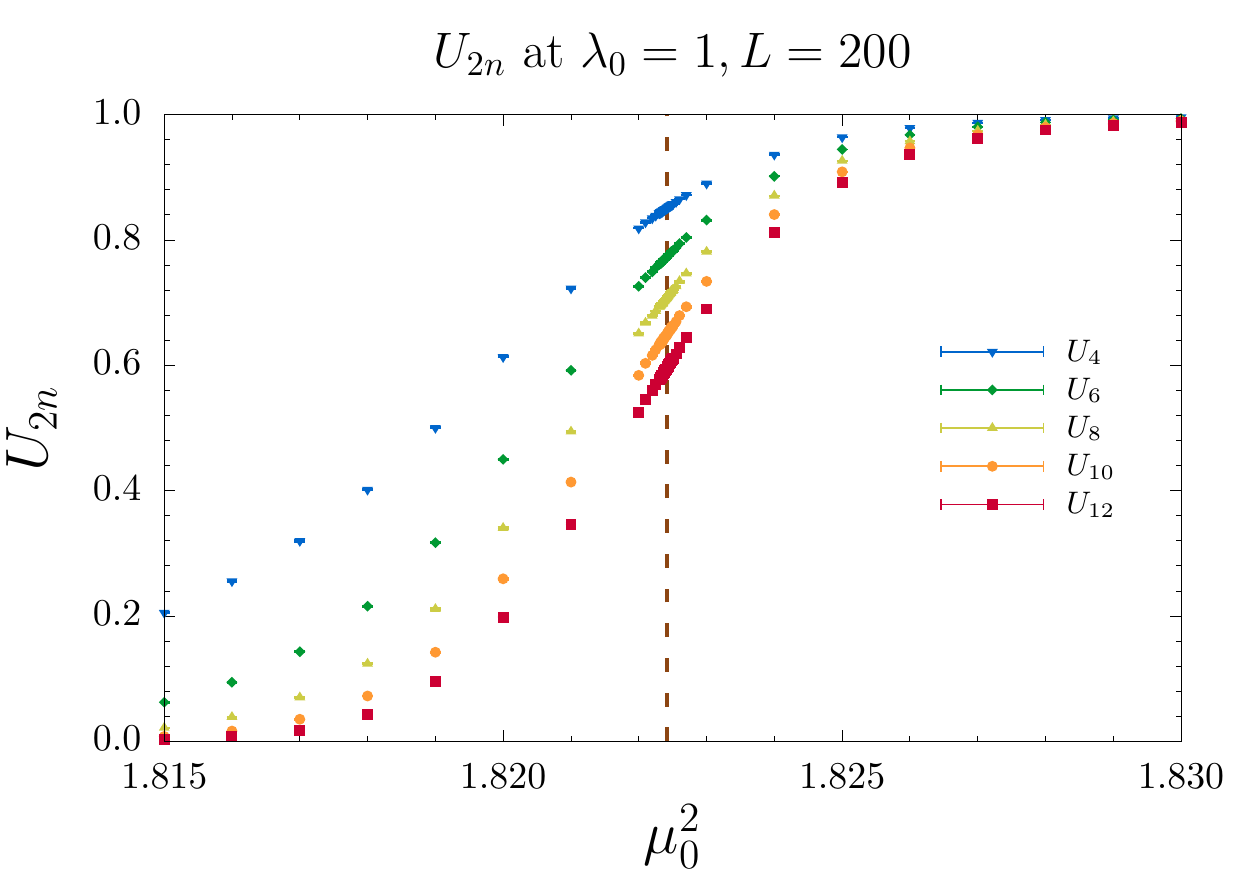}
\includegraphics[width=0.49\textwidth]{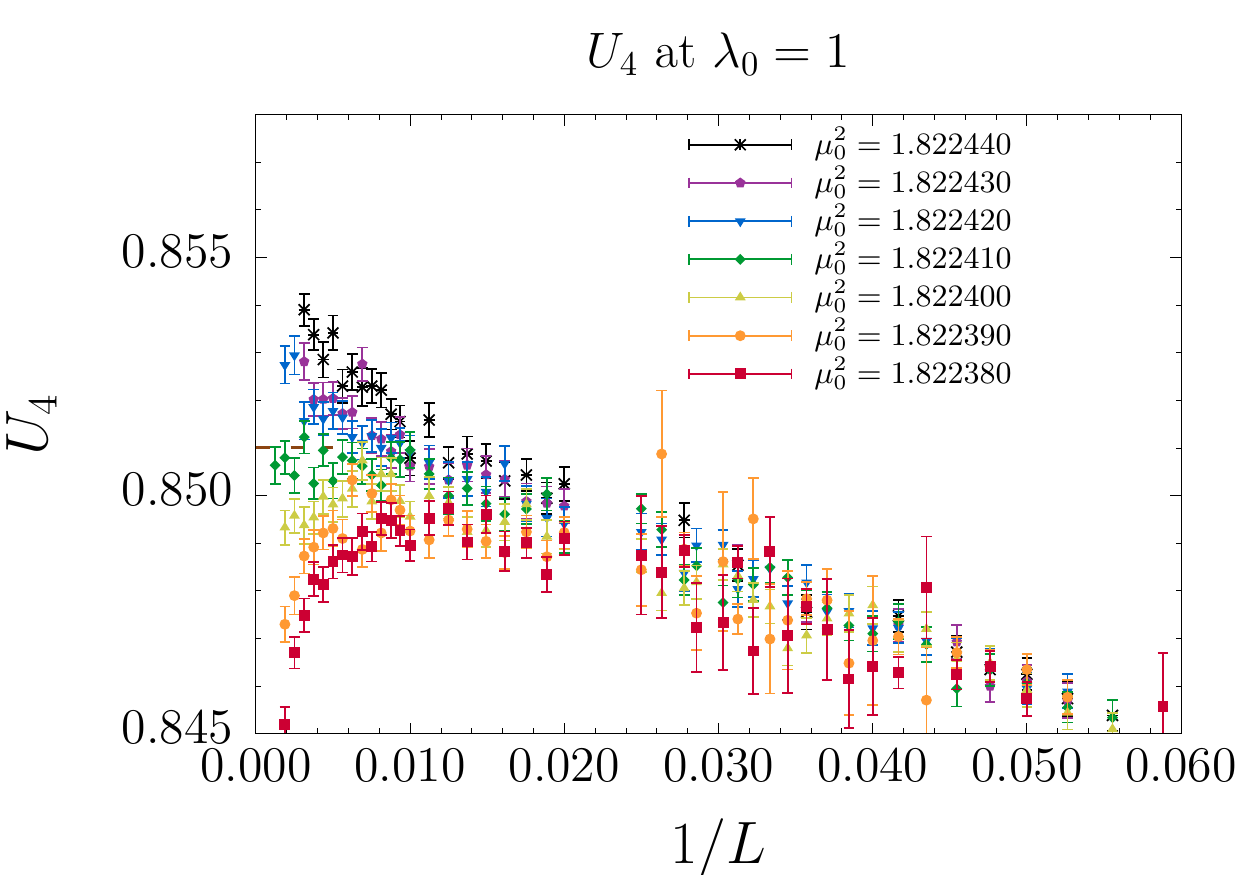}
\caption{\label{fig:Q_vs_musq} The left panel shows Binder cumulants up to
twelfth order \textit{vs}.\ $\mu_0^2$ for fixed simplicial complex size, $L=200$
and $\lambda_0=1$.  The vertical dashed line indicates $\mu_a^2 \approx
\mu_\mathrm{cr}^2$.  The right panel shows the cumulant $U_4$ \textit{vs}.\
simplicial complex size $L$ in the pseudocritical region holding  $\lambda_0 =
1$ fixed.  Each point represents a calculation and
connecting lines indicate constant $\mu_0^2$. Continuum estimate:
$U_{4}^* = 0.8510207(63)$.}
\end{figure}

In the left panel of~\figref{fig:Q_vs_musq}, we show the cumulants for a fixed size
of the simplicial complex ($L = 200$) as we sweep through the critical region
by varying $\mu^2_0$ holding $\lambda_0 = 1$ fixed.  Higher order cumulants are
statistically noisier but since they pass through the critical region with a
steeper slope \cite{Mon:1997} they have comparable statistical weight in
determining the critical coupling $\mu_a^2 \approx \mu_\mathrm{cr}^2$.  On the
right we show the cumulant $U_4$ \textit{vs}.\ simplicial complex size $L$ for
fixed values of $\mu^2_0$ in the critical region. As expected, the divergence
at large $L$ changes sign as $\mu^2_0$ passes through the critical region.

We will now proceed to describe our data fitting and analysis procedure.
To estimate quantities like $U_{4}^*$ as accurately as possible, we
use an iterative procedure to determine which ensembles, labeled by $(\mu_0^2,
\lambda_0, L)$, to include in the fits to various moments of magnetization. 
\setlength{\tabcolsep}{20pt}
%
\begin{wraptable}{r}{8.0cm}
\begin{tabular}{c|l}
$w$ & 0.03 \\
\hline\hline
$\alpha_3$ & $-22.9(4.7){\times}10^3$ \\
$\alpha_5$ & $12.4(3.3){\times}10^{10}$ \\
$a_{20}$ & 0.47895(11) \\
$\overline{a}_{21}$ & 0.1873(41) \\
$a_{40}$ & -0.39006(20) \\
$\overline{a}_{41}$ & -0.357(12) \\
$a_{60}$ & 1.3570(11) \\
$\overline{a}_{61}$ & 1.876(71) \\
$\overline{b}_{21}$ & -18(2106) \\
$\overline{b}_{41}$ & -23(2629) \\
$\overline{b}_{61}$ & 115(13187) \\
$c_{20}$ & -19.9(3.1) \\
$\mu^2_a$ & 1.82241324(70) \\
$\mu^2_b$ & 4(274) \\
\hline
$\chi^2/\mathrm{dof}$ & 1.001 \\
$\mathrm{dof}$ & 1195
\end{tabular}
\caption{\label{tab:fit_m06}Best fit values for the parameters
  described in~\secref{sub:FSS} and Appendix~\ref{app:cumulants} for data selection window
with the ratios  in~\eqref{eq:Window} bounded by $w = 0.03$.}
\end{wraptable}
%
We start by getting a rough fit to the data using some initial data selection
window $\mu^2_0 \in \left[ \mu^2_\mathrm{min}, \mu^2_\mathrm{max} \right]$ and
$L \in \left[ L_\mathrm{min}, L_\mathrm{max} \right]$ and then form the
following quantities
\bea
\delta a_{k1} &= \left| \frac{\overline{a}_{k1} \left( \mu^2_0 -
      \mu^2_a \right) \mathbb{L}^{d - \Delta_\epsilon}}{a_{k0}}
\right| \nn
\delta b_{k1} &= \left| \frac{\overline{b}_{k1} \left( \mu^2_0 - \mu^2_b \right)
\mathbb{L}^{d - \Delta_\omega}}{a_{k0}} \right|  \nn
\delta c_{20} &= \left| \frac{c_{20} \mathbb{L}^{2 \Delta_\sigma - d}}{a_{20}}
\right|   \nonumber 
\label{eq:Window}
\eea
Subsequently we adjust the range of data to be fit for each moment of magnetization
$m_n$ enforcing the data  cuts: $\delta a_{k1} , \; \delta b_{k1} , \;  \delta c_{20} \le
w$, where $w$ is the width of the window, for all $ k \le n$.  It makes sense that lower moments of
magnetization should have larger data selection regions since they vary 
more slowly in the critical region
relative to the higher moments, as indicated in~\figref{fig:Q_vs_musq}.  We then refit the new data selection and reselect
the data for the same $w$ using the new fit values and then iterate
until the process converges to a stable data set for that $w$.  We
expect that fit parameters like $a_{k0}$ will have systematic errors of
$\mathcal{O}(w^2)$ due to higher order terms in the Taylor series
expansion of the moments of the free energy which we did not include in our
fit.  So, we would like to take $w$ to be as small as possible but also
have enough data left to adequately constrain all of the important fit
parameters, particularly the $a_{k0}$.

\begin{figure}[ht]
\center
\includegraphics[width=0.49\textwidth]{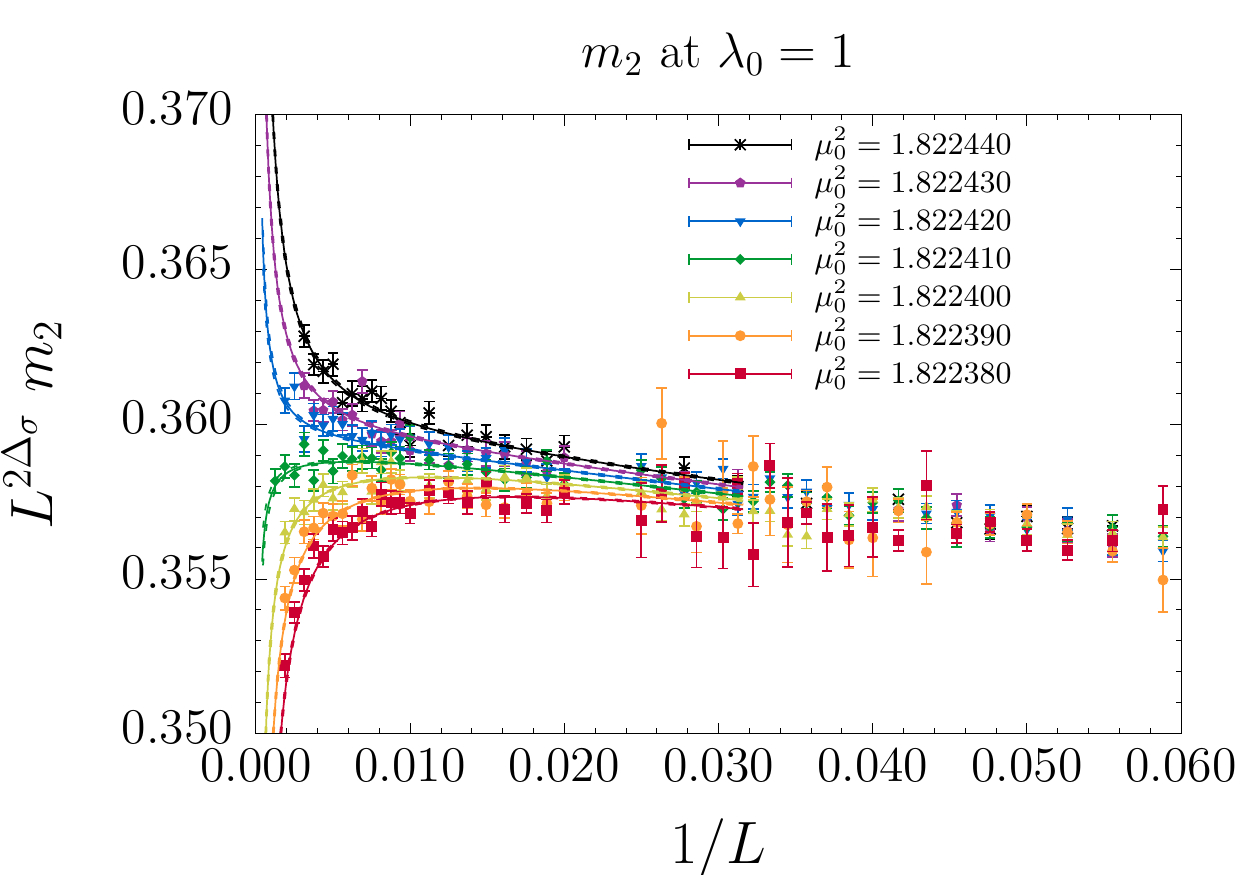}
\includegraphics[width=0.49\textwidth]{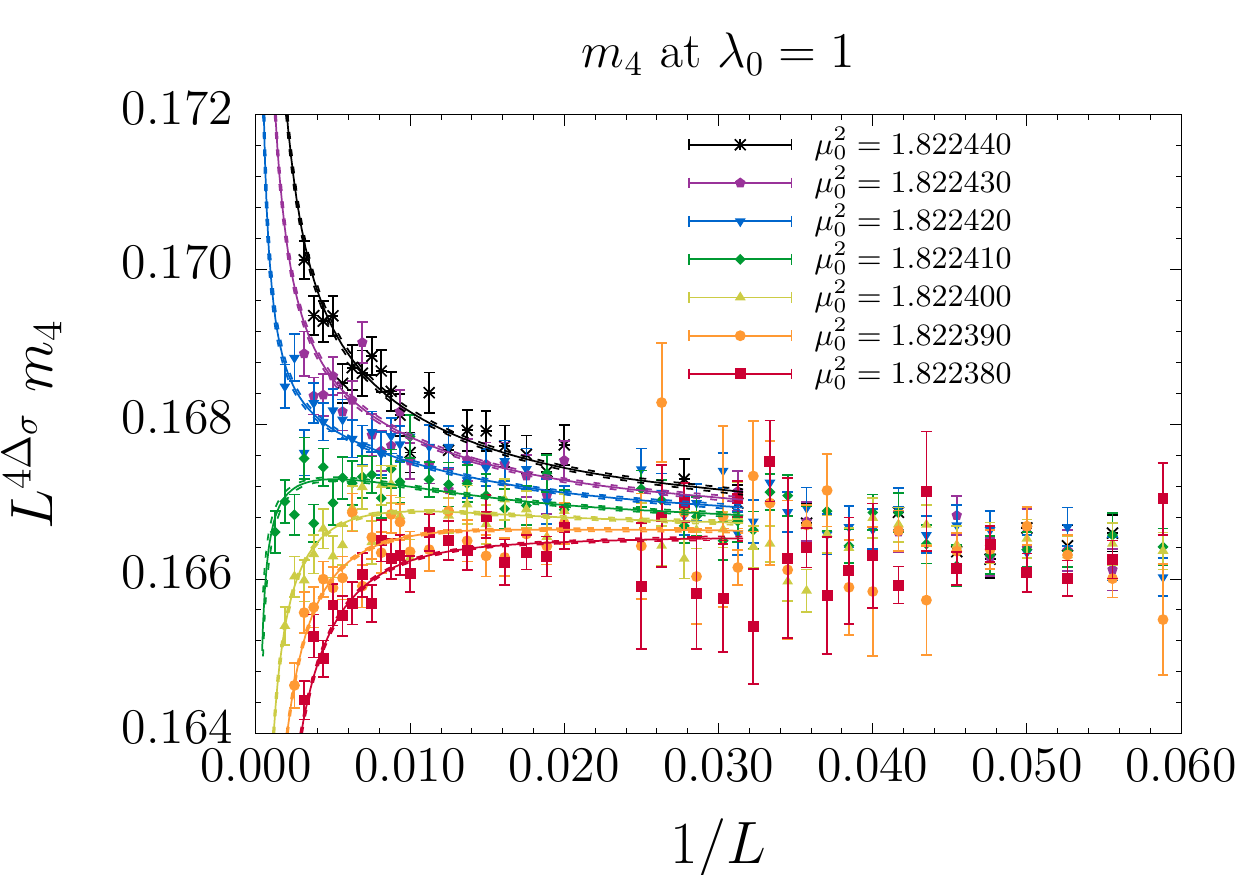}
\includegraphics[width=0.49\textwidth]{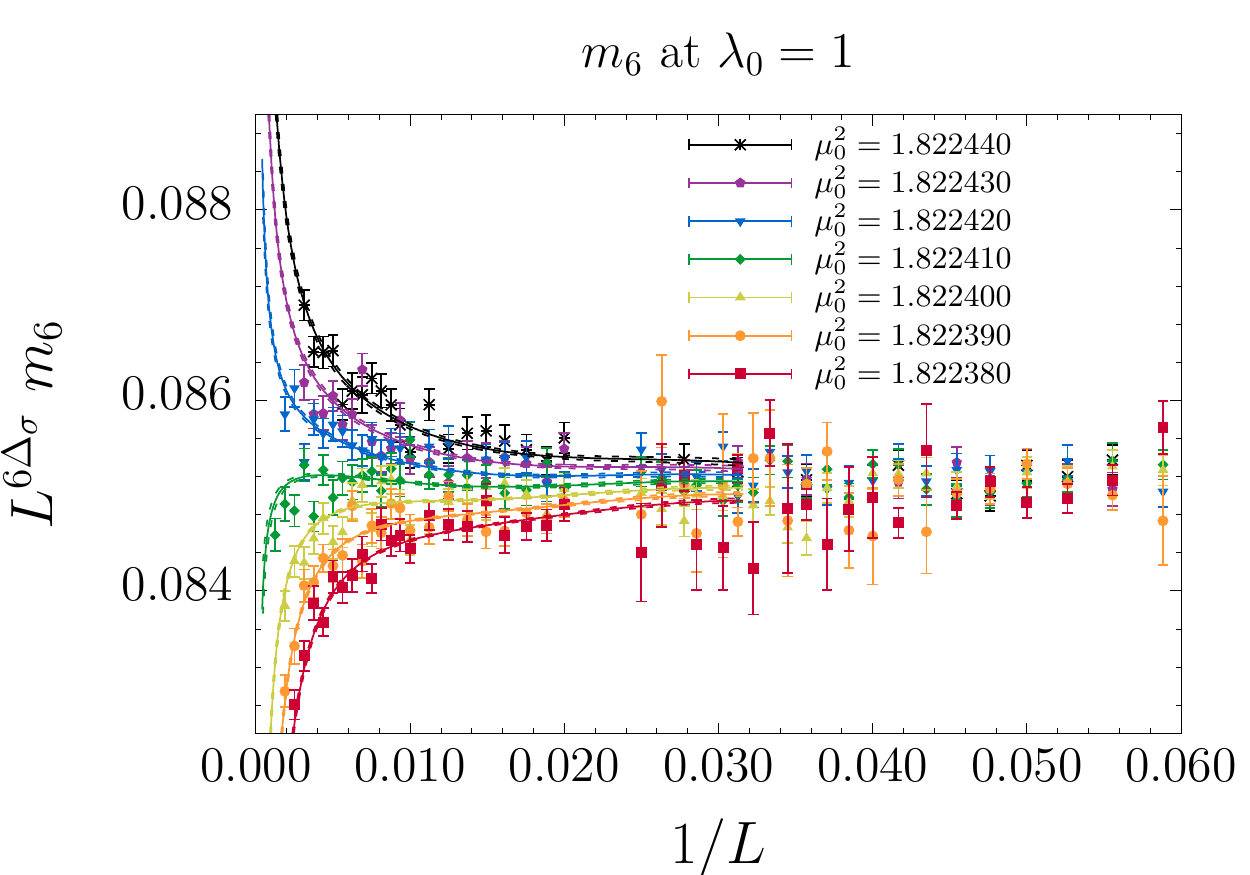}
\caption{\label{fig:fit_m06}A subset of the data used in a simultaneous fit to
the three lowest even moments for data selection parameter
$w=0.03$.}
\end{figure}

In~\figref{fig:fit_m06} we see a subset of the data selected with
$w=0.03$ and curves with error bands computed from the best fit values
of parameters shown in~\tabref{tab:fit_m06}.  Using the values shown in the
table for $w=0.03$, we get
$U_{4,cr}= 0.85020(58)(90)$ and $U_{6,cr} =
0.77193(37)(90)$ where the first error is statistical from the fit and the second is systematic. 

 In summary, the comparison of our Monte Carlo estimate on the QFE
 $\mS^2$ with the exact solution are the following:
 \begin{align}
	&\mbox{Monte Carlo Values:}& \quad &U_{4,cr} =  0.85020(58)(90)&\quad &U_{6,cr} = 0.77193(37)(90)  \nn
	&\mbox{Analytic CFT Values:}&  \quad &U^*_{4} = 0.8510207(63) &\quad &U^*_{6} = 0.7731441(213)
\end{align}
We see $O(10^{-3})$ agreement within the error estimates. Although
even more stringent tests can be made, with the
development of faster parallel code and a more extensive use of 
FSS analysis,  we
feel this is substantial support for the  convergence of our  QFE method to 
the exact $c = 1/2$ minimal CFT. In the next section we
give further tests for the two-point and four-point correlation functions.

\section{\label{sec:numerical_correlators}Conformal Correlator on \texorpdfstring{$\mS^2$}{S2}}
With our results for the Binder cumulants, we are confident that 
	the QFE lattice action on $\mathbb{S}^2$ has the correct critical limit.
Our analysis of magnetization moments and cumulants involves
global averages over n-point correlators. Here we turn to correlators
to examine more closely other consequences of conformal
symmetry and to provide more stringent tests for our QFE lattice
action.  We  begin with the two-point functions $\mathbb{S}^2$, and
explain how, with rotational invariance, the exact analytic result
can best be tested against our numerical result via a Legendre
expansion.  We next move to the exact four-point correlator, which takes on
a relatively simple form as a sum of Virasoro blocks.  
Finally we use
this as a toy example to learn how to compute CFT data (dimensions and
couplings) from  the conformal block operator product expansion (OPE). In particular, we extract the
central charge $c$ from the energy momentum tensor contribution to the
scalar four-point function.


\subsection{Two-Point Correlation Functions}

In the continuum CFT on the Riemann sphere $\mathbb{S}^2$, the conformal
two-point function, already mentioned
in~\eqref{eq:2pt_function_sphere}, is
\be
\label{eq:2pt_function_sphere_embed}
g(\theta_{12}) =\< \phi( \hat{r}_1) \phi( \hat{r}_2)\> =  \frac{1}{|\hat{r}_1
  - \hat{r}_2|^{2\Delta}}  =   \frac{1}{ (2 - 2 \cos\theta_{12})^\Delta} \;,
\ee
where  $\hat{r}_1,\hat{r}_2$ are unit vectors in the embedding space
$\mathbb{R}^3$ defined earlier in~\eqref{eq:SphereInR3} and
$\theta_{12} $ is  the angle between them. 
The projection of the two-point function onto Legendre functions, $P_l(\cos \theta_{12})$, can be computed analytically,
\be
c_l^\mathrm{cont} = \int^1_{-1} dz  \left( \frac{2}{1 - z} \right)^{1/8} P_l(z)
\implies \frac{8}{7}, \; \frac{8}{105},  \; \frac{24}{805}, \;
\frac{408}{24955}, \; \frac{680}{64883}, \; \frac{22440}{3049501} ,
\cdots
\ee
integrating over $z = \cos \theta_{12}$ for $l = 0, 1, 2, \cdots$.

\begin{figure}[ht]
\includegraphics[width=0.48\textwidth]{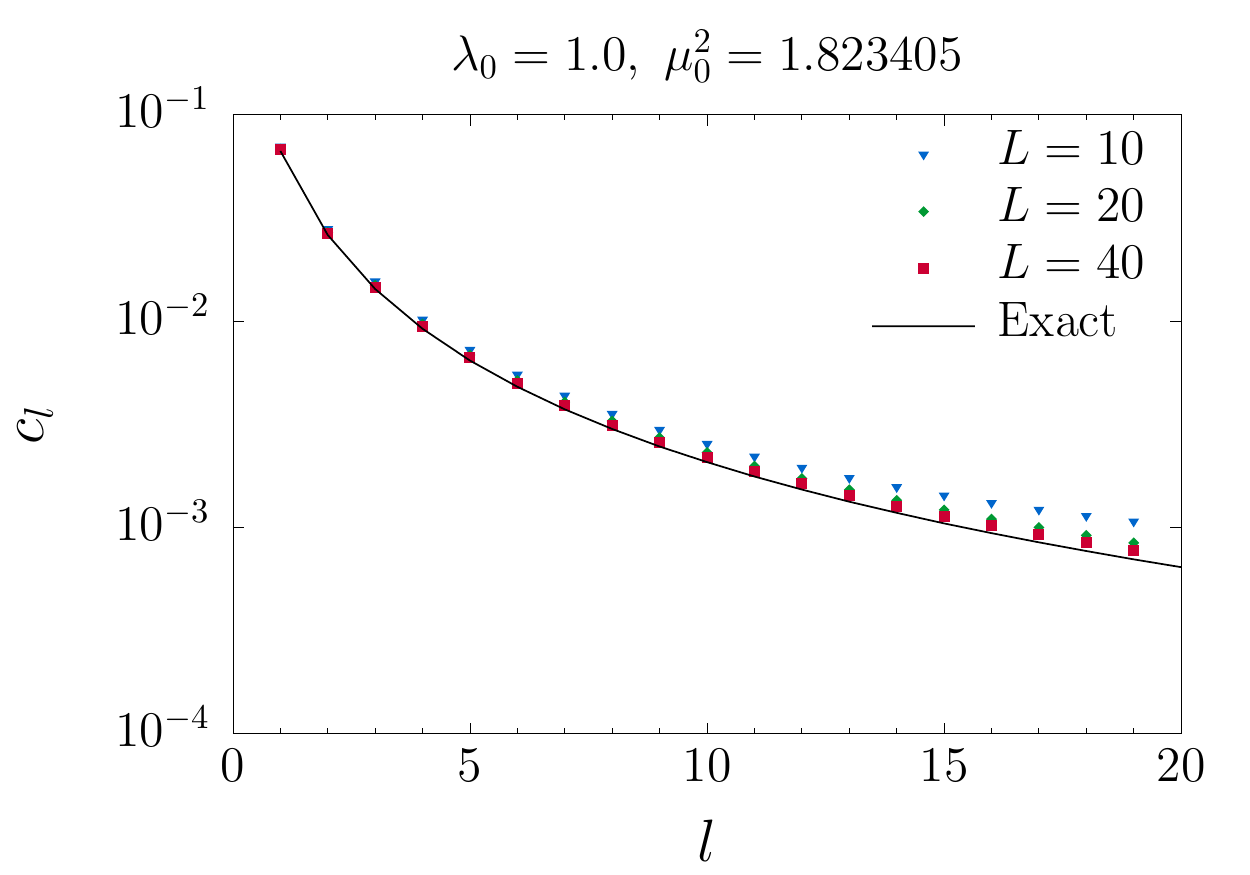}
\includegraphics[width=0.48\textwidth]{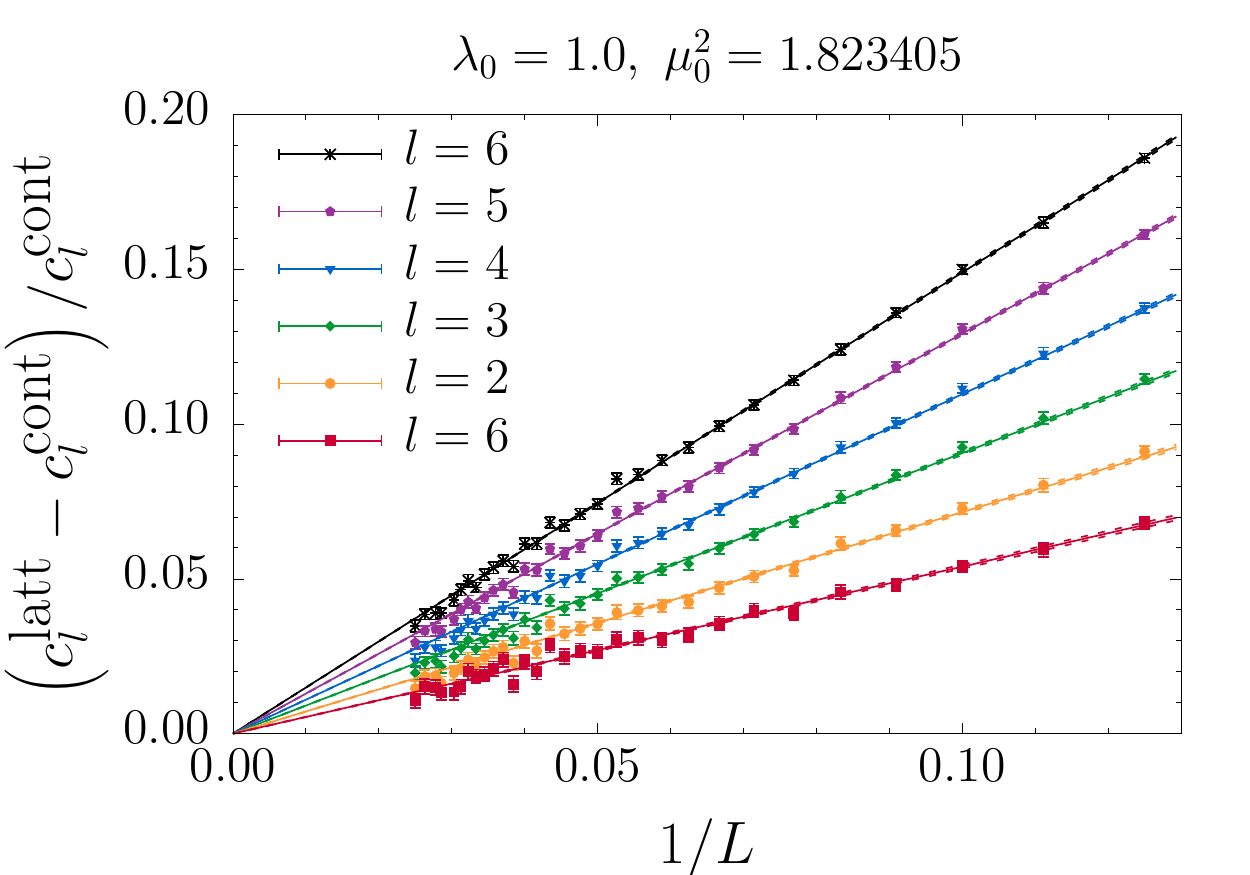}
\caption{\label{fig:MomentCorr}On the left, we show the simplicial two-point
functions $c_\ell(L)$ projected into Legendre coefficients for various values
of $L$, with the na\"ive dependence on $\ell$ scaled out for clarity of presentation. On the right, we show the relative difference between the
simplicial and continuum Legendre coefficients as we take the continuum limit
for fixed $\ell$.}
\end{figure}

In~\figref{fig:MomentCorr}, we compare the analytic value to the
moments of our correlation data. The fit is excellent showing only
slight deviations at high $l$ due to cut-off effects. Our
simplicial calculation of the two-point function is made at a fixed
value of $\mu^2_0$ closest to the pseudocritical values $\mu_a^2$
in~\tabref{tab:fit_m06}.  In~\figref{fig:MomentCorr}, the
normalization of the simplicial Legendre coefficients $c_\ell(L)$ are
chosen so that $c_0(L) = c_0^\mathrm{cont} = 8 / 7$. In the left
panel, we see that for any fixed value of $L$ the difference between
the simplicial and continuum values is small at small $\ell$ but
increases with increasing $\ell$.  Yet as $L$ increases, all
simplicial coefficients converge towards their continuum values.  The
right panel shows how the simplicial coefficients behave at fixed
$\ell$ as we approach the continuum limit.  The scaling curves shown
assume a na\"ive $1/L$ scaling of simplicial artifacts, which turns
out to be a good description of the data.

More generally in a CFT on the sphere, we will fit the two-point
function to determine
the operator dimension,  $\Delta_\sigma$. This can be done directly in co-ordinate space
by fitting to~\eqref{eq:2pt_function_sphere_embed} up to an overall
normalization, or numerically to the Legendre coefficients, 
$c_l^\mathrm{cont}(\Delta_\sigma) = \int^1_{-1} dz \left( 2/(1 - z)
\right)^{\Delta_\sigma} P_l(z)$, which obey the closed form recursion
relation,
\be
c_{l}^\mathrm{cont}(\Delta_\sigma)
=\frac{ l -1 +  \Delta_\sigma} 
 {l  + 1 - \Delta_\sigma } \;  c_{l-1}^\mathrm{cont} \quad, \quad
c_{0}^\mathrm{cont}(\Delta_\sigma) = \frac{1}{1 - \Delta_\sigma} \; .
\ee 
The overal normalization $c_{0}^\mathrm{cont}$ is of course proportional
to $m_2$ defined in~\eqref{eq:moments}. With our current simulations
of $\phi^4$ theory on $\mS^2$,  both methods agree with the
exact value $\Delta_\sigma = 0.125$ at the percent level.

From our analysis of the two-point function, we have a direct 
demonstration of how rotational symmetry is restored in the continuum limit.
Because the two-point function can be represented as an expansion in Legendre
functions, it is a function of only the angle between any two points on the
sphere and therefore rotationally invariant.  For any fixed finite $\ell$, no
matter how large, the simplicial coefficient $c_\ell(L)$ converges to the
continuum one as $L \to \infty$.


\subsection{\label{sub:fourpt}Four Point Functions}

The $c = 1/2$ minimal CFT has only three Virasoro primaries
${\bf 1}, \sigma, \epsilon$, with an OPE expansion, 
\be \sigma \times
\sigma = {\bf 1} + \epsilon \, ,\quad \epsilon \times \sigma =
\epsilon \, ,\quad \epsilon \times \epsilon = {\bf 1}\; .
 \ee 
 In our earlier paper~\cite{Brower:2016vsl}, we constructed a
 simplicial Wilson representation of Dirac Fermions $\mS^2$.  The map
 to the holomorphic, $\psi(z)$, and anti-holomorphic,
 $\bar \psi(\bar z)$, components of free Majorana fields on all \ndim{2}
 Riemann surfaces~\cite{Christe:1993ij} allowed us to define
 $\epsilon = \bar \psi \psi$
 and therefore compute the four-point functions:
 $\< \epsilon_1 \epsilon_2 \epsilon_3 \epsilon_4 \>$ and
 $\< \sigma_1 \epsilon_1 \sigma_2 \epsilon_1 \>$ without Monte Carlo
 simulations.  Here we test our QFE $\phi^4$ construction on $\mS^2$
 to compute the $\< \sigma_1 \sigma_2 \sigma_3 \sigma_4 \>$
 correlator.  In the continuum limit the leading behavior of the lattice field is the
 primary operator:
 $\phi_x \sim \sigma(x) + O(1/L^{\Delta_\sigma - \Delta'_\sigma}
 )$. The four-point function
 $\< \phi(\hat{r}_1) \phi(\hat{r}_2) \phi(\hat{r}_3) \phi(\hat{r}_4)
 \>$ is a function invariant under any of the 24 permutations of the
 four positions. This permutation invariance is easily enforced even
 on a finite lattice. In the continuum, the eight real coordinates can
 be reduced to five real coordinates using rotational invariance
 alone.

\begin{figure}[ht]
\includegraphics[width=0.49\textwidth]{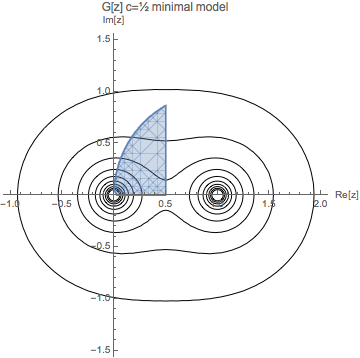}
\includegraphics[width=0.49\textwidth]{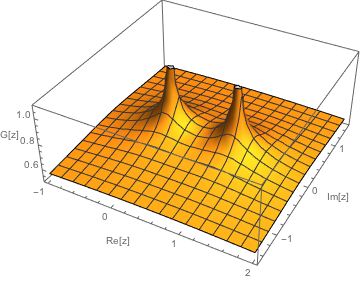}
\caption{\label{fig:G_visualization} We visualize the conformal function $G(z)$
in contour and surface renderings.  The outermost contour starts at $G(z) =
0.6$ and increases by steps of 0.1. Two symmetry planes are apparent:
Re($z$)=1/2 and Im($z$)=0. These are related to certain permutations of the
four-point function. The entire set of permeations
maps the blue zone exactly to the entire complex plane.}
\end{figure}

 Conformal symmetry allows for a further reduction to four real
coordinates, the two angular variables $\theta_{13}$ and $\theta_{24}$ defined as in~\eqref{eq:2pt_function_sphere_embed},
plus two real conformal cross ratios $u$ and $v$
\be
u \equiv \frac{|\hat{r}_1 - \hat{r}_2|^2 |\hat{r}_3 - \hat{r}_4|^2}
{|\hat{r}_1 - \hat{r}_3|^2 |\hat{r}_2 - \hat{r}_4|^2}  = |z|^2 , \quad
v \equiv \frac{|\hat{r}_1 - \hat{r}_4|^2 |\hat{r}_3 - \hat{r}_2|^2}
{|\hat{r}_1 - \hat{r}_3|^2 |\hat{r}_2 - \hat{r}_4|^2} = | 1 - z |^2  \;,
\ee
or equivalently a single complex variable $z$. 
\be
z  = \frac{(w_1 - w_2)(w_3 - w_4)}{(w_1 - w_3)(w_2 - w_4)} \quad \implies\quad 1- z = \frac{(w_2 - w_3) (w_1 - w_4)}{(w_1 - w_3) (w_2 -
  w_4) }
\ee
where $w$ is the complex variable for the $\mR^2$ prior to
the stereographic project to $\mS^2$ in~\eqref{eq:StereoProj}.
 In these variables, the four-point function can be written
\be
\label{eq:Gt}
 G(z) = \frac{\left\langle \phi_1 \phi_2 \phi_3 \phi_4
   \right\rangle}
{\left\langle \phi_1 \phi_3 \right\rangle
\ \left\langle \phi_2 \phi_4 \right\rangle} \; .
\ee
where the dependence on $\theta_{13}$ and $\theta_{24}$ cancels in the ratio.  
While $G(z)$ has the exchange symmetries in the {\em t-channel} ($1
\leftrightarrow 3$, $2 \leftrightarrow 4$), combining this properly with the 
M\"obius map $z \rightarrow 1 -z, z
\rightarrow 1/z$ gives 24 copies,  transforming the blue 
fundamental zone  on the left-hand side of~\figref{fig:G_visualization} to the entire complex z plane. 
In the specific case of the $c=1/2$ minimal model, the explicit form of
$G(z)$ is known in polar coordinates $z = r e^{i\theta}$ as
\bea
G(u,v) = v^\Delta g(u,v) 
&=& \frac{1}{2|z|^{1/4} |1 -z|^{1/4}} \left[ |1 + \sqrt{1 -z}| + |1 - \sqrt{1 -z}| \right]\nn
&=& \frac{1}{2|z|^{1/4} |1 -z|^{1/4}} \sqrt{2 + 2 \sqrt{(1 - z)(1 - \bar z)} + 2 \sqrt{z \bar z}} \;.\label{eq:fourptexact}
\eea
The first form is a sum of the two Virasoro blocks   and the second
form, given in Ref.~\cite{Deng:2003kiw}, 
is the explicitly crossing symmetric form.
For visualization, we plot $G(z)$ in~\figref{fig:G_visualization}.

Given that independent field configurations on our simplicial complexes can be
generated in nearly $\mathcal{O}(L^2)$ time, the dominant cost of computing the
four-point function would na\"ively be computing the product of four fields
which scales as $\mathcal{O}(L^8)$ and would be intractable for large $L$
close to the continuum limit.  We can reduce the cost by only sampling a subset of values of the
four-point function on each independent configuration we generate. This is achieved by drawing
$\mathcal{O}(L^2)$ quartets of random points on the simplicial complex. This balances the
computational cost equally between generating independent field configurations
and sampling the four-point function.

We then determine the product of the four simplicial fields, the unique
coordinates $(\theta_{13}, \theta_{24}, z)$ under the permutations of the
positions, and finally an estimate of $G(z)$ by dividing our product of
simplicial fields by the continuum two-point function,~\eqref{eq:2pt_function_sphere}, which is justified by our success of the
previous section in showing that the simplicial two-point function converges correctly in the continuum limit,
\be
G(z) = \frac{\left\langle \phi_1 \phi_2 \phi_3 \phi_4 \right\rangle}{
  g(\theta_{13}) g(\theta_{24})
} \to \frac{\left\langle \phi_1 \phi_2 \phi_3 \phi_4 \right\rangle}{
  \left\langle \phi_1 \phi_3 \right\rangle
  \left\langle \phi_2 \phi_4 \right\rangle
} \quad \mathrm{as} \quad L \to \infty \; .
\ee
For convenience when studying the four-point function, we bin the data by coordinate on the complex plane. We partition the unit disk $|z| \le 1$ on the complex plane into a radial grid of
$N^2_0$ bins of equal area.  The angular width of each bin is $\Delta\theta = 2 \pi
/ N_0$ and we choose $N_0 \bmod 4 = 0$ so that the bins are centered at $\theta_n = 0,
\pi/2, \pi, 3 \pi/2$.  This arrangement is ideal for using standard discrete
cosine transforms (DCT-I) for computing Fourier coefficients.  For this study
we chose $N_0=64$.

\begin{figure}[ht]
\begin{center}
\includegraphics[width=0.5\textwidth]{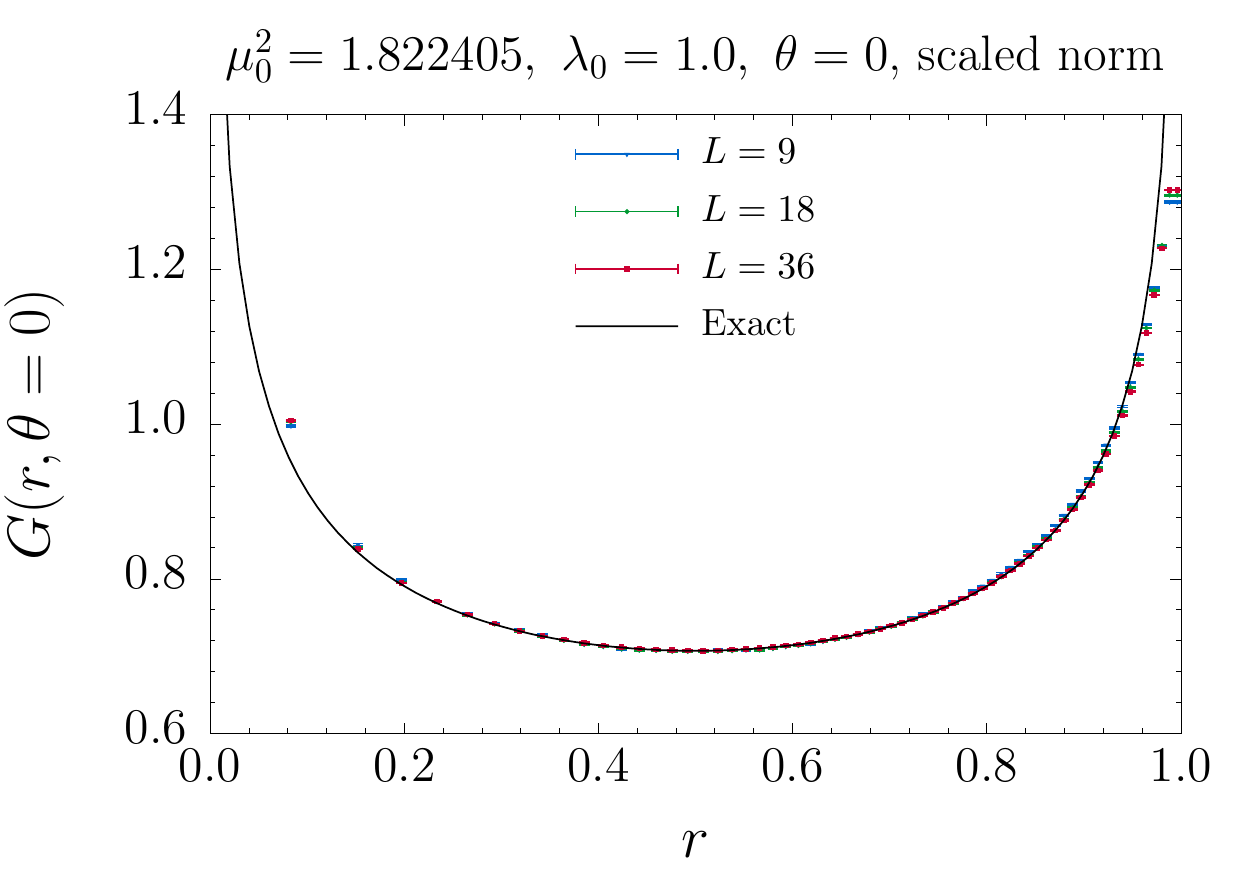}
\end{center}
\caption{\label{fig:G_theta_zero_vs_r}Results for the conformal part of the QFE
scalar four-point function of the Ising universality class on $\mathbb{S}^2$.
$G(r,\theta)$ should only depend on a single complex coordinate
$z=r e^{i\theta}$. We have computed the four-point function in the
entire complex plane in $\phi^4$ theory and show the function for $r \in
[0,1]$ at fixed $\theta=0$ as it converges to the continuum
result. }
\end{figure}

We present a subset of our results for the conformal portion of the
four-point function,~\eqref{eq:fourptexact},
in~\figref{fig:G_theta_zero_vs_r}. Here, we consider the fixed slice
$r \in [0,1]$ for $\theta = 0$. We see that, even for
moderately small values of $L$, the measured four-point function
converges well to the analytic, continuum result.

\subsection{\label{sub:OPE}Operator Product Expansion}

In~\secref{sub:fourpt}, we calculated the simplicial approximation to the
scalar four-point function which, after taking a ratio with a product of two-point
functions respecting $t$-channel symmetry, is described by a conformal
function $G(z)$ invariant under $z \to 1 - z$ and $z \to \bar{z}$.  We computed
the function everywhere in the unit circle, which can then be extended to the whole
complex plane utilizing the permutation symmetry of the full four-point
function. Our representation of the function $G(z)$ is a tabulation of values
on an equal area grid in polar coordinates $(r,\theta)$.

In the \textit{conformal bootstrap}  the  expansion around $z=0$ is used based on the operator
product expansion (OPE).  It is interesting to ask how well our Monte
Carlo simulation on $\mS^2$ can determine the data in an OPE expansion, 
 \begin{equation}
 \label{eq:Gs}
G_s(z) = \frac{\left\langle \phi_1 \phi_2 \phi_3 \phi_4 \right\rangle}{
  g(\theta_{12}) g(\theta_{34})
} = |z|^{\Delta_\sigma} G(z) = \sum_{\Delta_\mathcal{O},\ell} \lambda_\mathcal{O}^2
g_{\Delta_\mathcal{O},\ell}(z) \; ,
 \end{equation}
where $G_s(z)$ is $s$-channel symmetric, \textit{i.e.}\ symmetric under
interchange $1 \leftrightarrow 2$ and $3 \leftrightarrow 4$,
$\Delta_\mathcal{O}$ are the scaling dimensions of conformal primary operators
$\mathcal{O}$, and $\ell$ labels the spin. The functions
$g_{\Delta_\mathcal{O},\ell}(z)$ are called \textit{conformal blocks} whose
functional form is completely determined by conformal symmetry.  In $d=2$,
there is an explicit representation of the conformal blocks in terms of hypergeometric functions,
\begin{eqnarray}
g_{\Delta,\ell}(z) & = & \frac{1}{2} \left[ z^h \, {}_2F_1(h,h;2h;z)\right]
\left[ {\bar{z}}^{\bar{h}} \, {}_2F_1(\bar{h},\bar{h};2\bar{h};\bar{z}) \right] \;, \\*
&& h = \frac{\Delta + \ell}{2} , \quad \bar{h} = \frac{\Delta - \ell}{2} \nonumber \; .
\end{eqnarray}
%
%
The expansion in Fourier modes is given by
\begin{equation}
g_{\Delta,\ell}(r,\theta) = \sum_{m=0}^\infty \cos(m\theta) \left[
r^\Delta \sum_{|n-n^\prime+\ell |=m}
\frac{(h)_n^2 (\bar h)_{n^\prime}^2}{(2 h)_n (2\bar h)_{n^\prime}}
\frac{r^{n+n^\prime}}{n! {n^\prime}!}
\right] \;,
\end{equation}
using the Pochhammer symbol $(a)_n = \Gamma(a+n) /  \Gamma(a)$.

With a conserved energy momentum tensor $T$, there are special terms in the OPE
expansion of the scalar four-point function which have integer powers: $1 +
\lambda_T^2 \, g_{T,2}(r,\theta)$.  The identity operator term is normalized to
unity by convention.  The conformal block for the energy momentum tensor has a
closed form expression,
\begin{eqnarray}
g_{T,2}(z) &=& -3 \left( 1 + \frac{1}{z} \left( 1 - \frac{z}{2} \right) \log(1-z) \right)
\, + \, \mathrm{c.c.} \\*
&=& \frac{1}{2} r^2 \cos(2\theta) + \frac{1}{2} r^3 \cos(3\theta) + \frac{9}{20} r^4 \cos(4\theta)
+ \cdots \nonumber  \; ,
\end{eqnarray}
and the OPE coefficient $\lambda_T^2$ is related to the central charge of the CFT,
\begin{equation}
\lambda_T^2 = \frac{\Delta_\sigma^2 d^2}{C_T (d-1)^2} \to \frac{1}{16 C_T}
\, \mbox{\quad\quad for~} \, d=2\;,
\end{equation}
where $C_T = 2 c = 2 (1/2) = 1$ for the $d=2, c=1/2$ minimal model.

\begin{figure}
\includegraphics[width=0.48\textwidth]{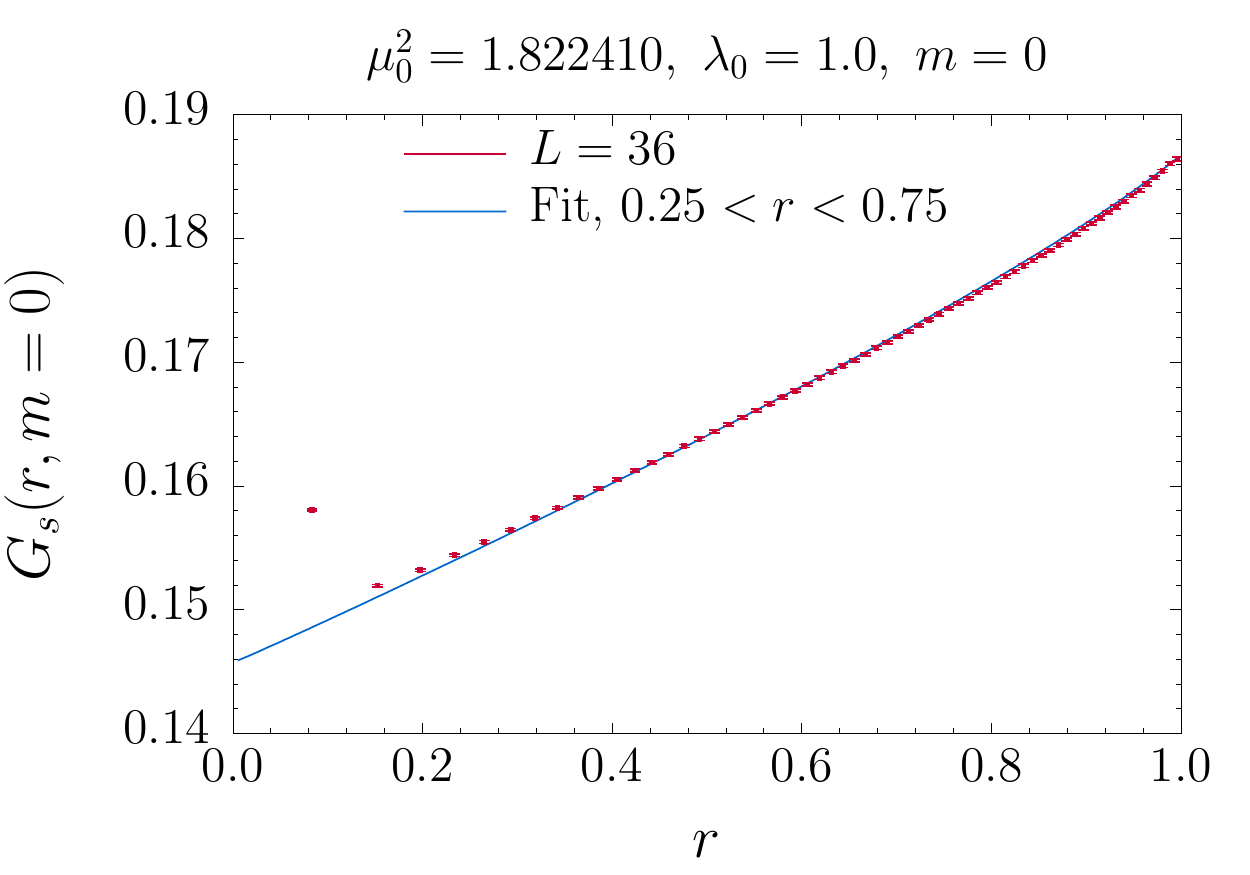}
\includegraphics[width=0.48\textwidth]{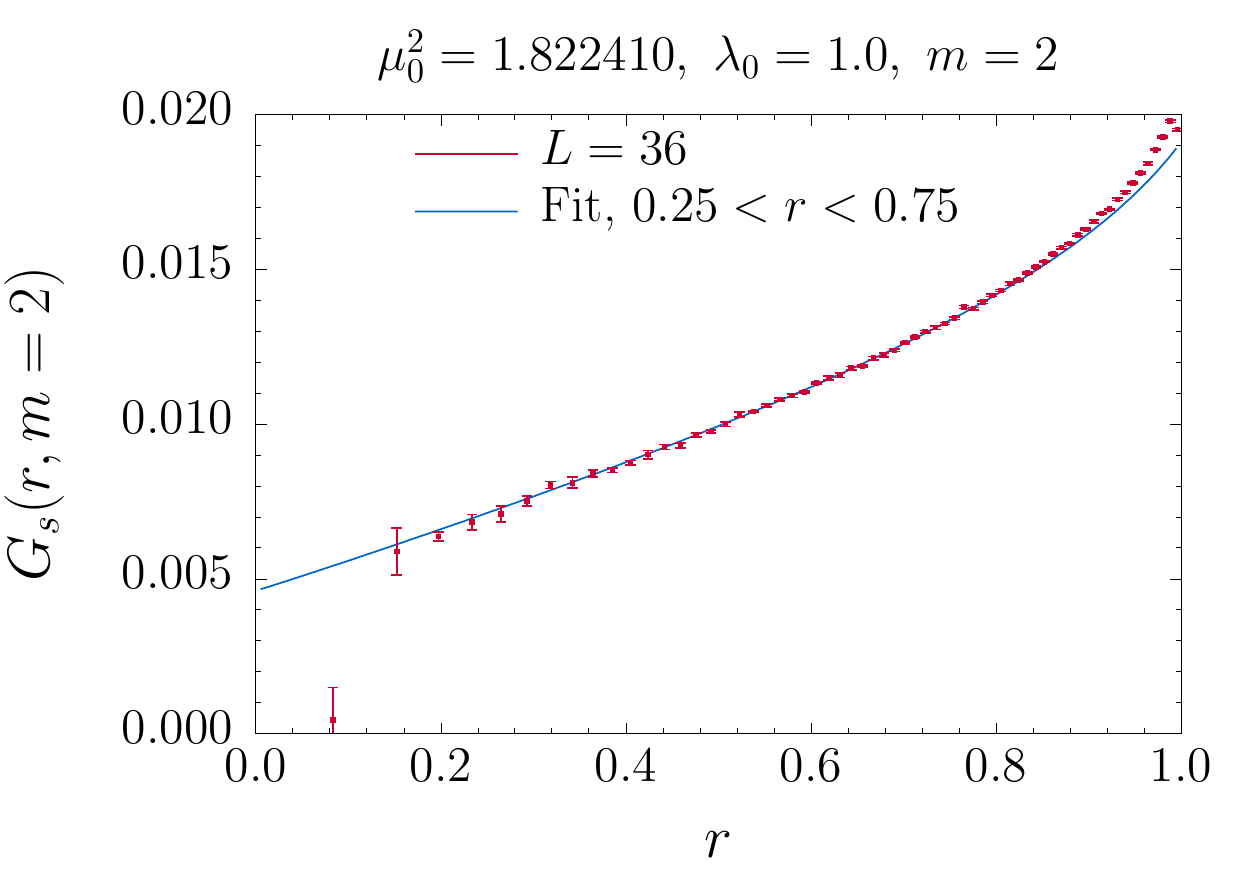}
\caption{\label{fig:Gs_fixed_m_vs_r}On the left, a fit to the
$m=0$ Fourier component of the $s$-channel symmetric conformal part of the
four-point function.  On the right, a fit to the $m=2$ Fourier component.
A simultaneous fit to both components can be used to fix the normalization of
the four-point function and determine the central charge.}
\end{figure}

We can extract an estimate of the central charge $c$ from our simplicial
calculation of the scalar four-point function by modeling the result with the
first few terms in the OPE expansion
\begin{equation}
G_s(r,\theta) \propto 1 + \lambda_\epsilon^2 \, g_{\epsilon,0}(r,\theta)
+ \lambda_T^2 \, g_{T,2}(r,\theta)
\end{equation}
We used a DCT-I to Fourier transform the data, simultaneously fitting $m=0,2$ to
four parameters: the overall normalization, $\lambda_\epsilon^2$,
$\lambda_T^2$ and $\Delta_\epsilon$.  We choose a single value of $\mu^2=1.822410$ which is close to
the pseudocritical value, and we restrict the fitting range in $r$ around 0.5
where the data shows the least discretization error.  A fit for $L=36$ and
$0.25 \le r \le 0.75$ is shown in~\figref{fig:Gs_fixed_m_vs_r}.  A summary
of several fits is shown in~\tabref{tab:Gs_fixed_m_fit}.  As expected, the
fit values converge towards the continuum CFT results
$(\Delta_\epsilon = 1, \lambda_\epsilon^2 =
1/4, c = 1/2)$  as $L$ increases towards the continuum limit and as the fit
range in $r$ is confined to a narrower region around $r=0.5$ where the
discretization artifacts appear smallest.

\setlength{\tabcolsep}{10pt}
\begin{table}
\begin{center}
\begin{tabular}{|c|c|c||c|c|c|c|}
\hline
$\mu^2$ & $s$ & $r_\mathrm{min} \le r \le r_\mathrm{max}$ & norm & $\Delta_\epsilon$ &
  $\lambda_\epsilon^2$ & $c$ \\
\hline\hline
1.82241 & 9 & $0.25 \le r \le 0.75$ & 0.2900 & 1.075 & 0.2536 & 0.4668 \\
1.82241 & 9 & $0.30 \le r \le 0.70$ & 0.2901 & 1.075 & 0.2533 & 0.4704 \\
1.82241 & 9 & $0.35 \le r \le 0.65$ & 0.2902 & 1.077 & 0.2533 & 0.4738 \\
1.82241 & 9 & $0.40 \le r \le 0.60$ & 0.2902 & 1.016 & 0.2427 & 0.4747 \\
\hline
1.82241 & 18 & $0.25 \le r \le 0.75$ & 0.2051 & 1.068 & 0.2563 & 0.4866 \\
1.82241 & 18 & $0.30 \le r \le 0.70$ & 0.2051 & 1.056 & 0.2544 & 0.4878 \\
1.82241 & 18 & $0.35 \le r \le 0.65$ & 0.2051 & 1.050 & 0.2535 & 0.4904 \\
1.82241 & 18 & $0.40 \le r \le 0.60$ & 0.2051 & 1.046 & 0.2526 & 0.4884 \\
\hline
1.82241 & 36 & $0.25 \le r \le 0.75$ & 0.1457 & 1.031 & 0.2528 & 0.4926 \\
1.82241 & 36 & $0.30 \le r \le 0.70$ & 0.1458 & 1.026 & 0.2519 & 0.4932 \\
1.82241 & 36 & $0.35 \le r \le 0.65$ & 0.1458 & 1.018 & 0.2508 & 0.4931 \\
1.82241 & 36 & $0.40 \le r \le 0.60$ & 0.1458 & 1.007 & 0.2486 & 0.4933 \\
\hline
\end{tabular}
\end{center}
\caption{\label{tab:Gs_fixed_m_fit}A summary of several fits to the $m=0,1,2,3$
Fourier components of the $s$-channel symmetric conformal part of the
four-point function. 
}int
\end{table}

\section{\label{sec:conclusion}Conclusion}

We have set up the basic formalism for defining a scalar field theory
on non-trivial Riemann  manifolds and tested it numerically in the
limited context of the \ndim{2} $\phi^4$ theory by comparing it against the $c =1/2$ 
Ising CFT at   the Wilson-Fisher fixed point.  This test is at a
sufficient accuracy compared with the exact result to encourage us
that we are able to define  this strongly coupled quantum field theory
on a curved manifold, in this case $\mS^2$. Faster parallel code could
 push these tests to much higher precision.  While this may be
pursued in future works, our goal here was to sketch a  framework for lattice quantum field theories on
a non-trivial Riemann manifold as a quantum extension of 
FEM which we refer to as Quantum Finite Elements (QFE). Whether
or not this has generally applicability remains, of course, to be
proven.

Our construction first relies on theoretical issues in the
application of FEM that, in spite of the vast literature, are not to our
knowledge proven in enough generality. Nonetheless, the documented
experience with FEM for PDEs does argue that this framework 
likely extends when properly formulated to convergence to the classical limit of
any renormalizable quantum field theory on a smooth Euclidean
Riemann manifold~\cite{Brower:2016vsl}. A rather non-trivial extension of the
classical finite element/DEC method for the 
lattice Dirac Fermion and non-Abelian gauge fields on a simplicial
Riemannian manifold  is provided  in Ref.~\cite{Brower:2016vsl}  and
Ref.~\cite{Christ:1982ck} respectively. The local spin and gauge
invariance on each site is achieved by compact spin connections
and compact gauge links. This we believe provides the basic tools for
classical simplicial constructions.

Second, and most importantly, we show that reaching the continuum
for a quantum FEM Lagrangian requires a modification of the action by 
counter terms to accommodate UV effects.  We conjecture that for a
super-renormalizable theory with a finite number of divergent
diagrams, a new QFE lattice action can be constructed by a natural
generalization of $\phi^4$ theory example presented here that removes
the scheme dependence of the FEM simplicial lattice UV cut-off. The
detailed generalization of this conjecture needs to be investigated
with many more examples and most likely  alternative approaches.

Third, once we have successfully constructed the QFE action with UV
counter terms consistent with the continuum limit to all orders
in perturbation theory, we conjecture that it represents a correct
formulation of the non-perturbative quantum field theory on a curved
Riemann manifold.  This parallels the conventional wisdom for LFTs 
on hypercubic lattices that when lattice symmetries are
protected (Lorentz, chiral, SUSY) only a finite set of
fine tuning parameters are need to reach the continuum theory.
Further research is need to establish the range of applicability of
our QFE proposal. More examples are obviously need to support this
proposal.

There are many more direct extensions of this formulation that we are
currently entertaining.  The first is to apply QFE to the \ndim{3}
$\phi^4$ theory in both radial quantization and the \ndim{3}
projective sphere, $\mS^3$. Our current software relies on a serial
code and the very efficient Brower-Tamayo~\cite{Brower:1989mt}
modification of the cluster algorithm. We are in the process
of developing parallel code that will allow for large lattices and
higher precision, which will also be a necessity in going beyond
2d and scalar fields. Although new software requires substantial
effort, we believe that all of the fundamental data parallel concepts
and advanced algorithms utilized in lattice QCD will still be
applicable here.  We note there are advantages to studying CFTs on spherical $\mS^d$
simplicial lattices. There are {\bf no finite volume
  approximations}. The entire $\mR^d$ is mapped to the sphere.  For
radial quantization the only finite volume effect is the one radial dimension
along the $\mR \times \mS^{d-1}$ cylinder. With periodic boundary
conditions, the finite extent of the cylinder is also an interesting
parameter for the study of CFTs at finite temperature. Radial quantization on $\mR \times \mS^{d-1}$ allows direct
access to the Dilatation operator to compute operator dimensions and
the OPE expansion. Even our \ndim{2} example is worth further
investigation at higher precision, and with an emphasis on new studies
enabled by a fully non-perturbative formulation of the $\phi^4$ theory
on $\mS^2$.  For example, we can consider new ways to compute the renormalization flow
of the central charge from the UV to the IR.  Further research is on
going to hopefully justify this optimism.

\section*{Acknowledgments}
We would like to thank  Steven  Avery, Peter Boyle, Norman Christ,
Luigi Del Debbio, Martin L\"uscher,
Ami Katz, Zuhair Kandker and Matt Walters for valuable
discussions. R.C.B.  and G.T.F. would like to
thank the Aspen Center and Kavali Institutes for their hospitality and
R.C.B. thanks the Galileo Gallilei Institute's CFT workshop and the Higgs Centre for
their hospitality during the completion of this manuscript. G.T.F. and
A.D.G. acknowledges support under contract number DE-SC0014664 and thanks Lawrence Livermore National Laboratory and Lawrence Berkeley National Laboratory for hospitality during the completion of this work. R.C.B. and E.S.W. acknowledge support by DOE grant DE-SC0015845
and  T.R. and C.-I T. in part by the Department of Energy under
contact DE-SC0010010-Task-A. T.R. is also supported by the University of Kansas Foundation Distinguished Professor Grant of Christophe Royon.

\bibliographystyle{unsrt}
\bibliography{bib/QFE}

\begin{thebibliography}{10}

\bibitem{Appelquist:2013sia}
Thomas Appelquist, Richard Brower, Simon Catterall, George Fleming, Joel Giedt,
  Anna Hasenfratz, Julius Kuti, Ethan Neil, and David Schaich.
\newblock {Lattice Gauge Theories at the Energy Frontier}.
\newblock In {\em {Community Summer Study 2013: Snowmass on the Mississippi
  (CSS2013) Minneapolis, MN, USA, July 29-August 6, 2013}}, 2013.

\bibitem{Brower:2012vg}
R.C. Brower, G.T. Fleming, and H.~Neuberger.
\newblock {Lattice Radial Quantization: 3D Ising}.
\newblock {\em Phys.Lett.}, B721:299--305, 2013.

\bibitem{Brower:2012zd}
Richard Brower, Claudio Rebbi, and David Schaich.
\newblock {Hybrid Monte Carlo simulation on the graphene hexagonal lattice}.
\newblock {\em PoS}, LATTICE2011:056, 2011.

\bibitem{Appelquist:2016viq}
Thomas Appelquist, Richard~C. Brower, George~T. Fleming, Anna Hasenfratz,
  Xiao-Yong Jin, Joe Kiskis, Ethan~T. Neil, James~C. Osborn, Claudio Rebbi,
  Enrico Rinaldi, David Schaich, Pavlos Vranas, Evan Weinberg, and Oliver
  Witzel.
\newblock {Strongly interacting dynamics and the search for new physics at the
  LHC}.
\newblock {\em Phys. Rev.}, D93(11):114514, 2016.

\bibitem{Kuti:2015awa}
Julius Kuti.
\newblock {The Higgs particle and the lattice}.
\newblock {\em PoS}, KMI2013:002, 2015.

\bibitem{Luscher:1982wf}
M.~Luscher.
\newblock {Dimensional Regularization in the Presence of Large Background
  Fields}.
\newblock {\em Annals Phys.}, 142:359, 1982.

\bibitem{Jack:1983sk}
I.~Jack and H.~Osborn.
\newblock {Background Field Calculations in Curved Space-time. 1. General
  Formalism and Application to Scalar Fields}.
\newblock {\em Nucl. Phys.}, B234:331--364, 1984.

\bibitem{Jack:1984vj}
I.~Jack and H.~Osborn.
\newblock {General Background Field Calculations With Fermion Fields}.
\newblock {\em Nucl. Phys.}, B249:472--506, 1985.

\bibitem{Buchbinder:1989zz}
I.~L. Buchbinder, S.~D. Odintsov, and I.~L. Shapiro.
\newblock {Renormalization Group Approach to Quantum Field Theory in Curved
  Space-time}.
\newblock {\em Riv. Nuovo Cim.}, 12N10:1--112, 1989.

\bibitem{Brower:2012mn}
Richard~C. Brower, George~T. Fleming, and Herbert Neuberger.
\newblock {Radial Quantization for Conformal Field Theories on the Lattice}.
\newblock {\em PoS}, LATTICE2012:061, 2012.

\bibitem{Brower:2014daa}
Richard~C. Brower, Michael Cheng, and George~T. Fleming.
\newblock {Improved Lattice Radial Quantization}.
\newblock {\em PoS}, LATTICE2013:335, 2014.

\bibitem{Regge:1961px}
T.~Regge.
\newblock {GENERAL RELATIVITY WITHOUT COORDINATES}.
\newblock {\em Nuovo Cim.}, 19:558--571, 1961.

\bibitem{StrangFix200805}
Gilbert Strang and George Fix.
\newblock {\em An Analysis of the Finite Element Method 2nd Edition}.
\newblock Wellesley-Cambridge, 2nd edition, 5 2008.

\bibitem{2005math8341D}
M.~{Desbrun}, A.~N. {Hirani}, M.~{Leok}, and J.~E. {Marsden}.
\newblock {Discrete Exterior Calculus}.
\newblock {\em ArXiv Mathematics e-prints}, August 2005.

\bibitem{Arnold2006}
Douglas~N. Arnold, Richard~S. Falk, and Ragnar Winther.
\newblock Finite element exterior calculus, homological techniques, and
  applications.
\newblock {\em Acta Numerica}, 15:1, may 2006.

\bibitem{Miller:2013gy}
Warner~A. Miller, Jonathan~R. McDonald, Paul~M. Alsing, David Gu, and
  Shing-Tung Yau.
\newblock {Simplicial Ricci Flow}.
\newblock {\em Commun. Math. Phys.}, 329:579--608, 2014.

\bibitem{Crane:2018}
Keenan Crane.
\newblock {\em Discrete Differential Geometry: An Applied Introduction}.
\newblock 2018.

\bibitem{Rychkov:2016iqz}
Slava Rychkov.
\newblock {EPFL Lectures on Conformal Field Theory in $D \ge 3$ Dimensions}.
\newblock 2016.

\bibitem{Brower:2016vsl}
Richard~C. Brower, Evan~S. Weinberg, George~T. Fleming, Andrew~D. Gasbarro,
  Timothy~G. Raben, and Chung-I Tan.
\newblock {Lattice Dirac Fermions on a Simplicial Riemannian Manifold}.
\newblock {\em Phys. Rev.}, D95(11):114510, 2017.

\bibitem{spanier2012algebraic}
E.H. Spanier.
\newblock {\em Algebraic Topology}.
\newblock Springer New York, 2012.

\bibitem{Christ:1982ci}
N.~H. Christ, R.~Friedberg, and T.~D. Lee.
\newblock {Weights of Links and Plaquettes in a Random Lattice}.
\newblock {\em Nucl. Phys.}, B210:337, 1982.

\bibitem{Christ:1982ck}
N.~H. Christ, R.~Friedberg, and T.~D. Lee.
\newblock {GAUGE THEORY ON A RANDOM LATTICE}.
\newblock {\em Nucl. Phys.}, B210:310, 1982.

\bibitem{Christ:1982zq}
N.H. Christ, R.~Friedberg, and T.D. Lee.
\newblock {Random Lattice Field Theory: General Formulation}.
\newblock {\em Nucl.Phys.}, B202:89, 1982.

\bibitem{1802.04506}
Mamdouh~S. Mohamed, Anil~N. Hirani, and Ravi Samtaney.
\newblock Numerical convergence of discrete exterior calculus on arbitrary
  surface meshes, 2018.

\bibitem{Carroll:1997ar}
Sean~M. Carroll.
\newblock {Lecture notes on general relativity}.
\newblock 1997.

\bibitem{Sen:2000ez}
Samik Sen, Siddhartha Sen, James~C. Sexton, and David~H. Adams.
\newblock {A Geometric discretization scheme applied to the Abelian
  Chern-Simons theory}.
\newblock {\em Phys. Rev.}, E61:3174--3185, 2000.

\bibitem{Delaunay:1934}
Boris~N. Delaunay.
\newblock {Sur la sph{\`e}re vide. A la m{\'e}moire de Georges Vorono{\"i}}.
\newblock {\em Bull. Acad. Sci. URSS, VII Ser.}, 1934:793--800, 1934.

\bibitem{Brower:2016moq}
Richard~C. Brower, George Fleming, Andrew Gasbarro, Timothy Raben, Chung-I Tan,
  and Evan Weinberg.
\newblock {Quantum Finite Elements for Lattice Field Theory}.
\newblock In {\em {Proceedings, 33rd International Symposium on Lattice Field
  Theory (Lattice 2015)}}, 2016.

\bibitem{Rothe:1992nt}
H.~J. Rothe.
\newblock {Lattice gauge theories: An Introduction}.
\newblock {\em World Sci. Lect. Notes Phys.}, 43:1--381, 1992.
\newblock [World Sci. Lect. Notes Phys.82,1(2012)].

\bibitem{Reisz:1987da}
Thomas Reisz.
\newblock {A Power Counting Theorem for Feynman Integrals on the Lattice}.
\newblock {\em Commun. Math. Phys.}, 116:81, 1988.

\bibitem{Reisz:1987px}
T.~Reisz.
\newblock {Renormalization of Feynman Integrals on the Lattice}.
\newblock {\em Commun. Math. Phys.}, 117:79, 1988.

\bibitem{Brower:1989mt}
R.C. Brower and P.~Tamayo.
\newblock {Embedded Dynamics for $\phi^4$ Theory}.
\newblock {\em Phys.Rev.Lett.}, 62:1087--1090, 1989.

\bibitem{Luscher:2010iy}
Martin Lüscher.
\newblock {Properties and uses of the Wilson flow in lattice QCD}.
\newblock {\em JHEP}, 08:071, 2010.
\newblock [Erratum: JHEP03,092(2014)].

\bibitem{Hasenfratz:1980kn}
Anna Hasenfratz and Peter Hasenfratz.
\newblock {The Connection Between the Lambda Parameters of Lattice and
  Continuum QCD}.
\newblock {\em Phys. Lett.}, 93B:165, 1980.
\newblock [,241(1980)].

\bibitem{Mon:1997}
K.K. Mon.
\newblock {Higher-order-magnetization-cumulant universality of the
  two-dimensional Ising model}.
\newblock {\em Phys.Rev.}, B55:38--40, 1997.

\bibitem{Deng:2003kiw}
Youjin Deng and Henk W.~J. Bl{\"o}te.
\newblock {Conformal invariance and the Ising model on a spheroid}.
\newblock {\em Phys. Rev.}, E67(3):036107, 2003.

\bibitem{Kadanoff:1969}
Leo~P. Kadanoff.
\newblock {Correlations along a Line in the Two-Dimensional Ising Model}.
\newblock {\em Phys.Rev.}, 188:859--863, 1969.

\bibitem{Luther:1975wr}
A.~Luther and I.~Peschel.
\newblock {Calculation of critical exponents in two-dimensions from quantum
  field theory in one-dimension}.
\newblock {\em Phys.Rev.}, B12:3908--3917, 1975.

\bibitem{Polyakov:1984yq}
Alexander~M. Polyakov, A.A. Belavin, and A.B. Zamolodchikov.
\newblock {Infinite Conformal Symmetry of Critical Fluctuations in
  Two-Dimensions}.
\newblock {\em J.Statist.Phys.}, 34:763, 1984.

\bibitem{Belavin:1984vu}
A.A. Belavin, Alexander~M. Polyakov, and A.B. Zamolodchikov.
\newblock {Infinite Conformal Symmetry in Two-Dimensional Quantum Field
  Theory}.
\newblock {\em Nucl.Phys.}, B241:333--380, 1984.

\bibitem{Dotsenko:1984nm}
V.S. Dotsenko and V.A. Fateev.
\newblock {Conformal Algebra and Multipoint Correlation Functions in
  Two-Dimensional Statistical Models}.
\newblock {\em Nucl.Phys.}, B240:312, 1984.

\bibitem{Burkhardt:1987}
Theodore~W. Burkhardt and Ihnsouk Guim.
\newblock {Bulk, surface, and interface properties of the Ising model and
  conformal invariance}.
\newblock {\em Phys.Rev.}, B36:2080--2083, 1987.

\bibitem{Voronoi:1908}
Georges Vorono{\"i}.
\newblock {Nouvelles applications des param{\`e}tr�s continus {\`a} la
  th{\'e}prie des formes quadratiques}.
\newblock {\em J.Reine Angew.Math.}, 133:97--102, 1908.

\bibitem{Blote:1995zik}
H.~W.~J. Bl{\"o}te, E.~Luijten, and J.~R. Heringa.
\newblock {Ising universality in three dimensions: a Monte Carlo study}.
\newblock {\em J. Phys.}, A28(22):6289--6313, 1Deng:2003kiw995.

\bibitem{Wolff:1988uh}
Ulli Wolff.
\newblock {Collective Monte Carlo Updating for Spin Systems}.
\newblock {\em Phys.Rev.Lett.}, 62:361, 1989.

\bibitem{Metropolis:1953am}
N.~Metropolis, A.W. Rosenbluth, M.N. Rosenbluth, A.H. Teller, and E.~Teller.
\newblock {Equation of state calculations by fast computing machines}.
\newblock {\em J.Chem.Phys.}, 21:1087--1092, 1953.

\bibitem{Adler:1981sn}
Stephen~L. Adler.
\newblock {An Overrelaxation Method for the Monte Carlo Evaluation of the
  Partition Function for Multiquadratic Actions}.
\newblock {\em Phys.Rev.}, D23:2901, 1981.

\bibitem{Whitmer:1984he}
C.~Whitmer.
\newblock {OVERRELAXATION METHODS FOR MONTE CARLO SIMULATIONS OF QUADRATIC AND
  MULTIQUADRATIC ACTIONS}.
\newblock {\em Phys.Rev.}, D29:306--311, 1984.

\bibitem{Brown:1987rra}
Frank~R. Brown and Thomas~J. Woch.
\newblock {Overrelaxed Heat Bath and Metropolis Algorithms for Accelerating
  Pure Gauge Monte Carlo Calculations}.
\newblock {\em Phys.Rev.Lett.}, 58:2394, 1987.

\bibitem{Creutz:1987xi}
Michael Creutz.
\newblock {Overrelaxation and Monte Carlo Simulation}.
\newblock {\em Phys.Rev.}, D36:515, 1987.

\bibitem{Binder:2001ha}
K.~Binder and E.~Luijten.
\newblock {Monte Carlo tests of renormalization group predictions for critical
  phenomena in Ising models}.
\newblock {\em Phys.Rept.}, 344:179--253, 2001.

\bibitem{Christe:1993ij}
Philippe Christe and Malte Henkel.
\newblock {Introduction to Conformal Invariance and its Applications to
  Critical Phenomena}.
\newblock {\em Lect. Notes Phys.}, M16:1--260, 1993.

\end{thebibliography}

\appendix

\section{\label{app:FreeTheory}Free Theory on Sphere and Cylinder }

Here we collect a few analytic expressions for the free
scalar theory that are useful in establishing convergence of the
simplicial propagators to the exact continuum on the $\mS^2$ and
$\mR \times \mS^2$ manifolds.

\paragraph{Free scalar on a sphere:}
The Green's function on  the sphere is given 
by spectral sum,
\be
G(\theta,\mu)  = \sum^\infty_{l =0} \sum^{l}_{m=-l} \frac{ Y^*_{lm}(\hat r_1)
  Y_{lm} (\hat r_2)}{l(l +1) + 1/4 + \mu^2} 
= \frac{1}{4 \pi}\sum^\infty_{l =0} \frac{ (2 l + 1) P_l(cos\theta)}{(l+1/2)^2 + \mu^2} \, ,
\ee
where $\hat r_1\cdot \hat r_2 = \cos(\theta_{12}) = z $.
For generality we have added  a  dimensionless ``mass'' term
$m^2_0 = \mu^2+ 1/4$ to regulate the IR. The series can also be summed 
 to get
\be
G(\theta,\mu) =\frac{1}{4 \pi} [Q_{-1/2 + i \mu}(z) +Q_{-1/2 - i
  \mu}(z) ] \; .
\label{eq:S2Greens}
\ee
in term of associated Legendre functions. This Green's function
(\ref{eq:S2Greens})  can also be written in compact integral representation,
\be
 G(\theta,\mu) = \frac{1}{4 \pi} \int^\infty_{- \infty} dt e^{ \textstyle i \mu
  t} G(t, \theta)  = \frac{1}{2 \pi} \int^\infty_{0}\frac{\cos(\mu t)  dt }{\sqrt{2
    \cosh(t)- 2 \cos(\theta) }} \; ,
\ee
exhibiting clearly the logarithmic singularity
as $1 -  z \rightarrow \theta^2/2$. Expanding the integral  we find, 
\be
G(\theta,\mu) \simeq \frac{1}{4 \pi} [ \log[\frac{32}{1-
  \cos(\theta)}]- 14 \zeta(3)\mu^2] + O(\theta^2, \mu^4) \; ,
\ee
in agreement with~\eqref{eq:contprop2d} in \secref{sec:univlogdiv}.

\paragraph{Conformal propagator on a  cylinder:} 
Next, we consider the 3-d conformal propagator for  radial
quantization on $\mR \times \mS^2$,
\bea
G(t, \theta) &=&\<\phi(x_1) \phi(x_2)\> =  (r_1
r_2)^{\Delta}\frac{1}{|x_1 - x_2|^{2\Delta} } \nn
&=& \frac{1}{  [r_1/r_2 + r_2/r_1 - 2 \cos(\theta)]^{\Delta}} 
=\frac{1}{[2 \cosh(t) - 2 \cos(\theta) ]^{\Delta}}  \; ,
\label{eq:ConformalProp}
\eea
where $t =t_2 - t_1 =  \log(r_2/r_1)$.  
Note by setting $t =0$, this is exactly the conformal two point function on the
sphere~\eqref{eq:2pt_function_sphere} as expected by dimensional
reduction.  (In the rest of the appendix, we  also
drop the factor of $1/(4 \pi)$ to coincide with
the normalization convention in~\secref{sec:numerical_correlators}.)

However, if we return to the \ndim{3} free  theory, setting $\Delta = (d - 2)/2
= 1/2,$ the dimensional reduction  by setting $t = 0$ in the
\ndim{3} conformal propagator fails to give the \ndim{2} propagator. To understand this 
we recall that in 2d, the scalar field, $\phi(x)$ is not a primary operator. As
is well known in string theory, conformal primaries are given by
vertex operators $:\exp[i k \phi]:$.   Instead, to reach the  free 
Green's function from the cylinder, we proceed as follows. The multipole expansion of the Coulomb term is
\be
\frac{1}{|x_1 - x_2|} =  \frac{1}{\sqrt{r_1 r_2}} \sum_l (r_1/r_2)^{l + 1/2}  P_l(\cos
\theta) 
\ee
for $r_2 > r_1$ or  $r_2 > r_1$ ,  after interchanging $r_1$ and
$r_2$.   Thus the Green's function is
\be
G(t, \theta) = \sum_l [\theta(t)  e^{\textstyle -(l + 1/2)t } + \theta(-t) e^{\textstyle (l + 1/2)t } ]P_l(\cos
\theta) 
\ee
rescaling by  $\sqrt{r_1 r_2}$ as in~\eqref{eq:ConformalProp} above.
In Fourier space this becomes
\bea
\widetilde G(\omega, \theta)  = \int^\infty_{- \infty} dt e^{ \textstyle i \omega
  t} G(t, \theta) 
 &=& \int^\infty_0 dt e^{ \textstyle i \omega
  t} \sum_l e^{\textstyle -(l + 1/2)t } P_l(\cos \theta) \nn
&+&  \int^0_{- \infty} dt e^{  \textstyle i \omega
  t}  \sum_l e^{\textstyle (l + 1/2)t } P_l(\cos \theta)\; ,
\eea
so
\bea
\widetilde G(\omega, \theta)  &=& \sum^\infty_{l=0} [\frac{1}{ l + 1/2 +i \omega} +
\frac{1}{ l + 1/2 - i \omega}]  P_l(\cos \theta)  \nn
&=&  \sum_l \frac{(2 l + 1) P_l(\cos \theta) }{ (l + 1/2)^2 + \omega^2} \; .
\eea
By fixing the frequency $\omega^2 = \mu^2$ we regain the
free propagator on the sphere. We take the value $\mu = 0$ as our standard
IR regulated propagator which has an appealing interpretation as 
constant line source in 3d.   Of course in practice on the lattice  for
our \ndim{3} simulations, we  will introduce a finite cylinder of
length $T = a L_t$, with periodic boundary conditions.
This corresponds to the thermal propagators, with periodic $t \in [0,T]$ represented by a discrete sum,
\be
G(t, \theta) =  \int \frac{d\omega}{2 \pi} e^{ \textstyle - i \omega
  t}\widetilde
G(\omega,\theta)
\rightarrow \frac{1}{T}\sum_n e^{ \textstyle - i \omega_n
  t}\widetilde
G(\omega_n,\theta)
\ee
over frequencies, $\omega_n = 2 \pi n/T$.


\section{\label{app:cumulants}Finite Size Scaling for Cumulants and Moments}

This follows the renormalization group arguments first presented in
\cite{Blote:1995zik} and reviewed in \cite{Binder:2001ha}. We begin
with the  free energy~\eqref{eq:FreeEnergy},
\be
\mathcal{F}\left(g_\sigma, g_\epsilon, \{g_\omega, \dots\}, a\right) =
\log Z \quad , \quad
Z = \int \left[ D\phi \right] \;  e^{\textstyle -S + h \int d^dx 
  \phi(x)}\;
\ee
expressed not as a function of the bare couplings $h,
\mu^2_0, \lambda_0$ but rather the relevant couplings $g_\sigma, g_\lambda$ and the
irrelevant couplings $\left\{ g_\omega, \dots \right\}$ of the Wilson-Fisher
fixed point.  For the lattice spacing,  we assume $a \sim 1/L$.
Under a change of scale by a factor $\ell$, the free energy renormalizes
\be
\mathcal{F}\left(g_\sigma, g_\epsilon, \{g_\omega, \dots\}, a\right)
= \ell^{-d} F\left( \ell^{y_\sigma} g_\sigma, \ell^{y_\epsilon} g_\epsilon,
\{\ell^{y_\omega} g_\omega, \dots\}, \ell a \right)
+ G\left( g_\sigma, g_\epsilon \right)
\ee
where $y_\mathcal{O} \equiv d - \Delta_\mathcal{O}$, $F$ is the singular, or scaling, part
of the free energy and $G$ is the regular part.  In what follows, we will continue to use
$y_\mathcal{O}$ instead of $\Delta_\mathcal{O}$ for compactness of notation.

If we choose ${\mathbb  L}  \equiv \sqrt[d]{\Omega_d}$, where $\Omega_d$ is the volume
of the $d$-dimensional manifold, and differentiate the free energy $k$ times with respect
to $g_\sigma$ and then take the limit $g_\sigma \to 0$ we get
\be
\label{eq:RG_moments}
\frac{\partial^k \mathcal{F}}{\partial g_\sigma^k}
= \mathcal{F}^{(k)}\left( g_\epsilon, \{g_\omega, \dots\}, 1/{\mathbb L} \right) =
{\mathbb L}^{k y_\sigma - d} F^{(k)}\left( {\mathbb L}^{y_\epsilon} g_\epsilon,
\{{\mathbb L}^{y_\omega} g_\omega, \dots\}, 1 \right) +
G^{(k)}(g_\epsilon) \; .
\ee

We compute derivatives of the free energy with respect to the bare parameter
$h$ and then take the limit $h \to 0$, to  get cumulants of the
magnetization
\be
\label{eq:FE_cumulants}
\kappa_k = \lim_{h \to 0} \frac{1}{\Omega_d^{k-1}}
\frac{\partial^k \mathcal{F}}{\partial h^k} \; ,
\ee
and put  them in the form of Binder cumulants,
\be
U_{2n} \propto \lim_{h \to 0} \frac{\kappa_{2n}}{\kappa_2^n}  \; ,
\ee
with the normalization to be determined as described in Sec.~\ref{sub:moments}.
We note that the symmetry $\phi \to - \phi$ guarantees that all odd cumulants
vanish, $\kappa_{2n+1} = 0$ and that $h$ is an odd-function of $g_\sigma$.
To take advantage of the RG scaling properties of the free energy near the
critical point, we must rewrite the cumulants carefully using the chain rule.
To make the following manageable, we define $\alpha_{2n+1}$
and understand that all expressions are in the limit $g_\sigma, h \to 0$, 
\begin{eqnarray}
\alpha_{2n+1} & \equiv & \lim_{h \to 0} \frac{\partial^{2n+1}
                         g_\sigma}{\partial h^{2n+1}}  \; ,\\
\label{eq:kappa2}   
{\mathbb L}^{2 \Delta_\sigma} \kappa_2 & = & \alpha_1^2 \left[ F^{(2)} + {\mathbb L}^{-d+2\Delta_\sigma} G^{(2)} \right] \; ,\\
\label{eq:kappa4}
{\mathbb L}^{4 \Delta_\sigma} \kappa_4 & = & \alpha_1^4 \left[ F^{(4)} + {\mathbb L}^{-3d+4\Delta_\sigma} G^{(4)} \right]
  + \alpha_1 \alpha_3 \left[ {\mathbb L}^{-2d+2\Delta_\sigma} F^{(2)} + {\mathbb L}^{-3d+4\Delta_\sigma} G^{(2)} \right] \; ,\\
\label{eq:kappa6}
{\mathbb L}^{6 \Delta_\sigma} \kappa_6 & = & \alpha_1^6 \left[ F^{(6)} + {\mathbb L}^{-5d+6\Delta_\sigma} G^{(6)} \right]
  + \alpha_1^3 \alpha_3 \left[ {\mathbb L}^{-2d+2\Delta_\sigma} F^{(4)} + {\mathbb L}^{-5d+6\Delta_\sigma} G^{(4)} \right]
  \nonumber \\*
  && + ( \alpha_3^2 + 3 \alpha_1 \alpha_5)
  \left[ {\mathbb L}^{-4d+4\Delta_\sigma} F^{(2)} + {\mathbb
     L}^{-5d+6\Delta_\sigma} G^{(2)} \right] \\
\label{eq:kappa8}
{\mathbb L}^{8 \Delta_\sigma} \kappa_8 & = &  \alpha_1^8 \left[ F^{(8)} + {\mathbb L}^{-7d+8\Delta_\sigma} G^{(8)} \right]
  + 56 \alpha_1^5 \alpha_3
  \left[ {\mathbb L}^{-2d+2\Delta_\sigma} F^{(6)} + {\mathbb
                                             L}^{-7d+8\Delta_\sigma}
                                             G^{(6)} \right] \; ,
  \nonumber  \\*
  && + ( 280 \alpha_1^2 \alpha_3^2 + 56 \alpha_1^3 \alpha_5 )
  \left[ {\mathbb L}^{-4d+4\Delta_\sigma} F^{(4)} + {\mathbb L}^{-7d+8\Delta_\sigma} G^{(4)} \right] \nonumber \\*
  && + ( 56 \alpha_3 \alpha_5 + 8 \alpha_1 \alpha_7 )
  \left[ {\mathbb L}^{-6d+6\Delta_\sigma} F^{(2)} + {\mathbb
     L}^{-7d+8\Delta_\sigma} G^{(2)} \right] \; 
\end{eqnarray}
At this point, the pattern is clear and the only difficulty extending the calculation to higher cumulants
is  computing the associated chain rule factors which are relatively straight-forward to compute
using a program like \textit{Mathematica}.  For example,  \texttt{D[f[g[h]],\{h,8\}]} will give all the terms
needed for $\partial^8 \mathcal{F} / \partial h^8$, remembering to set to zero odd derivatives of
$\mathcal{F}$ and even derivatives of $g$.

Before we continue, we point out an important physical concept.  At
finite ${\mathbb L}$, the difference between the lattice scalar
operator $\phi$ and the conformal primary operator $\sigma$ is
indicated by the presence of higher derivative terms $\alpha_{2n+1}$
in the cumulant expressions.  But, in the scaling limit
${\mathbb L} \to \infty$ all those terms vanish and the cumulants are
dominated by the moments of the free energy with respect to the coupling
$g_\sigma$ of the conformal primary operator $\sigma$.  In this sense,
the simplicial operator $\phi(x)$ becomes the primary operator
$\sigma(x)$ in the scaling limit.

We can now apply the important physical principle that close to the critical
point we can expand the derivatives of the free energy in a Taylor series
\begin{eqnarray}
\label{eq:RG_moment_F}
F^{(k)} & = &  a_{k0} + a_{k1} \left( g_\epsilon - g_\epsilon^* \right) {\mathbb L}^{y_\epsilon}
  + \cdots
  + b_{k1} \left( g_\omega - g_\omega^* \right) {\mathbb L}^{y_\omega}
  + \cdots. \\
\label{eq:RG_moment_G}
G^{(k)} & = & c_{k0} + c_{k1} \left( g_\epsilon - g_\epsilon^* \right) + \cdots
\end{eqnarray}
Substituting these expressions into those for the cumulants defined in~\eqrefrange{eq:kappa2}{eq:kappa8}
nearly gives us an expression we can fit to computed data.
Of course, we cannot directly vary the renormalized couplings $(g_\sigma,
g_\epsilon, g_\omega, \cdots)$ but instead can vary our bare couplings $(h,
\mu^2_0, \lambda_0)$ defined in our cutoff theory.  
Near the Wilson-Fisher fixed point we can relate the two sets of couplings by expanding 
\bea
\left( \begin{array}{c}
g_\epsilon - g_\epsilon^* \\
g_\omega - g_\omega^*
\end{array} \right) & = & \left( \begin{array}{cc}
R_{\epsilon\mu} & R_{\epsilon\lambda} \\
R_{\omega\mu} & R_{\omega\lambda}
\end{array} \right) \left( \begin{array}{c}
\mu^2_0 - \mu_*^2 \\
\lambda_0 - \lambda_*
\end{array} \right) + \cdots
\eea
So, substituting~\eqref{eq:coord_change} first into~\eqrefrange{eq:RG_moment_F}{eq:RG_moment_G} leads to expressions that can be used to model cumulant data
computed in a cutoff theory close to the Wilson-Fisher fixed point.

Often it is more convenient in the cutoff theory to compute moments instead of cumulants.
The expansion of moments in terms of cumulants is well known so we only write
the first few even moments here
\begin{eqnarray}
m_2    & = & \kappa_2 \\
m_4    & = & \kappa_4 + 3 \kappa_2^2 \\
m_6    & = & \kappa_6 + 15 \kappa_4 \kappa_2 + 15 \kappa_2^3 \\
m_8    & = & \kappa_8 + 28 \kappa_6 \kappa_2 + 35 \kappa_4^2
             + 210 \kappa_4 \kappa_2^2 + 105 \kappa_2^4 \\
m_{10} & = & \kappa_{10} + 45 \kappa_8 \kappa_2 + 210 \kappa_6 \kappa_4
             + 630 \kappa_6 \kappa_2^2 + 1575 \kappa_4^2 \kappa_2
             + 3150 \kappa_4 \kappa_2^3 + 945 \kappa_2^5 \\
m_{12} & = & \kappa_{12} + 66 \kappa_{10} \kappa_2 + 495 \kappa_8 \kappa_4
             + 1485 \kappa_8 \kappa_2^2 + 462 \kappa_6^2
             + 13860 \kappa_6 \kappa_4 \kappa_2 
             \nonumber \\*
       &   & + 13860 \kappa_6 \kappa_2^3 + 5775 \kappa_4^3 + 51975 \kappa_4^2 \kappa_2^2
             + 51975 \kappa_4 \kappa_2^4 + 10395 \kappa_2^6\;.
\end{eqnarray}

\end{document}